\documentclass[reprint,amssymb, amsmath,pof]{revtex4-2}
\usepackage{graphicx}
\graphicspath{{figures/}}
\usepackage{dcolumn, bm}
\usepackage[svgnames]{xcolor}

\usepackage[colorlinks=true,linkcolor=DarkBlue,urlcolor=DarkBlue,citecolor=FireBrick]{hyperref}
\usepackage{hyphenat}
\usepackage[caption=false]{subfig}
\usepackage{bbding}  %
\usepackage[final, protrusion=true,expansion=true]{microtype}

\begin{document}
\title{Experimental and Computational Investigation of a Fractal Grid Wake}%

\author{A. Fuchs}
\author{W. Medjroubi}
\author{H. Hochstein}
\author{G. G\"ulker}
\author{J. Peinke}

\affiliation{Institute of Physics and ForWind, University of Oldenburg, Küpkersweg 70, 26129 Oldenburg, Germany}

\date{\today}

\begin{abstract}
Fractal grids generate turbulence by exciting many length scales of different sizes simultaneously rather than using the nonlinear cascade mechanism to obtain multi-scale structures, as it is the case for regular grids. The interest in these grids has been further building up since the surprising findings stemming from the experimental and computational studies conducted on these grids. This work presents experimental wind tunnel and computational fluid dynamics (CFD) studies of the turbulent flow generated by a space-filling fractal square grid. The experimental work includes Particle Image Velocimetry (PIV) and hot-wire measurements. In addition, Delayed Detached Eddy Simulations (DDES) with a Spalart-Allmaras background turbulence model are conducted using the open-source package OpenFOAM. This is the first time DDES simulations are used to simulate and characterize the turbulent flow generated by a fractal grid. Finally, this article reports on the extensive statistical study and the direct comparison between the experimentally and numerically acquired time series to investigate and compare one-point- and two-point statistics. Our goal is to validate our computational results and provide enhanced insight into the complexity of the multi-scale generation of turbulence using a fractal grid with a low number of fractal iterations. In particular, we investigate the different turbulent structures and their complex interaction in the near-grid region or the production regime of the fractal grid flow.
\end{abstract}

\maketitle

\section{Introduction}
\label{intro}
Grid-generated turbulence has been investigated for more than 70 years, being the most common method to create turbulence in wind tunnels under controlled conditions. The turbulence generated by the so-called \textit{regular} grids is nearly homogeneous and isotropic downstream from the grid \cite{Batchelor_1948_final, Batchelor_1948_initial}. Nevertheless, the Reynolds numbers and the turbulent length scales obtained are quite low ($Re_{\lambda} =u' \lambda / \nu < 100$, where $Re_{\lambda}$ is the Reynolds number based on the Taylor microscale $\lambda$, the root mean square of the velocity fluctuations $u'=\sqrt{\left\langle \widetilde{u}^2\right\rangle}$ and the kinematic viscosity $\nu$ ). This constitutes a handicap when it is desired to investigate features of high Reynolds number turbulence. Several other grids were proposed to overcome this problem, including the use of mechanically moving parts called \textit{active grids} \cite{Gad_el_Hak_1974, Mydlarski_1996, Kang_2003}. In \cite{Makita_1991}, the use of an active grid resulted in turbulence with higher Reynolds number ($Re_{\lambda}=390$), higher turbulence intensities, and larger integral length scales. Moreover, the resulting energy spectrum had a wider inertial subrange, about two orders of magnitude in wavenumber, compared to a regular grid.

Hurst and Vassilicos \cite{Hurst_2007} proposed another method to generate higher Reynolds numbers and high turbulence intensities, which does not involve mechanically moving grid components and is easier to use and construct. In \cite{Hurst_2007}, the authors characterized the turbulence generated by three types of \textit{fractal grids} (cross, I and square shaped grids), using 21 different grids in wind tunnel experiments. The main idea behind fractal grids is to use grids that consist of a multi-scale collection of obstacles and openings based on a specific pattern, which is repeated in increasingly numerous copies at smaller scales. Fractal therefore emphasizes that these grids have multiple scales that are self-similar. This property helps reduce the parameters necessary to specify the design of these multi-scale grids. Besides the easy way to generate high Re-number turbulence, the topic of fractal grid turbulence has attracted considerable attention due to the specific features of the wake flow.

The concept behind fractal grids is to generate turbulence by \textit{directly} exciting a wide range of length scales \textit{at once}, rather than using the nonlinear cascade mechanism to obtain multi-scale structures, as it is the case for regular grids. These flow structures of different scales influence each other and show very different properties compared to all previously documented turbulent flows. Interesting results were obtained using Fractal Square Grids (FSG) with low blockage ratios (around $25\%$). These results include obtaining higher Reynolds numbers, even in small and conventional-size wind tunnels, compared to regular grids. In addition, it was shown that there exists an extended region between the fractal grid and a downstream distance, where the turbulence intensity increases progressively to reach a maximum value at a distance $x=x_{peak}$ \cite{Hurst_2007}. This region is called the production region. The turbulence intensity then decays downstream from $x_{peak}$, following what was thought first to be an exponential law \cite{Hurst_2007, Seoud_2007} but was shown later on to be a fast power law \cite{Valente_2011_JFM}. These findings contradict what is typically reported for regular grid turbulence, namely that the turbulence decays directly at the lee of the grid following a power-law \cite{Jayesh_Warhaft_1992}. Another interesting characteristic of fractal grid turbulence is that longitudinal and lateral turbulent integral length scales and Taylor microscales were reported to vary slowly downstream of the grid \cite{Hurst_2007}.  

A more detailed study of the \textit{decay region} mentioned in \cite{Hurst_2007} was conducted by Seoud and Vassilicos \cite{Seoud_2007}, which confirmed the findings in terms of turbulent length scale behavior and turbulence intensity decay law. Moreover, it was shown that the ratio of the longitudinal length scale and Taylor microscale remains approximatively constant in the decay region. In \cite{Seoud_2007, Vassilicos_Mazellier_2010, Valente_2011_JFM, vassilicos2015dissipation} this striking behavior was set into the context of a non constant energy dissipation rate $C_\varepsilon$. Contrary to what is predicted by the Richardson-Kolmogorov phenomenology, the authors could collapse one-dimensional energy spectra at different wake positions using only one length scale and $u'$. They also presented assessments of the homogeneity and isotropy of the turbulence produced by FSG \cite{Hurst_2007}.

Vassilicos and Mazelier \cite{Vassilicos_Mazellier_2010} related $x_{peak}$ to the \textit{wake-interaction} length scale $x_{*}$, which is the ratio of the length of the biggest square bar and its thickness. $x_{*}$ is found to be the appropriate length scale to characterize first and second-order statistics of the turbulent flow generated by FSG. The authors also showed that the turbulence statistics are non-homogeneous and non-Gaussian in the production region but become homogeneous and Gaussian in the decay region. They suggested that the FSG can be used for flow control and to enhance turbulent mixing properties by understanding and determination of how $x_{peak}$ and the generated turbulence intensities depend on fractal grid geometry. 

Several Direct Numerical Simulations (DNS) were conducted to investigate the flow generated by fractal grids \cite{Laizet_TSFP2009,Laizet_2009JMM, Laizet_2010CAF,Suzuki_2010_DNS, Suzuki_2013_DNS}. Although these simulations qualitatively reproduced some of the underlying characteristics of fractal grid turbulence, quantitative comparisons with experimental data was not possible for mainly three reasons. First, the simulations were conducted at very small Reynolds numbers compared to the experiments. As an example, in \cite{Laizet_2010CAF} the Reynolds number based on the effective mesh size (see Section \ref{geometryFG}) is $Re_{M_{eff}}=4430$ for the simulations and $Re_{M_{eff}}=20800$ for the experiments, which is a reduction with a factor of 5. Second, fractal grids with a smaller iteration number were simulated, mainly 3 iterations, compared to the available experimental studies where the iterations vary from 4 to 6. Finally, the extent of the computational domain in the streamwise direction was limited due to the high computational cost associated with DNS. As all the flow details have to be resolved, the meshes used in the different DNS simulations are huge in terms of the number of nodes. Therefore, substantial computational power is needed to keep the computational time realistic. In the following, we summarize two important DNS reference papers.    

Laizet and Vassilicos \cite{Laizet_TSFP2009} conducted DNS simulations using the numerical code \textit{Incompact3d} based on sixth-order compact schemes for spatial discretization and second-order Adams-Bashforth schemes for temporal discretization. They modeled a regular and a fractal grid (with the same effective mesh size) using Immersed Boundary Method (IBM) and their total numerical mesh of 765 million points. They could recover the production and decay regions as well as the decay behavior of the turbulence intensity, although they did not test the simulation results to determine the nature of the decay law. Moreover, the turbulence recovered in the wake of the fractal grid was found to be not homogeneous, which contradicts the findings in the experiments \cite{Hurst_2007, Vassilicos_Mazellier_2010}. They related this observation to the small iteration number of their fractal grid ($N=3$) and the domain size chosen to be insufficient for reaching homogeneity.  

In a more recent paper, Laizet and Vassilicos \cite{Laizet_2011FTC} used a new version of \textit{Incompact3d}, which scales better when used with large numbers of computational cores. They simulated one regular and three square fractal grids (with different aspect ratios). The simulations required the use of 3456 computational cores. The DNS results show that the turbulence produced by the fractal grids is more intermittent than the one produced by regular ones. Nevertheless, the extent of the computational domain and number of fractal iterations ($N=3$) were still insufficient to obtain homogeneous turbulence in the decay region. Therefore, they emphasize the need for more simulations with extended computational domain and bigger iterations ($N>3$) and experiments with $N=3$ to be able to compare the results with DNS. The authors also showed that the value of $x_{peak}$ introduced in \cite{Vassilicos_Mazellier_2010} does not apply for their grids, which have a smaller N, and introduced a new $x_{peak}$ formula based on N and the blockage ratio $\sigma$.  

IIn the present paper, the wake of the flow through an $N=3$ fractal square grid is investigated numerically and experimentally. The experimental data was acquired from PIV and hot-wire anemometry. Our main purpose is to validate our numerical simulations against the experimental results and to characterize the turbulence generated by fractal grids in terms of turbulent kinetic energy decay law and one-point statistical quantities such as turbulence intensity, third and fourth central moment and probability density functions of the velocity fluctuations. Furthermore, the characterization of such turbulent fields will assess the capabilities of DDES in reproducing the experimental results. A further goal of this investigation is to compare experimental and numerical data of the wake of a fractal grid with the experimental examination of a regular grid with the same mesh size and blockage ratio.
Our simulations and experiments also involve an extended domain to obtain more homogeneous and isotropic turbulence than DNS obtained so far. 

This paper is organized as follows. Section 2 introduces the geometrical aspects of the fractal square grid and of a regular grid having the same mesh size and blockage ratio. The regular grid has been investigated only experimentally and is used here for comparison purposes. The experimental setup and details of the PIV and hot-wire anemometry used are presented in Section 3. Section 4 is dedicated to the numerical setup, including a brief overview of the turbulence modeling and the numerical solver. Section 5 discusses the low-order and higher-order statistics results.
Finally, in Section 6 the conclusions are presented with an outlook for further investigations.

\newpage
\section{Grid geometrical parameter}
\label{geometryFG}
\begin{figure}[h]
	\subfloat[]{\label{fig:parameter_frac}\includegraphics[width=0.24\textwidth]{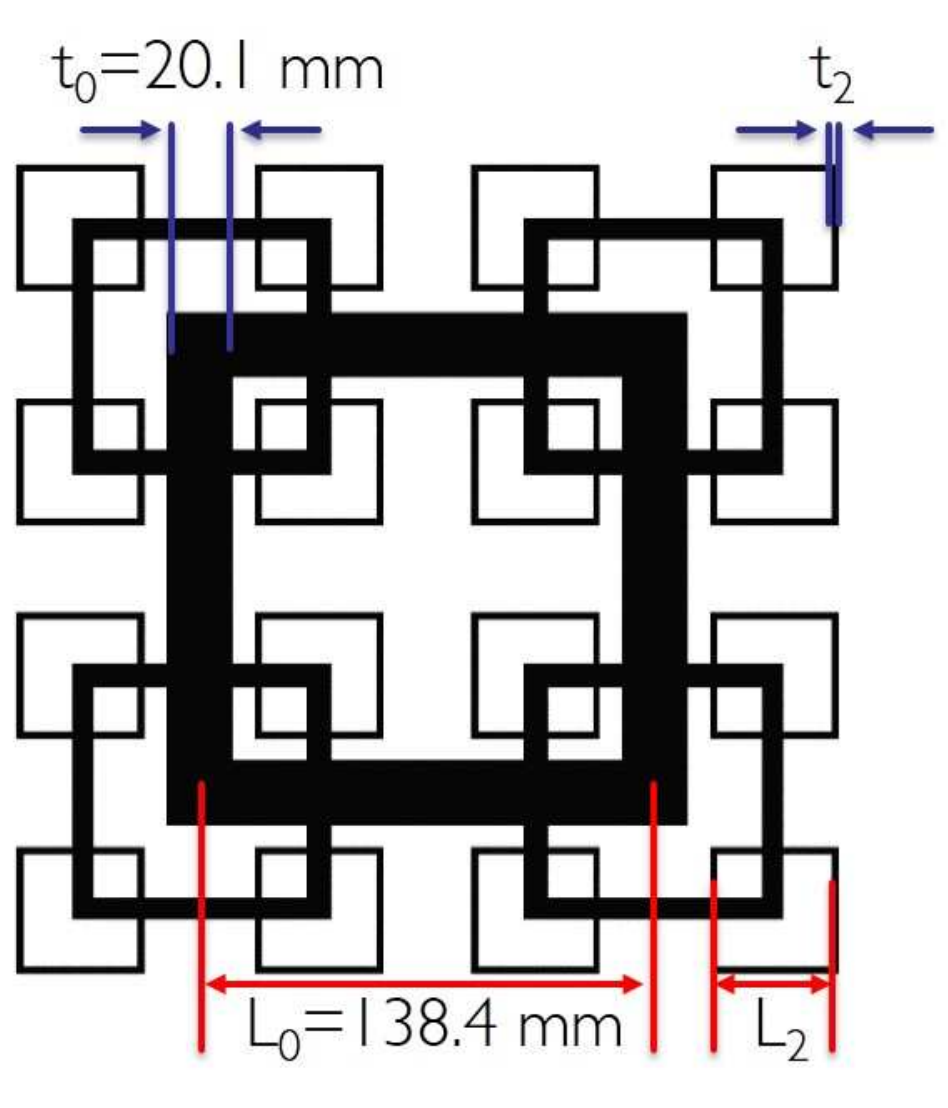}}  
	\subfloat[]{\label{fig:parameter_klassisch}\includegraphics[width=0.22\textwidth]{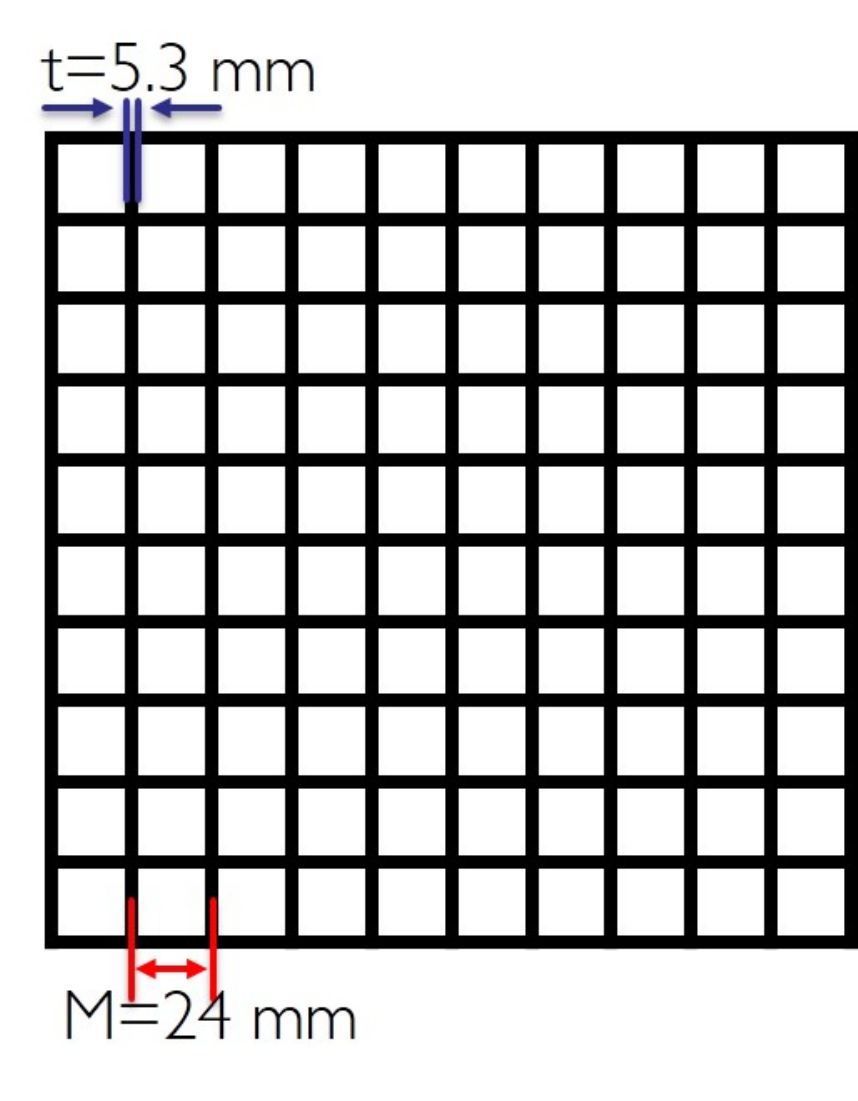}}  
	\caption{Illustration of (a) the used $N=3$  space-filling square fractal grid (FSG) geometry and (b) the regular grid used for comparison.}
	\label{fig:parameter}
\end{figure}

Figure \ref{fig:parameter} shows the grids used in the present work. They are placed at the inlet of the test section of a wind tunnel in the experiments and considered when implementing the corresponding numerical simulations (see also  Figure \ref{fig:N3_fg_domain}).
\vspace{-0.5cm}

In general, fractal grids are constructed from a multi-scale collection of obstacles based on a single pattern repeated in increasingly numerous copies with different scales. As presented in \cite{Hurst_2007}, fractal grids can be constructed using different geometrical patterns (see Figure 1 in \cite{Laizet_2010CAF}). We selected a frequently used pattern for our fractal grid  \cite{Seoud_2007, Vassilicos_Mazellier_2010, Valente_2011_JFM}, to be able to set our results in the context of other research activities. The pattern of our fractal grid is based on a square shape with $N=3$ fractal iterations. The fractal iteration parameter is the number the square shape that is repeated at different scales. At each iteration $(j=0, ... , N-1)$, the number of squares is four times higher than in the iteration $j-1$. Each scale iteration $j$ is defined by a length $L_j$ and a thickness $t_j$ of the square bars constituting the grid. The thickness of the square bars in the streamwise direction is kept constant. The dimensions of the square patterns are related by the ratio of the length of subsequent iterations $R_L = \frac{L_j}{L_{j-1}}$ and by the ratio of the thickness of subsequent iterations $R_t = \frac{t_j}{t_{j-1}}$; respectively. The geometry of the fractal grid used in this paper is completely characterized by two further parameters, namely the ratio of the length of the first iteration to the last one $L_r=\frac{L_0}{L_{N-1}}$ and the ratio of the thickness of the first iteration to the last one $t_r=\frac{t_0}{t_{N-1}}$. Moreover, a fractal grid is said to be space-filling, when its fractal dimension $D_f = 2$ (see \cite{Vassilicos_Hunt_1991, Hurst_2007}, for the definition of $D_f$), which is the case when $R_{L}=0.5$.

Contrary to regular grids, fractal grids, especially fractal square grids, do not have a well-defined mesh size. However, an equivalent  \textit{effective}  mesh size was defined in \cite{Hurst_2007} as 
\begin{equation}
	M_{eff} = \frac{4 T^2}{P} \sqrt{1-\sigma},
	\label{eq:M_eff}
\end{equation}
where $T^2$ is the cross-section of the wind tunnel/simulation domain, $P$ is the perimeter of the fractal grid and $\sigma$ is the blockage ratio, which can be determined by a calculation using the following formula
\begin{equation}
	\sigma = \frac{A}{ T^2} =\frac{L_0t_0 \sum \limits^{N-1}_{j=0} 4^{j+1} R^j_L R^j_t - t^2_0 \sum \limits^{N-1}_{j=1} 2^{2j+1} R^{2j-1}_t}{T^2}.
	\label{eq:bloc}
\end{equation}
$A$ is the sum of the areas $A_j$ covered by each fractal iteration $j$ and corrected by the areas covered by two iterations. $A$ is therefore, the fractal grid's total area. It should be noted that the local blockage ratio varies strongly for different parts of the grids. A complete quantitative description of the fractal grid we used in this study is shown in table \ref{table:tab1}.
\begin{table*}
\caption{Geometrical properties of the fractal grid used in this study.} 
\centering 
\begin{tabular}{c| c| c| c| c| c| c| c |c |c} 
N & $\sigma$ /\% & $L_{0}$ /mm & $t_{0}$ /mm & $R_{L}$ & $R_{t}$ & $L_r$ & $t_r$ & $M_{eff}$ /mm & T/mm \\ %
\hline\hline 
3 & 38.2 & 138.4 & 20.1 & 0.52 & 0.36 & 3.7 & 7.7 & 24 & 250 \\ 
\end{tabular}
\label{table:tab1}
\end{table*}

Another important length, which seems to collapse several turbulent statistical quantities of the streamwise fluctuating velocity, is the position of maximum turbulence intensity $x_{peak}$, which also defines the production ($x<x_{peak}$) and decay regions ($x>x_{peak}$). It was introduced by Mazellier and Vassilicos \cite{Vassilicos_Mazellier_2010} and defined using the empirical formula 
\begin{equation}
	x_{peak} \approx  0.45 \frac{L_{0}^2}{t_{0}},
	\label{eq:x_peak}
\end{equation}
which relies upon $x_{peak}$ to the geometrical properties of the fractal grid. This means that the turbulence generated by fractal grids can be ``manufactured''. 

To account for the distance at which the different wakes generated by the fractal grid bars interact, \cite{Vassilicos_Mazellier_2010} introduced the \textit{wake-interaction} length scale $x_{*}$. It is defined with the formula $x_{*}=\frac{L_{0}^2}{t_{0}}$ and is in our case $x_{*}=953$ mm (see definitions of $L_{0}$ and $t_{0}$ in table \ref{table:tab1}). The definitions of the fractal grid effective mesh size eq. \eqref{eq:M_eff} and the blockage ratio eq. \eqref{eq:bloc} has the advantage of returning the effective mesh size when applied to a fractal grid so that a comparison can be made between regular and fractal grids based on $M_{eff}$. In this respect, we experimentally investigated a regular grid (shown in Figure \ref{fig:parameter_klassisch}) with the same mesh size ($M=M_{eff}$) and blockage ratio as the used fractal grid for comparison.

\section{Experimental Setup}
\label{Experimental Setup}
The experiments were conducted in a closed loop wind tunnel with test section dimensions of 200 cm x 25 cm x 25 cm (length x width x height) at the University of Oldenburg. The wind tunnel has a background turbulence intensity along the centerline of the complete test section of approximately $2 \%$ for $U_\infty \leq 10$ m/s. The inlet velocity was set to 10 m/s, corresponding to a Reynolds number related to the biggest grid bar length $L_0$ of about \mbox{$Re_{L_0}=U_\infty L_0 / \nu =83800$,} where $\nu$ is the kinematic viscosity. Optical access is provided through the wind tunnel side walls. The experimental data were acquired from hot-wire anemometry and Particle Image Velocimetry (PIV) measurements.

Constant temperature anemometry measurements of the velocity were performed using (\emph{Dantec 55P01} platinum-plated tungsten wire) single-hot-wire with a wire sensing length of about $l_w=2.0 \pm 0.1$ mm and a diameter of $d_w = 5$ $\mu m$ which corresponds to a length-to-diameter ratio of $l_w / d_w \approx 400$. A \emph{StreamLine} measurement system by \emph{Dantec} in combination with CTA Modules 90C10 and the \emph{StreamWare} version 3.50.0.9 was used for the measurements. The hot-wire was calibrated with \emph{Dantec Dynamics Hot-Wire Calibrator}. The overheat ratio was set to 0.8. In the streamwise direction, measurements were performed on the centerline between 5 cm $\leq x \leq $ 176 cm distance to the grid. The data was sampled with the frequency  $f_s= 60$ kHz with a \emph{NI PXI  1042} AD-converter and a total of 3.6 million samples were collected per measurement point, representing 60 seconds of measurement data. To satisfy the Nyquist condition, the data was low-pass filtered at frequency $f_l=30$ kHz.

In addition to the hot-wire measurements, the flow velocity is measured with two-component (2C) two-dimensional (2D) Particle Image Velocimetry (PIV). A double-pulsed Nd:YAG laser system illuminates the flow with a maximum energy output of 2x380 mJ pulse$^{-1}$ ($\lambda$ = 532 nm, \emph{Quantel Brilliant B Twin}). The laser system is set at a sampling frequency of 10 Hz and a laser pulse delay of 63 $\mu s$ (time between two image frames) in order to record statistically independent images. To illuminate the desired field of view, a laser light sheet is formed with an optical arrangement (spherical lens \mbox{($f=-100$ mm)} and cylindrical lens ($f=200$ mm)) to converge the sheet into a minimum thickness of around 2 mm. 
The light sheet plane is in the vertical mid-plane of the wind tunnel.
The flow was seeded with Di-Ethyl-Hexyl-Sebacat (DEHS) droplet aerosols generated by a \emph{Palas AGF 5} aerosol generator for atomizing liquids. 
The diameter of the mist of DEHS droplets used in the present study is about 0,3 $\mu m$ to 1 $\mu m$. At each measuring station (there are 15 stations in total) 3000 double images are recorded by a cooled digital 14 bit CCD Camera (PCO1600) with 1600x1200 pixel resolution. The camera looks perpendicular to the light sheet from the side and is synchronized with the laser sheet pulse at 10 Hz. The camera is fitted with a Nikon mikro Nikkor 55 mm objective lens set with an aperture setting of f/8. A minimal effect of peak locking was found for these experimental conditions.
The PIV analysis is done using commercial PIV software (\mbox{\emph{PIV View}} version 3.5.7) using an interrogation window of 2 mm$^2$. The vector fields are obtained by processing the PIV images using a recursive cross-correlation engine with a final interrogation window of size 16 x 16 pixels with 75\% overlap.

\section{Numerical Setup}
\subsection{Turbulence modelling}
The three-dimensional, incompressible Navier-Stokes equations describe the flow generated by a fractal grid. The equations are discretized and solved using a turbulence model. In this investigation, the Delayed Detached Eddy Simulation (DDES) \cite{Spalart_Strelets_1997} with a Spalart-Allmaras background turbulence model \cite{Spalart_Allmaras_1994}, commonly referred to as SA-DDES is used. DDES is a hybrid method stemming from the Detached Eddy Simulation method (DES) \cite{Spalart_Strelets_1997}, which involves the use of Reynolds Averaged Navier-Stokes Simulation (RANS) at the wall and Large Eddy Simulation (LES) away from it. This method combines the simplicity of the RANS formulation and the accuracy of LES, with the advantage of being less expensive in terms of computational time when compared with pure LES. DDES is an improvement of the original DES formulation, where the so-called "modeled stress depletion" (or MSD), is treated \cite{Menter_2004, Spalart_2006}. MSD occurs when the LES mode of the DES method is operating in the boundary-layer \cite{Spalart_2006}. In DDES, a shield function is derived, which guarantees that the boundary layer is solved by the RANS mode of the DES model. In the following, a brief description of SA-DDES is given. For more information refer to \cite{Spalart_Strelets_1997, Mockett_2009}.  

In the framework of eddy viscosity models, the Reynolds stresses are expressed in terms of the eddy viscosity $\nu_t$ and the strain-rate tensor $S_{ij}$ given by the formulation $-\left\langle u_{i}u_{j} \right\rangle=2 \nu_{t} S_{ij}$.
We consider the Spalart-Allmaras model, which is a RANS model based on the eddy viscosity approach \cite{Spalart_Allmaras_1994}, where a transport equation is defined for the modified turbulent viscosity $\tilde{\nu}$ as follows
\begin{equation}
\label{eq:sa1}
\begin{split}
\frac{\partial{\tilde{\nu}}}{\partial{t}} + u_j \frac{\partial{\tilde{\nu}}}{\partial{x_j}} & = c_{b1} (1 - f_{t2})\tilde{S} \tilde{\nu} - \left[c_{w1}f_w - \frac{c_{b1}}{\kappa^2} f_{t2} \right] \left( \frac{\tilde{\nu}}{d} \right) ^2 \\
&	+ \frac{1}{\sigma} \left[ \frac{\partial}{\partial{x_j}} \left( (\nu + \tilde{\nu}) \frac{\partial{\tilde{\nu}}}{\partial{x_j}} \right) + c_{b2} \frac{\partial{\tilde{\nu}}}{\partial{x_i}} \frac{\partial{\tilde{\nu}}}{\partial{x_i}} \right]. 
\end{split}
\end{equation}

\noindent $\tilde{\nu}$ is related to the turbulent eddy viscosity by the equation
\begin{equation}
\nu_t = \tilde{\nu} f_{\nu 1},
\end{equation}

\noindent and the function $f_{\nu 1}$ is defined as
\begin{equation}
f_{\nu 1}=\frac{\chi^3}{\chi^{3} + c^{3}_{\nu 1}},
\end{equation}

\noindent where $\chi=\frac{\tilde{\nu}}{\nu}$, $\nu$ being the kinematic viscosity. $\tilde{S}$ is defined as
\begin{equation}
\tilde{S}= f_{\nu 3}S + \frac{\tilde{\nu}}{\kappa^2 + d^2} f_{\nu 2},
\end{equation}

\noindent where $S$ is the magnitude of the vorticity, and the functions $f_{\nu 2}$ and $f_{\nu 3}$ are defined as follows
\begin{equation}
\begin{split}
f_{\nu 2}&= \frac{1}{(1+ \chi / c_{\nu 2})^3}, \\
f_{\nu 3} &= \frac{(1+ \chi f_{\nu 1})(1-f_{\nu 2})}{\chi},
\end{split}
\end{equation}

 \noindent where, $d$ is the distance to the closest wall. Finally, $f_{\omega}$ is a function, which also contains the distance $d$ and is defined as
\begin{equation}
f_{\omega}=g \left[ \frac{1+ c_{\omega 3}^6}{g^6 + c_{\omega 3}^6} \right] ^{\frac{1}{6}},
\end{equation} 

\noindent where $g$ and $r$ are defined as $g= r + c_{\omega 2} (r^6 - r)$ and $r= \frac{\tilde{\nu}}{\tilde{S} \kappa^2 d^2}$. The remaining model constants are defined in \cite{Spalart_Allmaras_1994}.

The DES model used in this contribution was obtained by replacing the distance variable $d$ with the distance $\tilde{d}$ defined by the expression
\begin{equation}
\tilde{d}= min(l_{RANS},l_{LES})=min\left( d,C_{DES}\Delta \right), 
\end{equation}

\noindent where, $C_{DES}=0.65$ is a constant and \mbox{$\Delta=max(\Delta_x, \Delta_y, \Delta_z)$} represents the grid size (the filter size). One of the well-known shortcomings of DES is the so-called Grid Induced Separation (GIS) which can lead to the reduction of the RANS Reynolds stresses.  
This problem occurs when the model operates in LES mode in the boundary-layer \cite{Spalart_2006}, which results in the separation point moving upstream (thus the name GIS). This results from the fact that in refined regions of the mesh, $l_{LES}$ becomes smaller and the switch from RANS to LES occurs prematurely. To overcome this problem, the "detached" version of the DES method was proposed in \cite{Spalart_2006, Menter_2004}. In DDES, the switching from RANS to LES mode is performed in a more complex manner, where an empirical shielding function $f_d$ is introduced. This results in the following expression for the DDES length scale
\begin{equation}
	\tilde{l}_{DDES}=l_{RANS} - f_{d}max(0, l_{RANS} - l_{LES}),
\end{equation}  

\noindent where $f_d$ tends towards $0$ inside the boundary-layer region and towards $1$ away from the boundary-layer. Moreover, $f_d$ is a continuous function that enables a smooth transition between the RANS and the LES regions.

\subsection{Mesh and Numerical Method}
\label{sec:Mesh and Numerical Method}
The numerical simulation was set up analogous to the experiments in order to compare the results consistently. The open source code OpenFOAM \cite{OpenFOAM:website} was used to solve the incompressible Navier-Stokes equations. OpenFOAM is based on a collocated formulation of the finite volume method, and it consists of a collection of libraries written in C++, which can be used to simulate a large class of flow problems. For more information, see the official documentation \cite{OpenFOAM:website}. The solver used in this investigation is the transient solver pimpleFoam, which is a merging between the PISO (Pressure implicit with splitting of operator) and SIMPLE (Semi-Implicit Method for Pressure Linked Equation) algorithms. A second-order central-differencing scheme is used for spatial discretization, and a backward, second-order time-advancing schemes were used. The solver is parallelized using the Message-Passing Interface (MPI), which is necessary for problems of this size. The SA-DDES model used in this work is already implemented in the official release of OpenFOAM. 

\begin{figure}[h]
	\centering
	\subfloat[]{\label{fig:N3_fg}\includegraphics[width=0.3\textwidth]{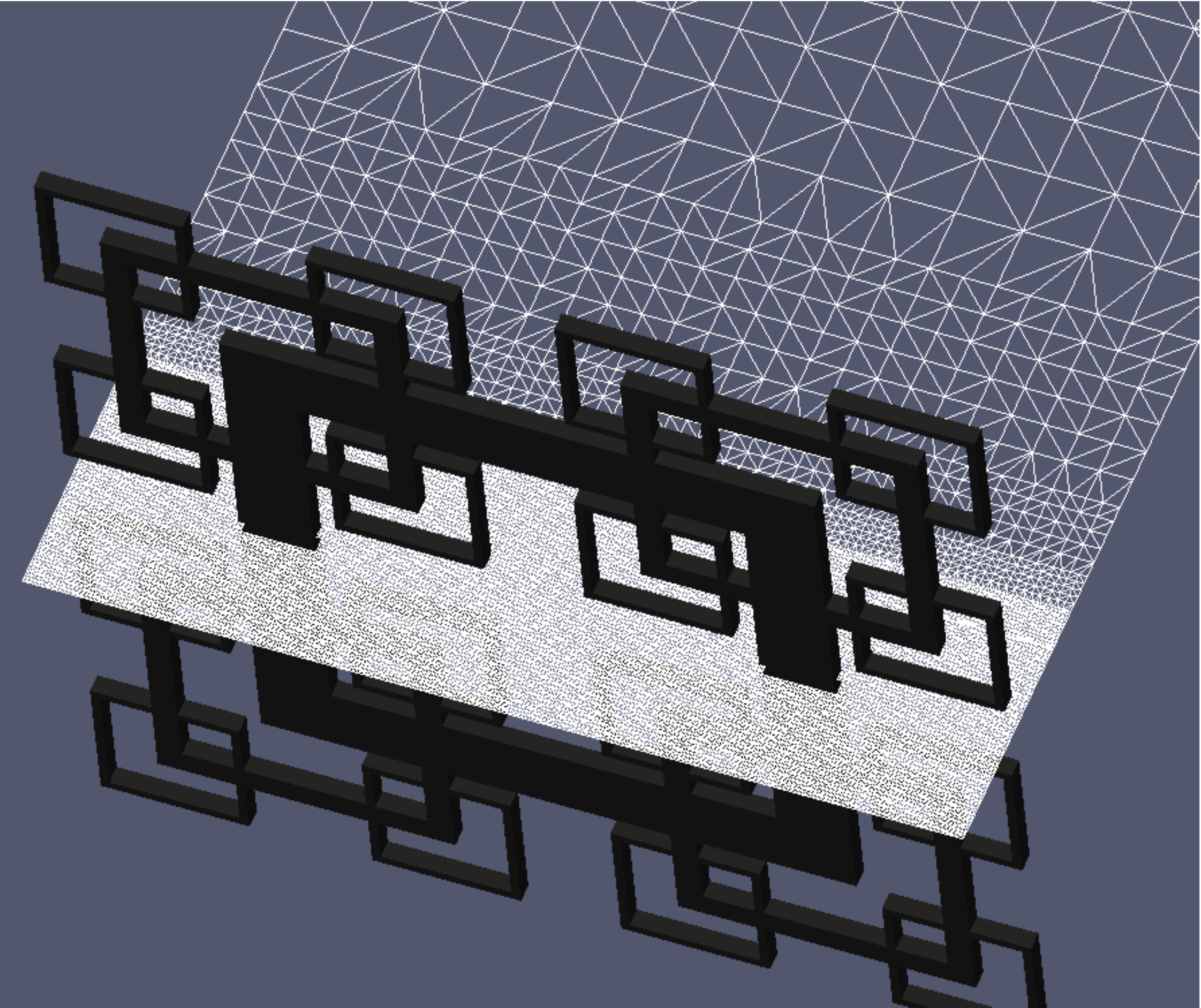}}\\
	\subfloat[]{\label{fig:N3_fg_domain}\includegraphics[width=0.4\textwidth]{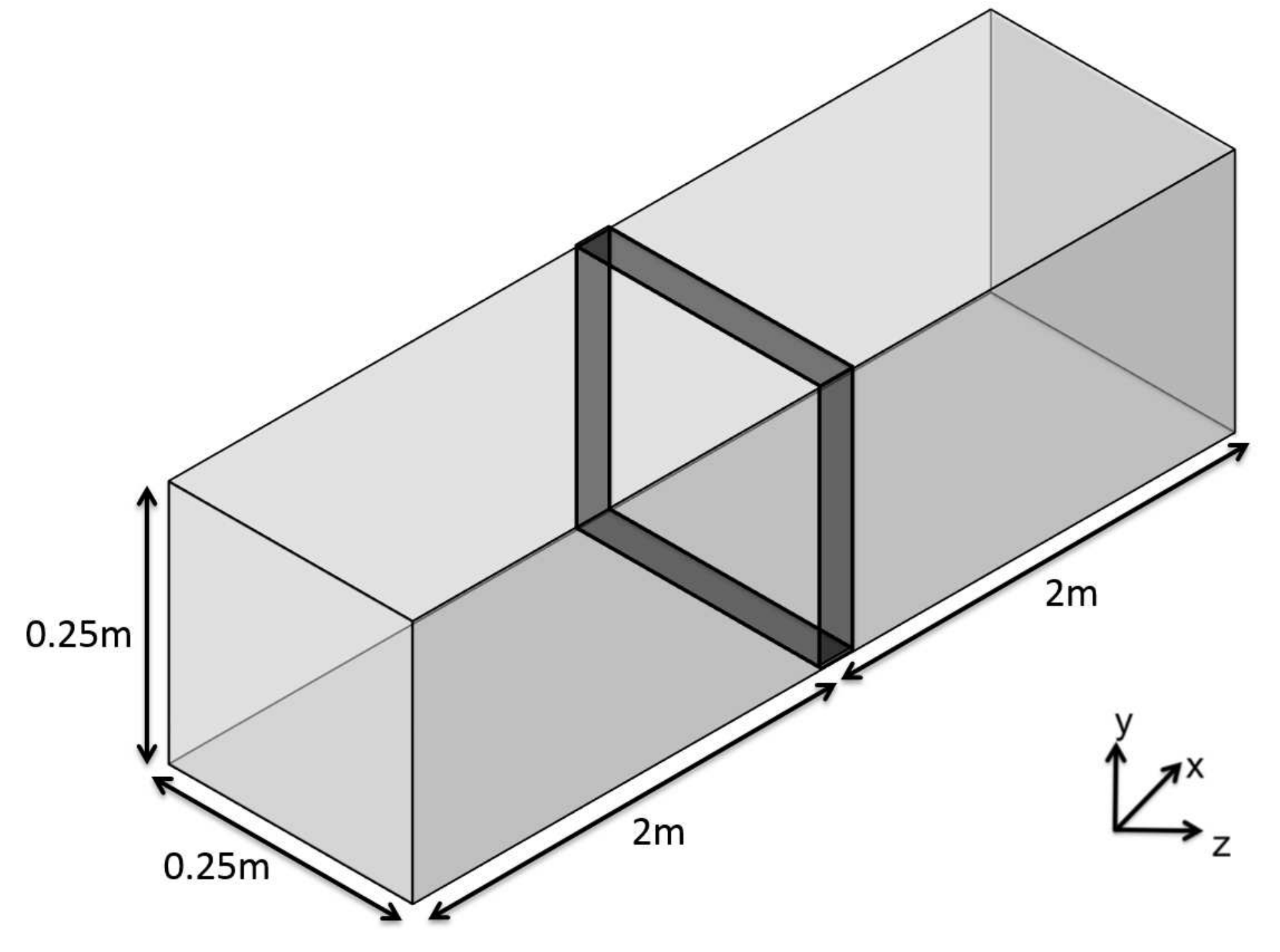}}  
	\caption{(a) Details of the computational mesh around the N3 fractal grid, generated using blockMesh and snappyHexMesh. (b) Schematic representation of the numerical domain and the system of coordinates. The fractal grid is positioned where the dark block is drawn.} 
\end{figure}

The numerical mesh was generated using the built-in OpenFOAM meshing tools blockMesh and snappyHexMesh \cite{OpenFOAM:website}. First, a background mesh constituted of hexahedral cells filling the entire computational domain is generated using blockMesh. The snappyHexMesh tool is then applied on the generated hexahedral mesh, which includes the fractal mesh. As a result, an unstructured mesh is obtained, where regions of interest in the wake are refined, as shown in Fig. \ref{fig:N3_fg}. Locally refining the mesh required the use of the parallel option of the snappyHexMesh tool. The mesh obtained is composed of a total of 24 million cells.

The fractal grid is simulated in a domain with similar dimensions as the real wind tunnel. The domain begins 2 m upstream of the fractal grid and covers a distance of 2 m downstream (see Fig. \ref{fig:N3_fg_domain}). The flow-parallel boundaries are treated as frictionless walls, where the slip boundary condition was applied for all flow variables. At the inflow boundary, Neumann boundary condition was used for the pressure and Dirichlet condition for the velocity. At the outflow boundary, the pressure was set to be equal to the static pressure and a Neumann boundary condition was used for the velocity. On the fractal grid, a wall function is used for the modified viscosity $\tilde{\nu}$, with the size of the first cell of the mesh in terms of the dimensionless wall distance is $y^{+} \sim 200$ \cite{whiteCFD}.   

For each simulation, 480 processors were used, and for each time step 4.5 $GB$ of data was collected for post-processing. The data sampling frequency was 60 $kHz$, chosen to match the experimental one. It took approximately 72 hours to simulate one second of data and a total of 16 seconds of numerical data was collected. The data was collected in the same positions as for the experimental study. The numerical simulations were conducted on the computer cluster of the ForWind Group \cite{FLOW01:website}.

\section{Results and discussion}
\label{Results}
\begin{figure}[h!]
	\centering
	\includegraphics[width=0.3\textwidth]{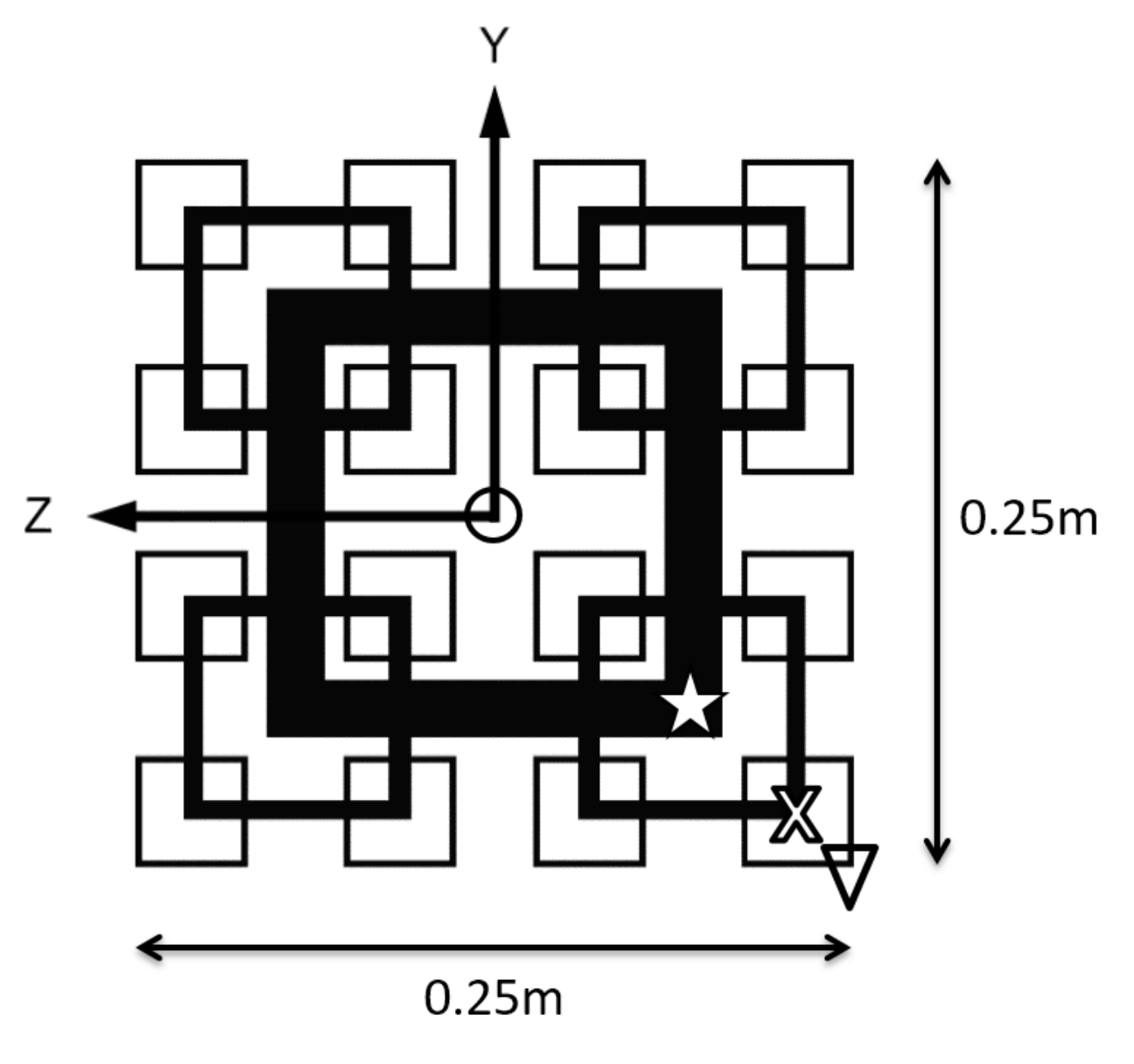}
	\caption{Positioning of the centerline $\bigcirc$ and three probes at different scales, where data is sampled from the CFD simulations. The probes are labeled from the thickest to the thinnest bar, as position \FiveStarOpen, $\times$ and $\triangledown$; respectively. For the experiments, data was collected at the centerline of the domain.}
	\label{fig:positions_probes}
\end{figure}

The velocity signals collected from both CFD and experiments are denoted as the instantaneous streamwise velocity $u(t)$ (or $u$), which is split into a average value $\left\langle u \right\rangle$ and a fluctuating velocity component $\widetilde{u}$.  
We denote $u'=\sqrt{\left\langle \widetilde{u}^2\right\rangle}$.
The instantaneous velocities in the horizontal and lateral directions are denoted as $v$, $w$ correspondingly. 
Next, the experimental results of the fractal grid will be compared to the results of the corresponding numerical investigation as well as the experimental results obtained for the above mentioned regular grid in terms of statistical turbulence properties and flow visualizations. 
The velocity from the computational results was sampled at the centerline and at three other positions, at 76 stations along the streamline direction. The three positions off the centerline are labeled in Figure \ref{fig:positions_probes} as \FiveStarOpen, $\times$ and $\triangledown$ and range from the largest to the smallest scales in the fractal grid; respectively. The hot-wire measurements were sampled at the centerline (at the same 76 stations) in the experiments for both the fractal and regular grid. In most figures, the streamwise distance to the grid $x$ is normalized by the biggest grid bar length $L_0$ following the conventional manner. For the regular grid $L_0$ is assumed to be equivalent to $M=24$ mm and for the fractal grid $L_0=138.4$ mm even though their effective mesh size is much smaller.
\vspace{-0.4cm}

\subsection{Results in terms of low order statistics\\ (one-point statistics)}
\label{chap:first order statistics}
\subsubsection{Instantaneous velocity magnitude distribution\\ in the x-y plane}
\label{chap:Instantaneous velocity magnitude distribution in the x-y plane}
\begin{figure*}
	\centering
	\subfloat[]{\label{fig:CL-instvel}\includegraphics[width=0.85\textwidth]{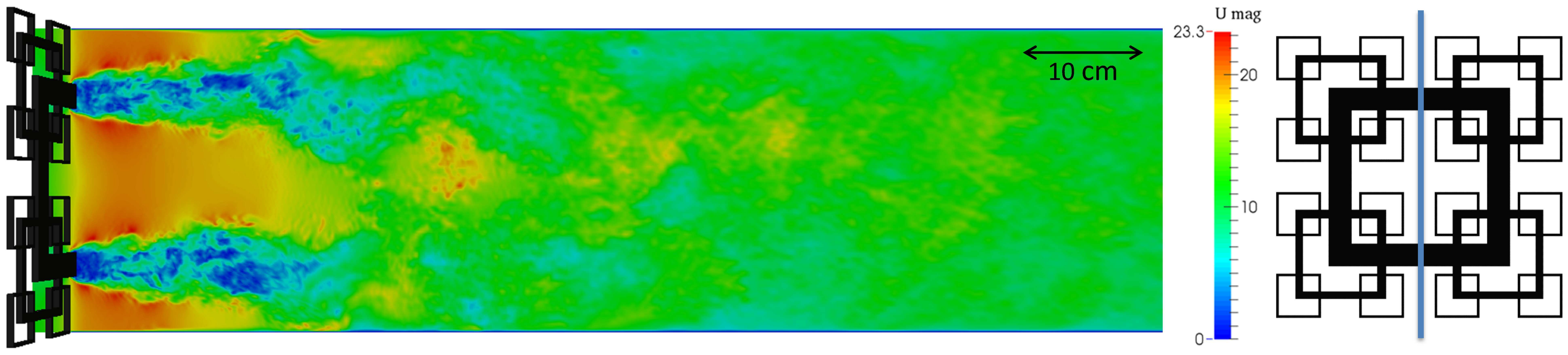}} \\
	\subfloat[]{\label{fig:lower-instvel}\includegraphics[width=0.85\textwidth]{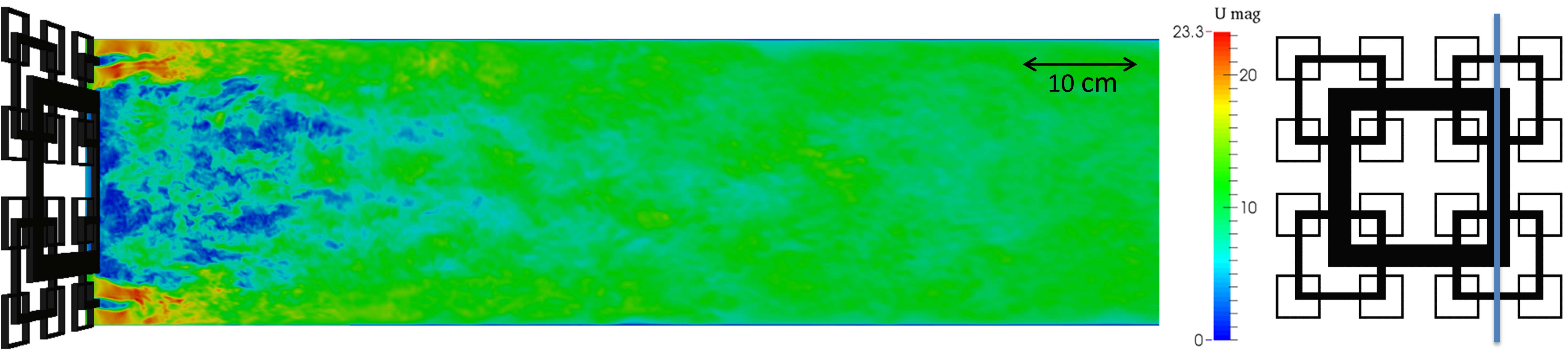}} \\
	\subfloat[]{\label{fig:middle-instvel}\includegraphics[width=0.85\textwidth]{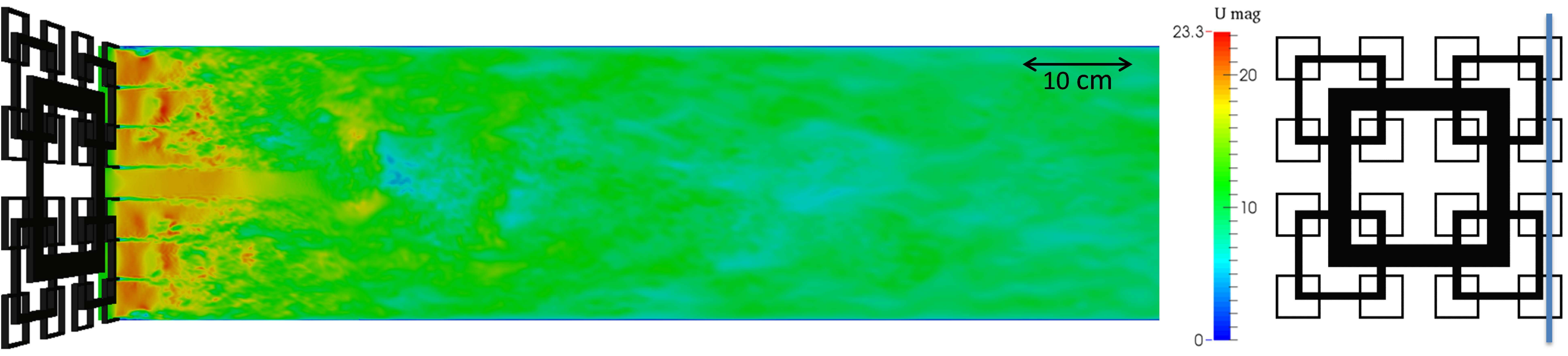}}
	\caption{Instantaneous velocity magnitude distribution in the x-y plane at three different levels (simulated data), as depicted on the right of each figure.}
	\label{fig:inst_velocity_cfd} 
\end{figure*}

Figure \ref{fig:inst_velocity_cfd} illustrates the turbulent flow at the lee of the fractal grid in the form of instantaneous snapshots (simulated data) of the magnitude of the velocity \mbox{$U_{mag}=\sqrt{u^2+v^2+w^2}$} at the x-y plane. The snapshots illustrated here were taken in 0 cm $\leq x \leq$ 90 cm distance to the grid at different z values, corresponding to different sections of the fractal grids, as indicated on the right side. 

\begin{figure*}
	\centering
	\includegraphics[width=0.75\textwidth]{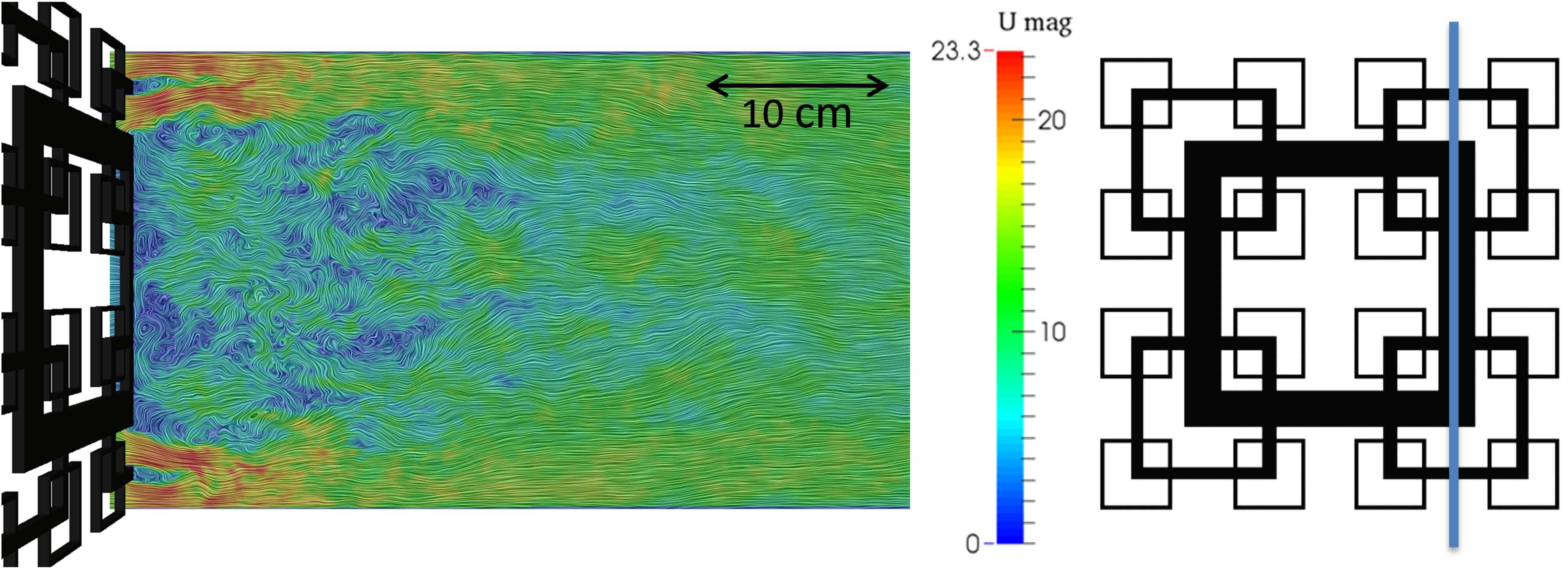}
	\caption{Distribution of the velocity (simulated data) using the Line Integral Convolution (LIC). Snapshots illustrated here were taken in 0 cm $\leq x \leq$ 45 cm distance to the grid.}
	\label{fig:LIC_middle}
\end{figure*}

The first observation is that three distinct wakes with different sizes exist, corresponding to the fractal grid's three iterations. The red color in Figure \ref{fig:inst_velocity_cfd} indicates high velocity values which correspond to jet-like wakes. These wakes result from openings in the fractal grid, as it is the case for example for the centerline opening (see Fig. \ref{fig:CL-instvel}). The blue color indicates low velocity values, which correspond to the wakes that are created in the vicinity of the fractal grid bars. Figure \ref{fig:inst_velocity_cfd} also clearly indicates that the wakes close to the grid generated a non-homogeneous distribution of the magnitude of velocity.  
As a further indication of the non-homogeneity of the turbulent flow, we plot the Line Integral Convolution (LIC) \cite{LIC1} in Fig. \ref{fig:LIC_middle} using the corresponding velocity vector field of Fig. \ref{fig:lower-instvel}. The LIC technique indicates recirculation regions existing in the near wakes of the fractal grid bars. This observation is also confirmed by the investigation of the mean velocity distributions, which will be presented in Sections \ref{chap:Mean velocity distribution in the x-y plane} and \ref{chap:Mean streamwise velocity distribution}. 
The fractal grid generates a flow with a complex wake interaction and mixing in the near wake of the fractal grid due to its multi-scale nature, which is visible from the previous visualizations of the instantaneous velocity magnitude and the LIC plot. This suggests sequential interactions between wakes from small-scale wakes all the way up to larger scale wakes. These sequential interactions of different wakes sizes are referred to in \cite{laizet2012fractal} as the space-scale unfolding (SSU) mechanism. This behaviour could be responsible for the prolonged production region observed in the experimental results \cite{Hurst_2007, Vassilicos_Mazellier_2010} (see Section \ref{sec:Streamwise turbulence intensity distribution}).

\subsubsection{Mean velocity distribution in the x-y plane}
\label{chap:Mean velocity distribution in the x-y plane}
\begin{figure*}
	\centering
	\subfloat[]{\label{fig:U_average_exp}\includegraphics[width=0.8\textwidth]{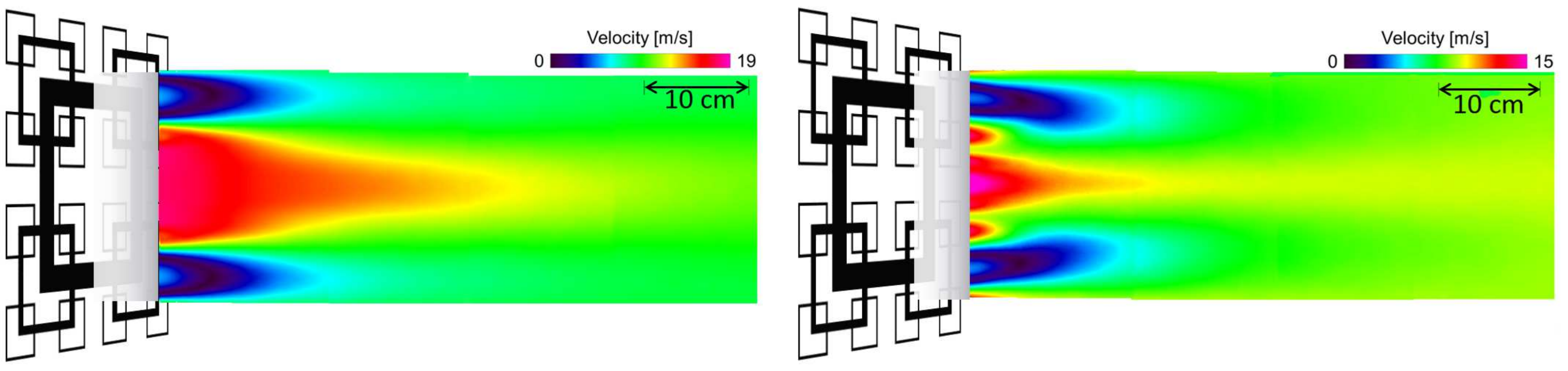}}\\
	\subfloat[]{\label{fig:U_average_cfd}\includegraphics[width=0.8\textwidth]{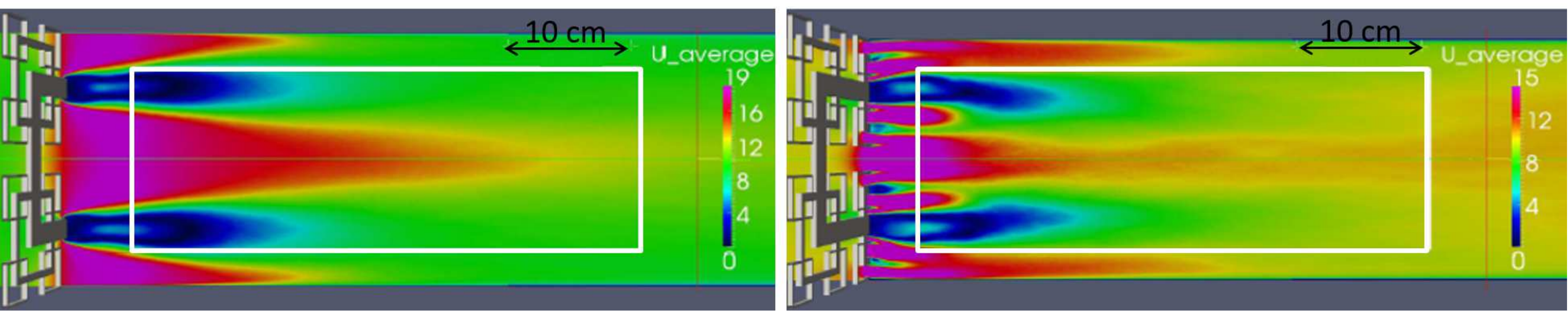}}  
	\caption{Distribution of the magnitude of the mean velocity for (a) the PIV results and (b) for the CFD results. The data is taken on a plane at the centerline for the Figures at the left and at a shifted plane of the grid for the Figures on the right. The white frames in (b) correspond to the obtained PIV data.}
	\label{fig:Umean_distribu_cfdExp:Umean_distribu_cfdExp}
\end{figure*}

To study the turbulent flow evolution downstream of the fractal grid, the distribution of the magnitude of the mean velocity is plotted in Figure \ref{fig:Umean_distribu_cfdExp:Umean_distribu_cfdExp}.
The data is taken on a x-y plane at the centerline for the Figures at the left and at a shifted plane of the grid for the Figures on the right. PIV measurements were not taken further upstream than 5 cm because of laser light reflections. The white frames in Figure \ref{fig:U_average_cfd} correspond to the obtained PIV data.

Figure \ref{fig:Umean_distribu_cfdExp:Umean_distribu_cfdExp} shows clearly a non-uniform distribution of the magnitude of the mean velocity near the grid and the presence of a jet-like flow induced by the large opening at the center of the fractal grid. This jet-like behavior was also observed in previous experimental \cite{Hurst_2007} and computational studies using DNS (at much lower Reynolds numbers than the experiments) \cite{Laizet_2011FTC}. The jet-like flow is surrounded by regions downstream from the grid bars where large mean velocity deficits, due to the non-uniform distribution of the local blockage, are clearly present. As the flow develops downstream, the jet-like behavior is smoothed out by the action of turbulent diffusion. 

The results presented in Figure \ref{fig:Umean_distribu_cfdExp:Umean_distribu_cfdExp} show that the computational values obtained for the mean velocity agree very well with the results obtained by PIV measurements. Both techniques reproduced the prolonged imprint of the fractal grid and the resulting combination of wake-like regions downstream from the grid bars and the jet-like flows induced by the openings of the grid. As a result of this observed behavior, strong mean flow gradients along the streamwise and transverse directions can be identified, which will be described in detail in Section \ref{sec:Large-scale anisotropy distribution in the x-y plane}.

\subsubsection{Streamwise evolution of the mean velocity}
\label{chap:Mean streamwise velocity distribution}
\begin{figure*}
	\centering
	\subfloat[]{\label{fig:Umean_expVScfd}\includegraphics[width=0.425\textwidth]{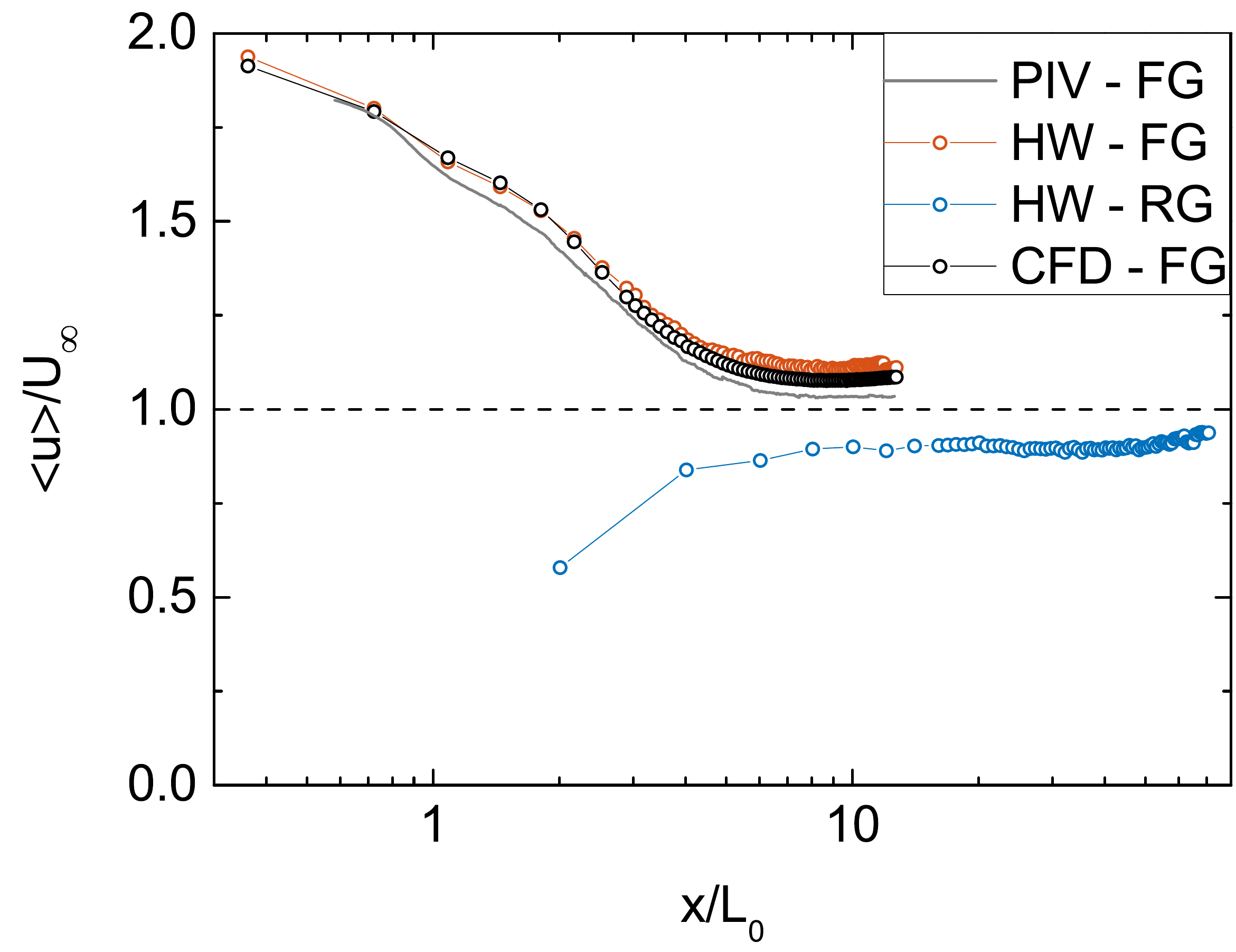}}  
	\subfloat[]{\label{fig:Umean_cfd_allscales}\includegraphics[width=0.425\textwidth]{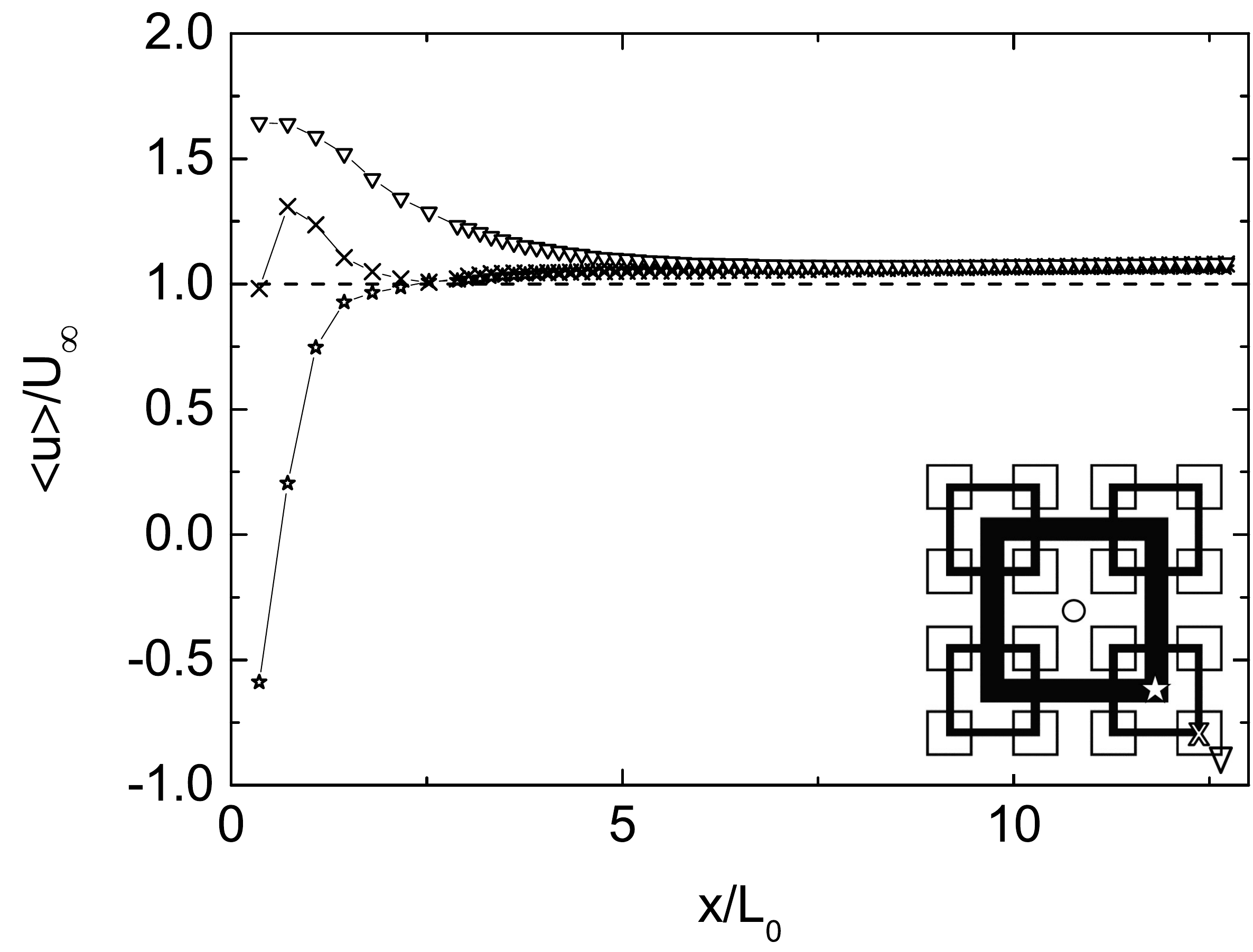}}  
	\caption{Streamwise evolution of the longitudinal mean velocity $\left\langle u \right\rangle$ normalized by inflow velocity $U_\infty$ as functions of the distance $x$ downstream of the grid (a) along the centerline from hot-wire measurements (HW) and computational data (CFD) (b) along off-centerline positions from computational data.}
	\label{fig:Umean_norm}
\end{figure*}

The evolution of the streamwise mean velocity component $\left\langle u \right\rangle$ as a function of the distance along the centerline normalized by the inflow velocity $U_\infty$ is plotted in Figure \ref{fig:Umean_expVScfd} for the fractal square grid and for the regular grid. Note that $U_\infty$ was set to 10 m/s without a grid placed at the inlet of the test section. 

Figure \ref{fig:Umean_expVScfd} clearly indicates a significant difference between the magnitude and the evolution of the longitudinal mean velocity along the centerline. For the regular grid, the results show that large mean velocity deficits exist very close to the grid. Furthermore, the mean velocity increases fast along the centerline with increasing distance. However, the mean velocity remains lower than the inflow velocity. For the fractal grid, Figure \ref{fig:Umean_expVScfd} displays the continuous decrease of the mean velocity along the centerline with increasing distance, from a value almost twice the prescribed inflow velocity $U_\infty$ directly behind the grid. This is related to the jet-like flow induced by the high opening at the center of the fractal grid (see Section \ref{chap:Mean velocity distribution in the x-y plane}). Note that, even in the far wake, the mean velocity along the centerline remains higher than the inflow velocity $U_\infty$. This observation is in good agreement with the values reported in \cite{Vassilicos_Mazellier_2010}. The results presented so far show that the computational values obtained for the mean velocity distribution and the evolution of the longitudinal mean velocity along the centerline agree well with the results obtained by PIV and hot-wire measurements. 
Our experimental and numerical study of $\frac{\left\langle u \right\rangle}{U_\infty}$ are consistent with findings from previous experimental studies that reported values slightly higher than $U_\infty$ in the far wake of the fractal square grid \cite{Hurst_2007, Seoud_2007, Vassilicos_Mazellier_2010}.
In Figure \ref{fig:Umean_cfd_allscales} the normalized mean velocity is plotted at three off-centerline positions indicated in the bottom right. The evolution behind positions $\times$ and $\triangledown$ show qualitatively the same tendency as the evolution along the centerline, however due to the higher blockage, the mean velocity in the immediate vicinity of the grid bars is significantly lower than the values along the centerline. 
In contrast, the flow behind the biggest grid bar (corresponding to position \FiveStarOpen as indicated in Figure \ref{fig:positions_probes}) differs from the other investigated locations. The evolution begins with large negative values of $\frac{\left\langle u \right\rangle}{U_\infty}$, confirming the existence of recirculation regions in the near wakes of the fractal grid bars as shown in the instantaneous velocity magnitude distribution in the x-y plane introduced earlier in Section \ref{chap:Instantaneous velocity magnitude distribution in the x-y plane}.

\subsubsection{Large-scale anisotropy}
\label{sec:Large-scale anisotropy distribution in the x-y plane}
\begin{figure}[h]
	\centering
	\includegraphics[width=0.48\textwidth]{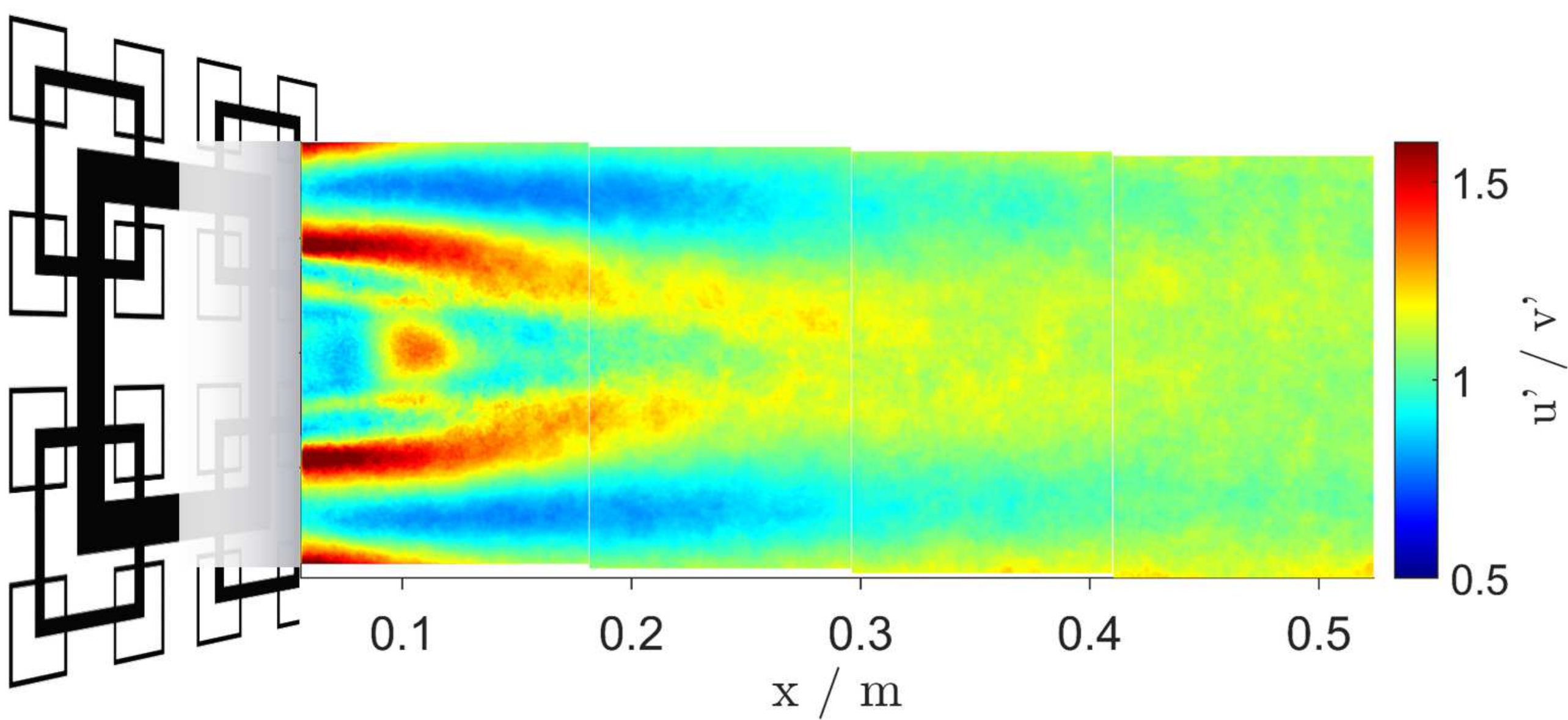}
	\caption{Distribution of large-scale anisotropy factor $\frac{u'}{v'}$ in the x-y plane along the centerline. The data is obtained by PIV measurements.}
	\label{fig:iso_piv}
\end{figure}

\begin{figure*}
	\centering
	\subfloat[]{\label{fig:homo_u}\includegraphics[width=0.425\textwidth]{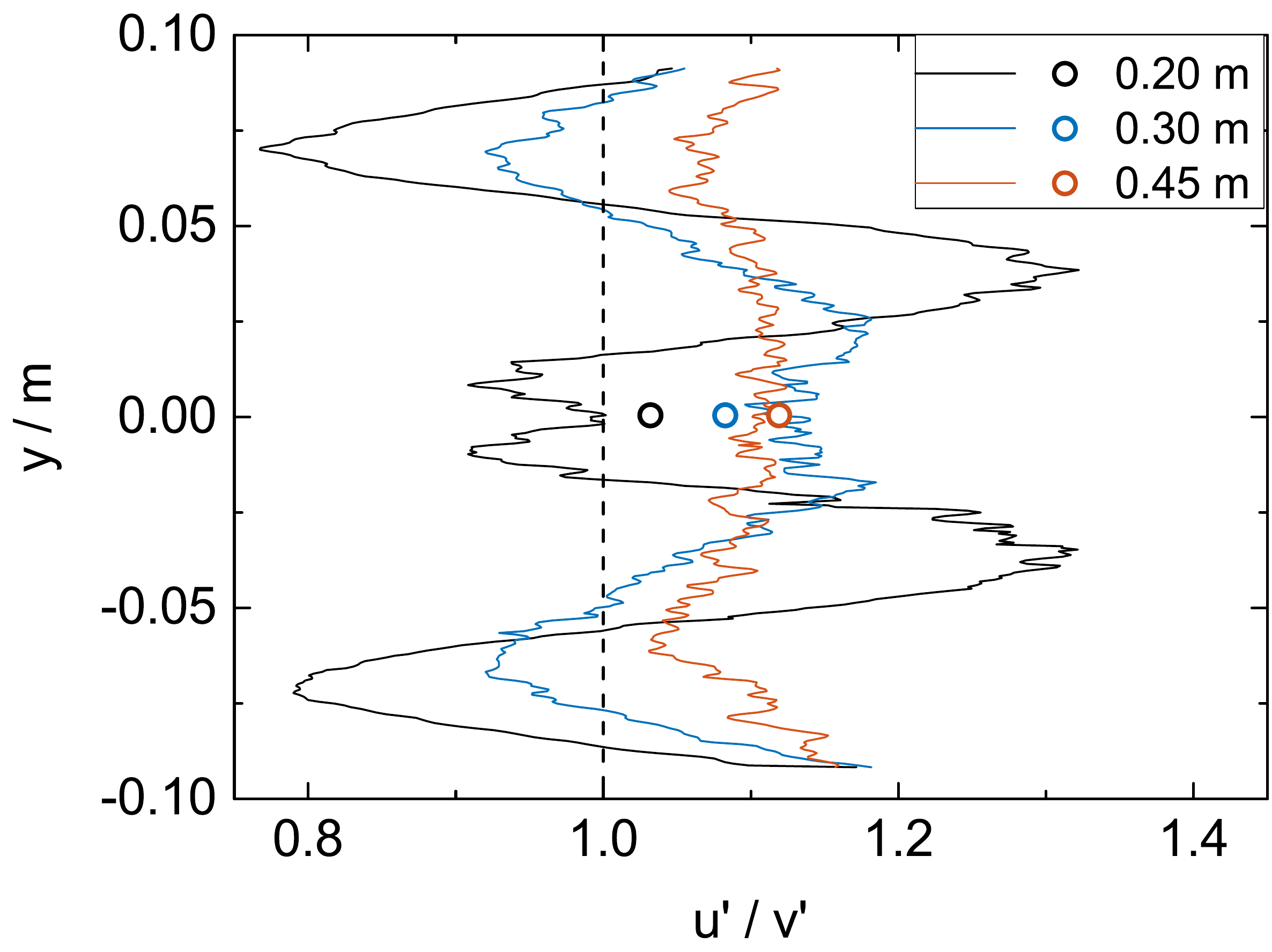}}  
	\subfloat[]{\label{fig:homo_v}\includegraphics[width=0.425\textwidth]{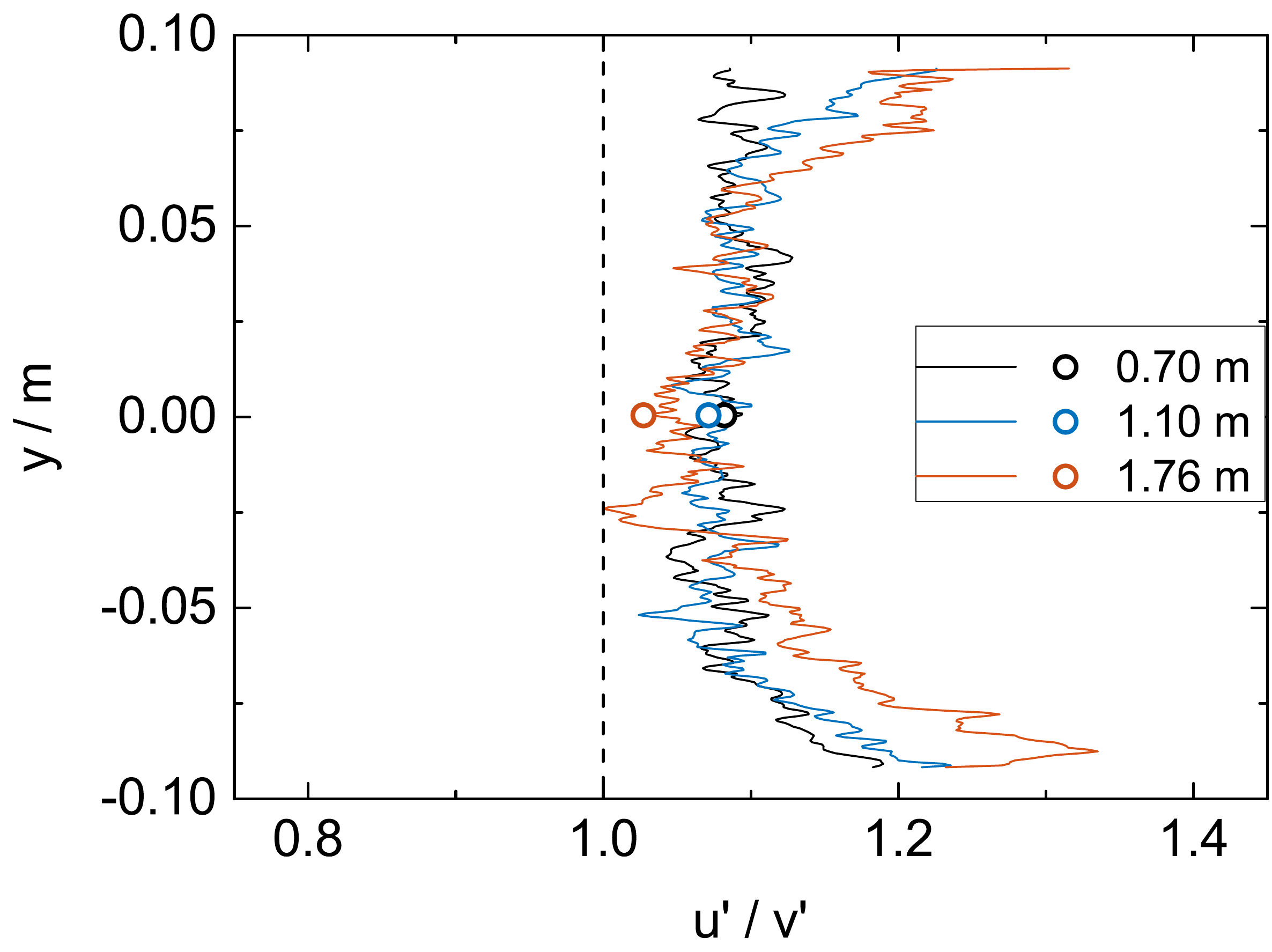}} 
	\caption{Distribution of the ratio of streamwise and transverse root mean square velocity fluctuations components in the x-y plane (a) at three different locations in the near-grid region (b) at three different locations in the far-grid region obtained from experimental data (PIV - solid lines) and corresponding computational data along the centerline ($\bigcirc$). The vertical dashed line represents $\frac{u'}{v'}=1$.}
	\label{fig:homog_cl}
\end{figure*}

From Figure \ref{fig:Umean_distribu_cfdExp:Umean_distribu_cfdExp} and \ref{fig:Umean_norm}, there are indications that the flow in the near-wake of the fractal grid is non-homogeneous and non-isotropic. This behavior, also common for regular grids \cite{Bellot_1966}, is due to the strong mean velocity flow gradients along the streamwise and the transverse directions present in the lee of the grid, that are caused by the inhomogeneous distribution of the blockage. As a further indication of the non-homogeneity and anisotropy of the turbulent flow generated by the fractal grid, we plotted in Figure \ref{fig:iso_piv} the distribution of the ratio $\frac{u'}{v'}$ or the so-called large-scale anisotropy factor in the x-y plane measured by PIV. 
We applied this criterion because in the investigation of regular grid turbulence, the simplest assessment of global isotropy is commonly evaluated by comparing the ratio of streamwise and transverse root mean square velocity fluctuations components. In addition in Figure \ref{fig:homog_cl} the large-scale anisotropy factor in the x-y plane at six different locations $x= [0.20, 0.30, 0.45, 0.70, 1.10, 1.76]$ m from experimental data (PIV) is plotted. Furthermore, we plotted in this Figure the corresponding value of the large-scale anisotropy factor along the centerline from CFD for comparison purposes.

Indeed Figure \ref{fig:iso_piv} and Figure \ref{fig:homog_cl} show a strong variation of the anisotropy factor along the streamwise and transverse direction. Especially in the near-grid region in the vicinity of the fractal grid bars, the flow is strongly anisotropic and then tends towards isotropy further downstream. 
Both Figures show that the imprints of the largest grid bars have a lasting influence for a significant downstream extent, as the inhomogeneity and anisotropy effects are present even at $x=0.25$ m. In this region, the anisotropy pattern shows fine structures, which can be seen in relation to SSU mechanism. In the far field, the vertical profiles become flatter when moving downstream and the large-scale anisotropy factor reaches values between 1 and 1.2 throughout the entire cross-sectional area. Note the exception of the region close to the horizontal wall, where the component perpendicular to the horizontal wall $v'$  is suppressed and therefore $\frac{u'}{v'}$ reaches much higher values (as shown in Figure \ref{fig:homo_v}). 
The evolution of $\frac{u'}{v'}$ downstream along the centerline obtained from CFD (only plotted at six different locations here) confirms that the isotropy factor is contained within the range of about 1 and 1.25, which is similar to those measured in case of regular grid turbulence \cite{valente2014non,nagata2008direct,kitamura2014invariants}.

The results presented here show the evolution from near-field large-scale anisotropy to a return to nearly large-scale isotropy in the far-field. Furthermore, our experimental and numerical investigation indicate that in the far-field of the lee of the fractal grid in the region around the centerline the turbulence is approximately homogeneous and isotropic. These findings are in agreement with those of previous experiments and CFD studies of fractal grids \cite{Hurst_2007, Vassilicos_Mazellier_2010, nagata2013turbulence,discetti2013piv}.

\subsubsection{Streamwise evolution of the turbulence intensity}
\label{sec:Streamwise turbulence intensity distribution}
\begin{figure*}
	\centering
	\subfloat[]{\label{fig:TU_expCFD}\includegraphics[width=0.425\textwidth]{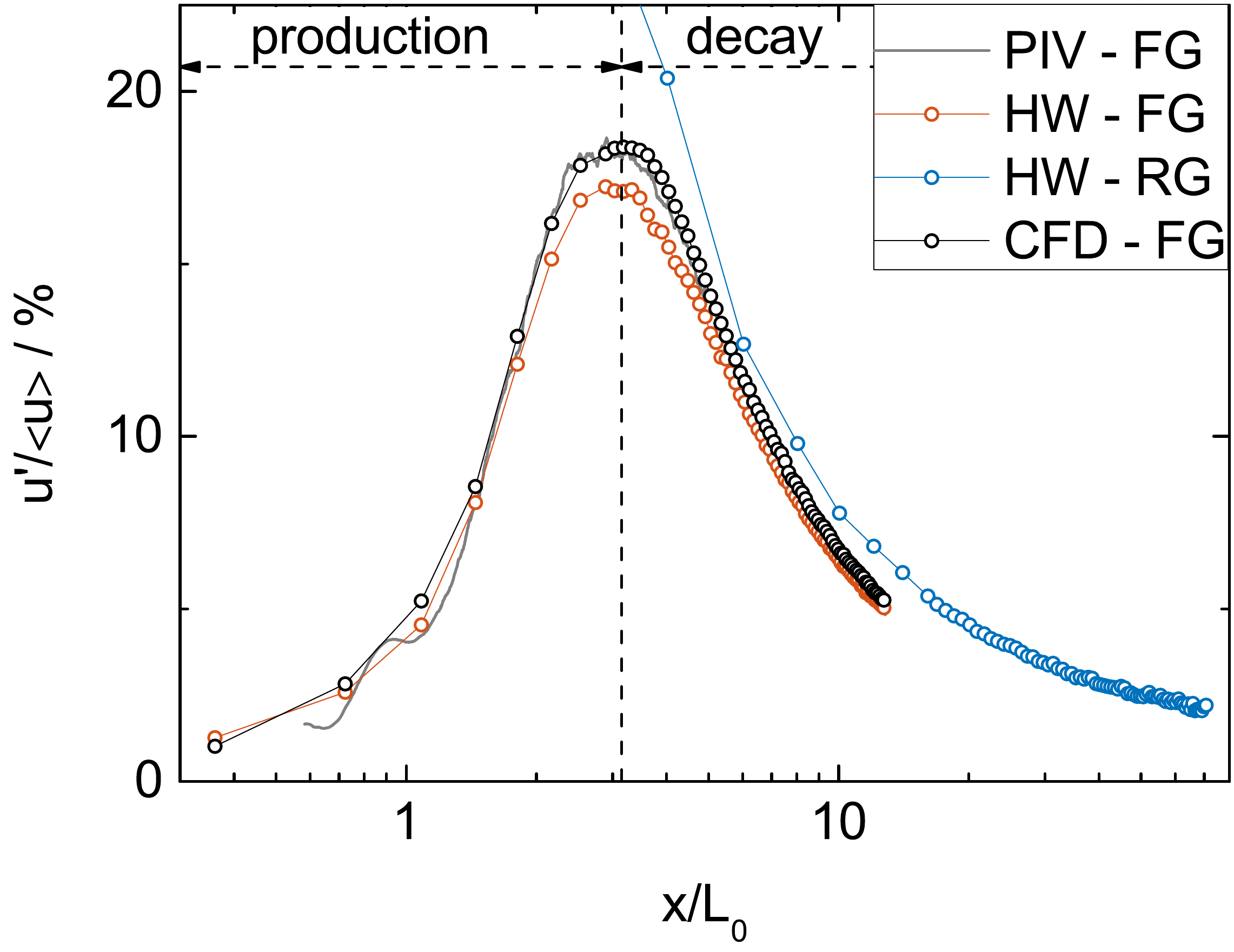}}  
	\subfloat[]{\label{fig:Tu_cfd_allscales}\includegraphics[width=0.425\textwidth]{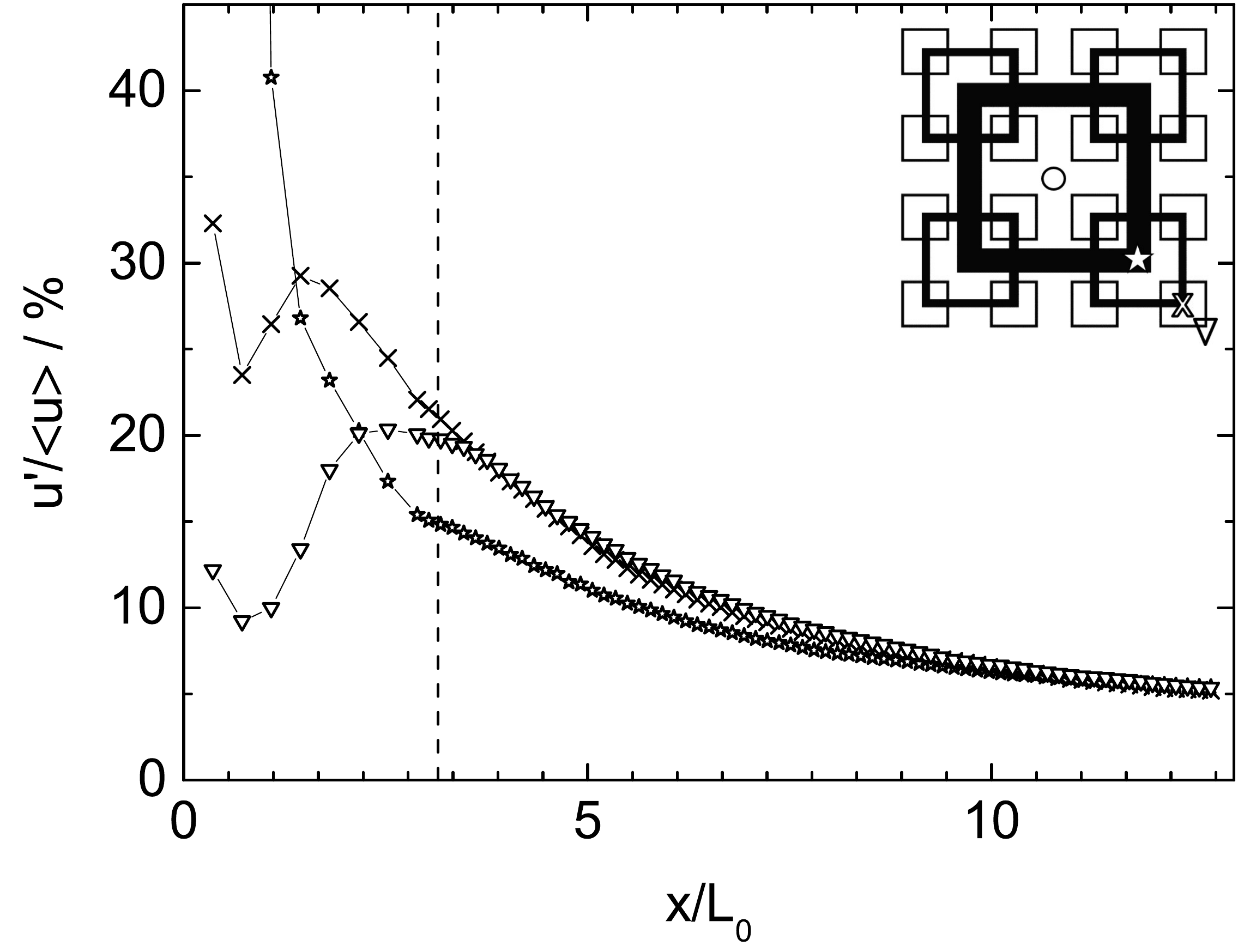}}  
	\caption{Streamwise evolution of the turbulence intensity $u'/U$ as functions of the distance $x$ downstream of the grid (a) along the centerline from hot-wire measurements (HW) and computational data (CFD) (b)  along off-centerline positions from computational data. In the case of the fractal grid, the vertical dashed line represents the position of the intensity peak on the centerline .}
	\label{fig:Tu}
\end{figure*}

In Figure \ref{fig:TU_expCFD}, the evolution of the streamwise turbulence intensity $u'/\left\langle u \right\rangle$ obtained computationally and experimentally is plotted as a function of the distance $x/L_0$ downstream along the centerline. 
Figure \ref{fig:TU_expCFD} indicates a significant difference in the turbulence intensity evolution and magnitude along the centerline. For the regular grid, the classical behavior is observed: Close to the grid, the turbulence intensity is characterized by a peak and decays with increasing distance. In contrast, the turbulence intensity evolution generated by the fractal grid begins with a lower value and continuously increases directly behind the grid until it reaches a maximum value at a certain distance defined as $x_{peak}$. This observation results from the sequential interaction of the different wakes originating from different grid bars present in the fractal grid. 
These wakes converge along the centerline at different distances to the grid and as a result, the value of turbulence intensity increases along the centerline until all wakes interact with each other (referred in \cite{laizet2012fractal} as the space-scale unfolding (SSU) mechanism). Behind the peak, the turbulence intensity then decays along the centerline towards the end of the test section or numerical domain. 
Therefore, Hurst and Vassilicos \cite{Hurst_2007, Vassilicos_Mazellier_2010} defined two distinct regions between the grid and the downstream distance, namely the production region at $x < x_{peak}$, where the turbulence progressively builds up and the flow is being homogenized (see Section \ref{sec:Large-scale anisotropy distribution in the x-y plane}) as the turbulence is convected downstream, and the decay region at $x > x_{peak}$,  where the turbulence continuously decays downstream. 
The computational values obtained for the turbulence intensities agree very well with the results obtained by hot-wire anemometry measurements, although the computational data show a slightly higher peak of turbulence intensity (18.4\% compared to the experimental one of 17.3\%). Additionally, the position \nohyphens{$x_{peak} \approx 0.44$ m} $ \approx  0.46 x_{*}$ for the CFD data differs slightly from the position $x_{peak} \approx 0.42$ m $ \approx  0.44 x_{*}$ obtained by the experimental data. 
Our results concerning the position $x_{peak}$ are in good agreement with the empirical formula (see eq. \eqref{eq:x_peak}) introduce by \cite{Hurst_2007}. Note that the definition of eq. \eqref{eq:x_peak} is valid for fractal grids with $N>3$ and low blockage ratios ($\sigma \leq 25\%$) \cite{Laizet_2011FTC}, which were extensively studied experimentally. The grid used in this investigation consists of 3 iterations and possesses a high blockage ratio ($\sigma=38.2\%$). 
Furthermore, the computational results show some small deviations from the experimental results in the decay region at $x > x_{peak}$, which will be examined in more detail in the next Section \ref{sec:Decay power law of the turbulence intensity}.

For a better understanding of the turbulent flow in the lee of the fractal square grid, we examined in Figure \ref{fig:Tu_cfd_allscales} the evolution of the streamwise turbulence intensity obtained by computational study as a function of the distance downstream along three different off-centerline positions (as indicated in Figure \ref{fig:positions_probes}). The results presented in Figure \ref{fig:Tu_cfd_allscales} indicate that the turbulence intensity evolution and magnitude strongly depend on the position behind the fractal grid. %
The evolutions behind the positions $\times$ and $\triangledown$ show qualitatively the same tendency as the evolution along the centerline. 
However due to the sequential interactions of the single wakes generated by the different bars of the fractal grid, as previously shown in Figure \ref{fig:inst_velocity_cfd}, the turbulence intensity evolution behind these two particular off-centerline positions peaks at shorter distances (\nohyphens{$x_{peak} \times \approx 0.25$ m} and \nohyphens{$x_{peak} \triangledown \approx 0.35$ m}) to the grid and at higher intensity values than the evolution along the centerline. 
In contrast, the flow behind the biggest grid bar (corresponding to position \FiveStarOpen) differs compared to the other investigated probe locations behind the fractal grid. The evolution of the turbulence intensity shows the characteristic shape of the evolution along the centerline behind a regular grid.
At this off-centerline position the maximum value of the turbulence intensity evolution is higher than $40 \%$. This is due to the high velocity fluctuations and the low mean velocity present when the sampling position is located directly behind the thickest bar. As already mentioned in the near wake of the thickest grid bar there exists recirculation regions with very low velocities and therefore the turbulence intensity reaches high values.

The results presented in this Section show that the turbulence intensity evolution along the centerline and the evolutions behind the examined off-centerline positions converge for $x > 2 x_{peak} \approx  x_{*}$. 
Furthermore, the intensity peak decreases when the sampling position changes from the largest to the smallest grid bar. 
Thus, our experimental and computational investigation confirms the characteristic shape observed in \cite{Hurst_2007, Vassilicos_Mazellier_2010}.  Moreover, in \cite{krogstad2012near} it has been shown that multi-scale grids generate clearly higher turbulence intensities compared to regular grids, even if the two grids have the same blockage ratio, what is in accordance with our results.

\newpage
\subsubsection{Decay power law of turbulent kinetic energy} 
\label{sec:Decay power law of the turbulence intensity}
\begin{figure*}
	\centering
	\subfloat[]{\label{fig:TU_expCFD_log}\includegraphics[width=0.425\textwidth]{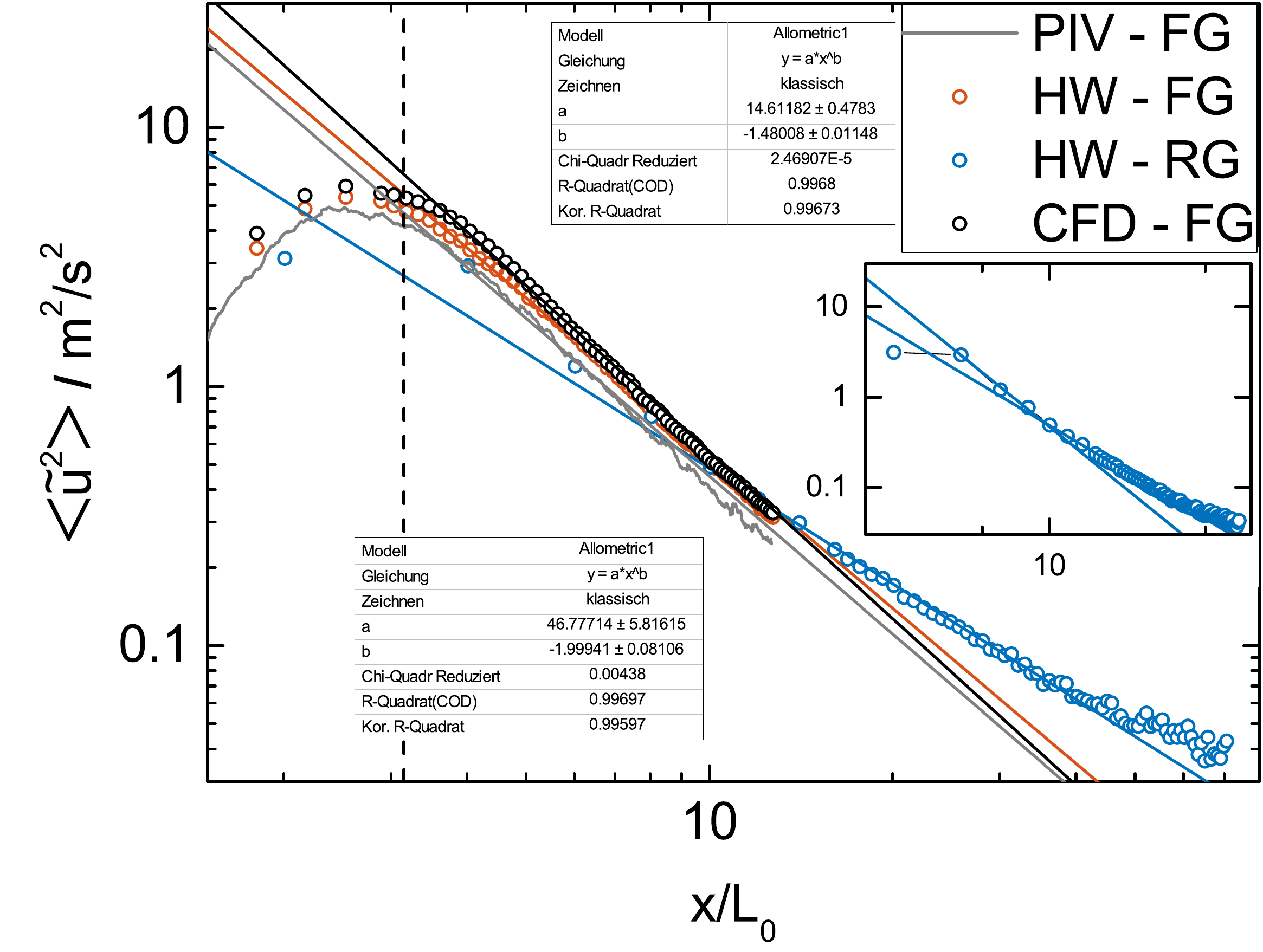}}  
	\subfloat[]{\label{fig:TU_simCFD_log}\includegraphics[width=0.425\textwidth]{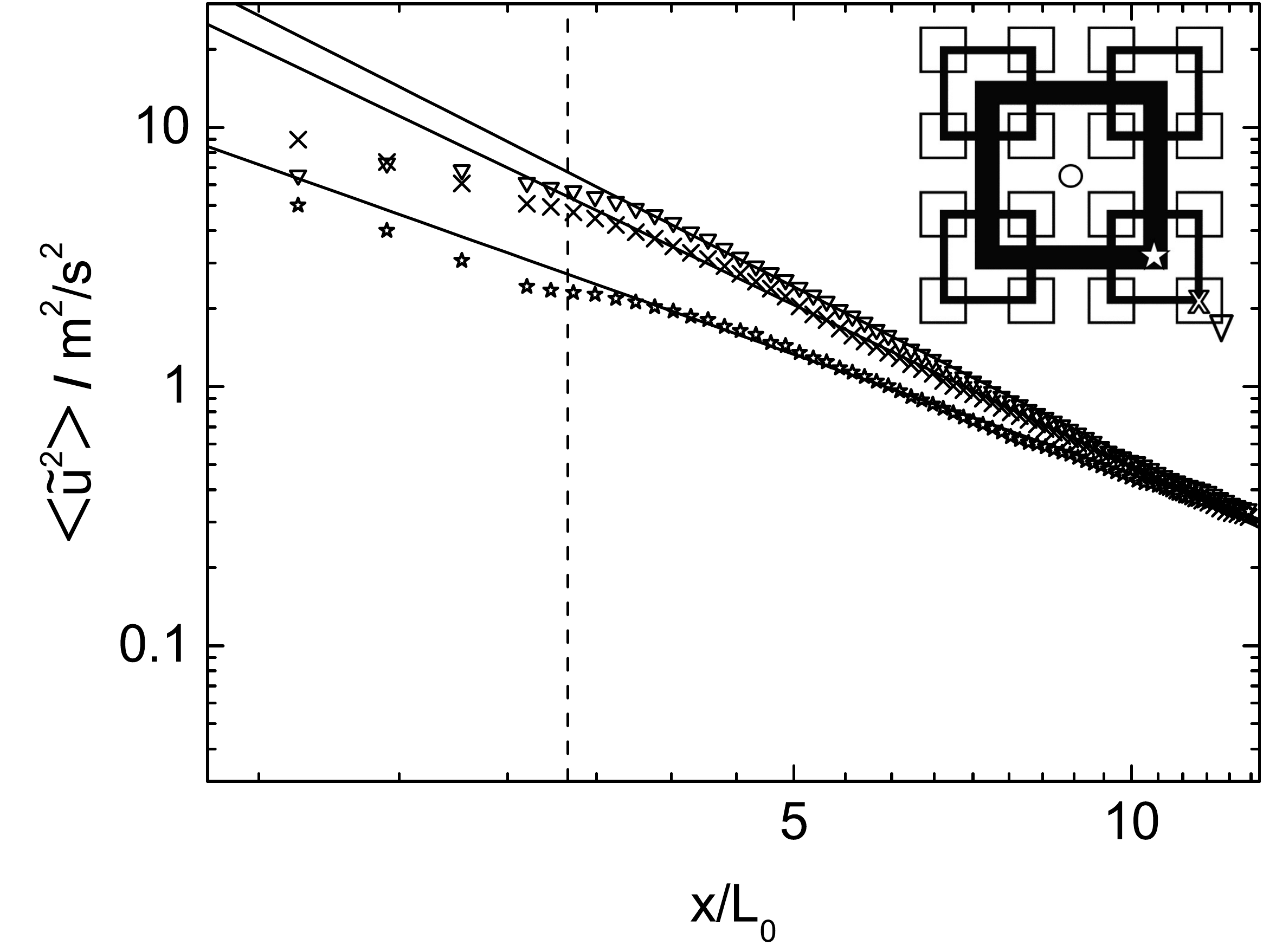}}  
	\caption{Streamwise evolution of the mean square of the streamwise velocity fluctuations as a function of the distance downstream on a log-log scale (a) along the centerline from hot-wire measurements (HW) and CFD (b) along off-centerline positions from computational data. In the case of the fractal grid the vertical dashed line represents the position of the turbulence intensity peak on the centerline. Solid lines indicate a least-squares error fit (in log-log axes). The corresponding fitted exponents and their associated errors using a virtual origin $x_0=0$ are listed in table \ref{table:tab2}.}
	\label{fig:Tu_log_decay}
\end{figure*}

The turbulent kinetic energy decay defined as \mbox{$\frac{1}{2}\langle q^2\rangle = \frac{1}{2} \left[ \langle \widetilde{u}^2\rangle + \langle \widetilde{v}^2\rangle + \langle \widetilde{w}^2\rangle \right]$} is going to be investigated in this Section. For homogeneous isotropic turbulence (HIT) the turbulent kinetic energy should decay as a power law $\langle q^2\rangle \propto (x-x_0)^{n}$, where $n$ is the power law exponent and $x_0$ is a virtual origin. 
In Figure \ref{fig:Tu_log_decay} we investigate the decay behavior of mean square of the streamwise velocity fluctuations (accordingly two-thirds of the turbulent kinetic energy) as a function of the downstream distance on a log-log scale obtained by the present computational and experimental study. For both CFD and experiments data $\langle \widetilde{u}^2\rangle$ is fitted, in the decay region, with power-law $\langle \widetilde{u}^2\rangle=A (x-x_0)^{n}$, where $A$ is a dimensional parameter. The solid lines in Figure \ref{fig:Tu_log_decay} indicate a least-squares error fit (in log-log axes) by using: 
$\ln \left( \langle \widetilde{u}^2\rangle \right)=\ln(A) + n  \ln \left(x-x_0 \right)$. To obtain comparable results, we assume that the position of the virtual origin coincides with the physical grid location, thus $x_0$ is forced to $x_0=0$. 
This method is typically used to obtain first-order estimates of the power law exponent because $n$ strongly depends on $x_0$ and it is widely recognized in the literature that an accurate determination of the virtual origin is very difficult. 
Table \ref{table:tab2} presents the results obtained by fitting the power law to $\langle \widetilde{u}^2\rangle$ in the decay region. In the case of the regular grid, we exclude data points in the region $x >1.3$ m, where the turbulence is dominated by the background turbulence of the wind tunnel. 
Figure \ref{fig:Tu_log_decay} and table \ref{table:tab2} show that the CFD data is in a fairly good agreement with the experimental results and that both the fractal and the regular grid generated turbulence are well approximated by the power law curve fit in the decay region. Thereby, the computational data is characterized by a slightly faster power law decay compared to the experimental acquired data (see table \ref{table:tab2}). In contrast to the regular grid with a power law exponent of -1.38 the fractal grid has a much lower value of about -2.09 with regard to experimental data and -2.20 for the simulated data. The results obtained by fitting the power law to $\langle \widetilde{u}^2\rangle$ relating to the regular grid agree with the empirical exponent reported in the literature \cite{Mohamed_1990}.
As already observed in previous experiments, for example of \cite{Hurst_2007,Valente_2011_JFM}, our experimental and CFD study confirms the unusually high power law exponent of the turbulent kinetic energy decay obtained for the fractal grid. In addition the results of the experimental study of \cite{hearst2014decay} show that there exist two distinct regions, namely the near-grid region, ($\frac{x}{L_0} < 20$) and the far-grid region ($\frac{x}{L_0} > 24$) where the turbulent kinetic energy decays at different rates. They observed an energy decay following a fast power law decay in the near-grid region and found that an usual power law decay region exists in the far-field. In our study, we investigated the region $0.3 \leq \frac{x}{L_0} \leq 12.8$, therefore our results agree very well with the findings mentioned above of the near-grid region.  
In Figure \ref{fig:TU_simCFD_log}, we examined the evolution of the mean square of the streamwise velocity fluctuations obtained by the computational study as a function of the distance downstream along three different off-centerline positions. The CFD data is again well approximated by the power law curve fit in the decay region. The power law fitting results indicate that behind the positions $\times$ and $\triangledown$ the decay exponents are significantly lower than the decay exponent behind the biggest grid bar (\FiveStarOpen).
Most interestingly, we see that the decay behind the positions $\times$ and $\triangledown$ resemble a lot the centerline results, where as the decay behind the biggest grid bar (\FiveStarOpen) is closer to the findings of the regular grid decay.

\begin{table*}
	\centering 
	\caption{Power law fitting results. $A$ is a dimensional parameter, $n$ is the power law exponent and $\sigma_{fit}$
		is the minimized sum of the squares of the residuals.}
	\begin{tabular}{c|| c |c |c |c| c| c} 
		& \multicolumn{2}{c| }{Experiment} &  \multicolumn{4}{c}{CFD-Simulation (fractal grid)} \\ 
		&  (regular grid) & (fractal grid)  & (centerline) & \FiveStarOpen & $\times$ & $\triangledown$\\
		\hline\hline 
		$A$ & 0.07 & 1.03 & 1.09 & 0.75 & 0.96 & 1.08 \\ 
		\hline 
		$n$ & -1.38 $\pm$ 0.02 & -2.09 $\pm$ 0.01 & -2.20 $\pm$ 0.01 & -1.56 $\pm$ 0.01 & -2.08 $\pm$ 0.01 & -2.18 $\pm$ 0.01 \\
		\hline
		$\sigma_{fit}$ & 0.025 & 0.009 & 0.007 & 0.009 & 0.007 & 0.011 \\
	\end{tabular}
	\label{table:tab2} 
\end{table*}

\subsection{Results in terms of higher order statistics (one-point statistics)}
\subsubsection{Skewness and kurtosis of the velocity fluctuations}
\label{sec:Skewness and flatness}
Next, the computational and experimental results are compared using higher-order statistics regarding the flow speed data itself (one-point statistics), namely, comparing the normalized third and fourth central moments of the fluctuating velocities $\widetilde{u}$. These higher-order statistics will provide important information about the shape of the $pdf\left(\widetilde{u}/u'\right)$ (probability density function), more precisely the symmetry and the Gaussian (or the non-Gaussian) nature of the turbulent velocity fluctuations. We study these important aspects of the turbulent flow in terms of skewness $S_{\widetilde{u}}$ (see eq. \eqref{eq:skewness}) and kurtosis $F_{\widetilde{u}}$ (see eq. \eqref{eq:flatness}).
\begin{equation}
	\label{eq:skewness}
	S_{\widetilde{u}} = \frac{\langle \widetilde{u}^3\rangle}{\langle \widetilde{u}^2\rangle^{\frac{3}{2}}} = \frac{\langle \widetilde{u}^3\rangle}{ u'^3}
\end{equation} 
\begin{equation}
	\label{eq:flatness}
	F_{\widetilde{u}} = \frac{\langle \widetilde{u}^4\rangle}{\langle \widetilde{u}^2\rangle^2} = \frac{\langle \widetilde{u}^4\rangle}{ u'^4}
\end{equation} 

\begin{figure*}
	\centering
	\subfloat[]{\label{fig:skewness}\includegraphics[width=0.425\textwidth]{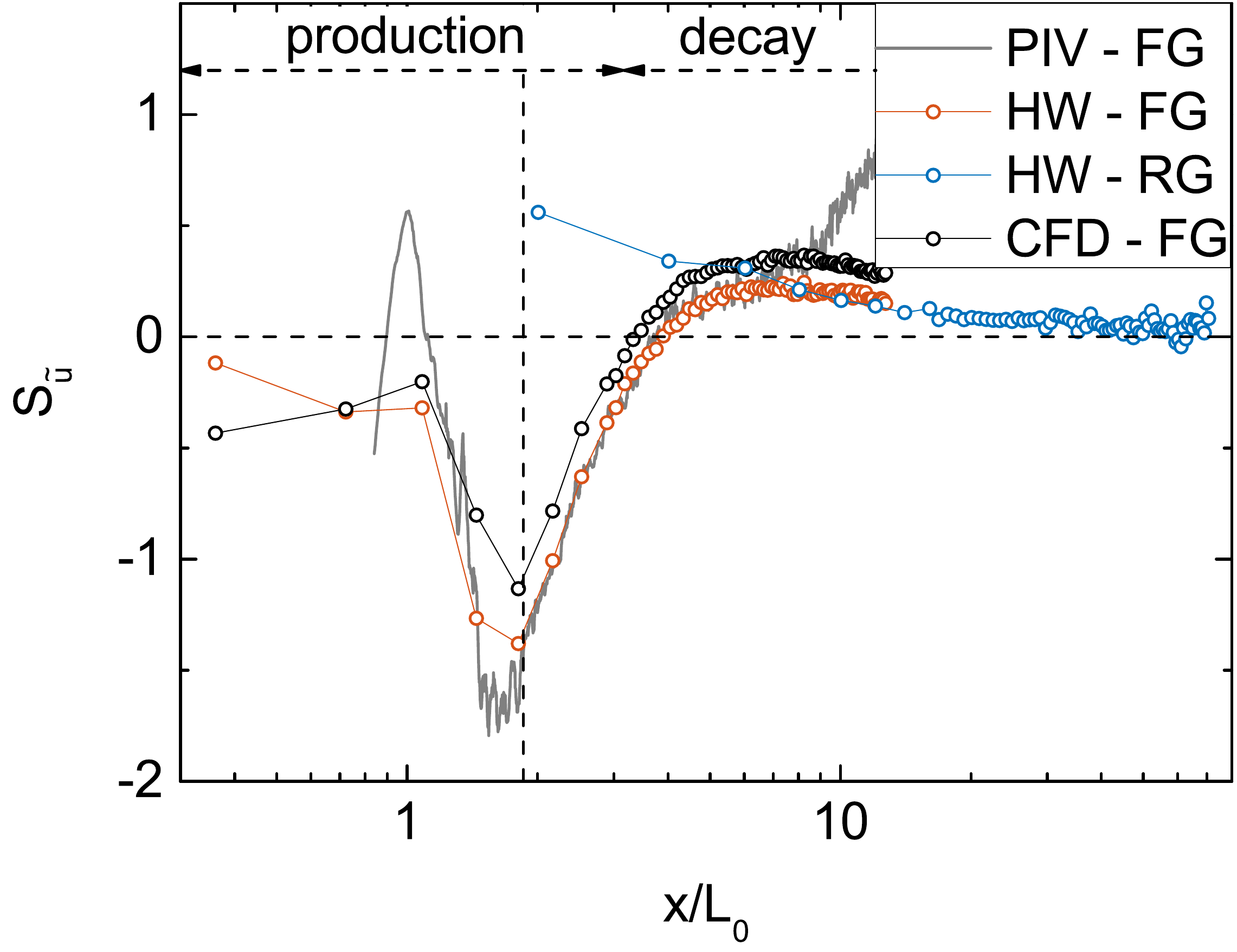}}  
	\subfloat[]{\label{fig:skewness_off}\includegraphics[width=0.425\textwidth]{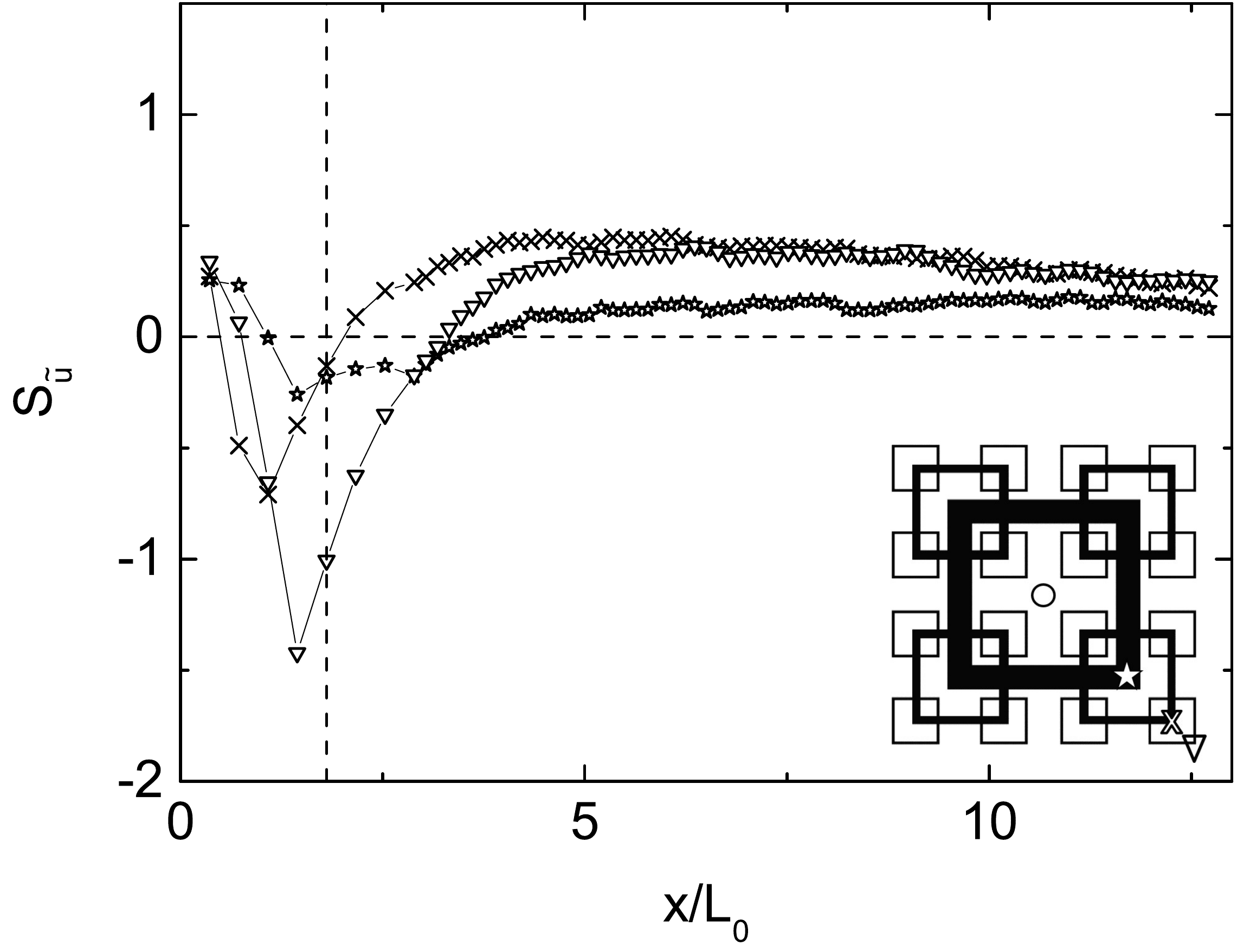}} \\ 
	\subfloat[]{\label{fig:flatness}\includegraphics[width=0.432\textwidth]{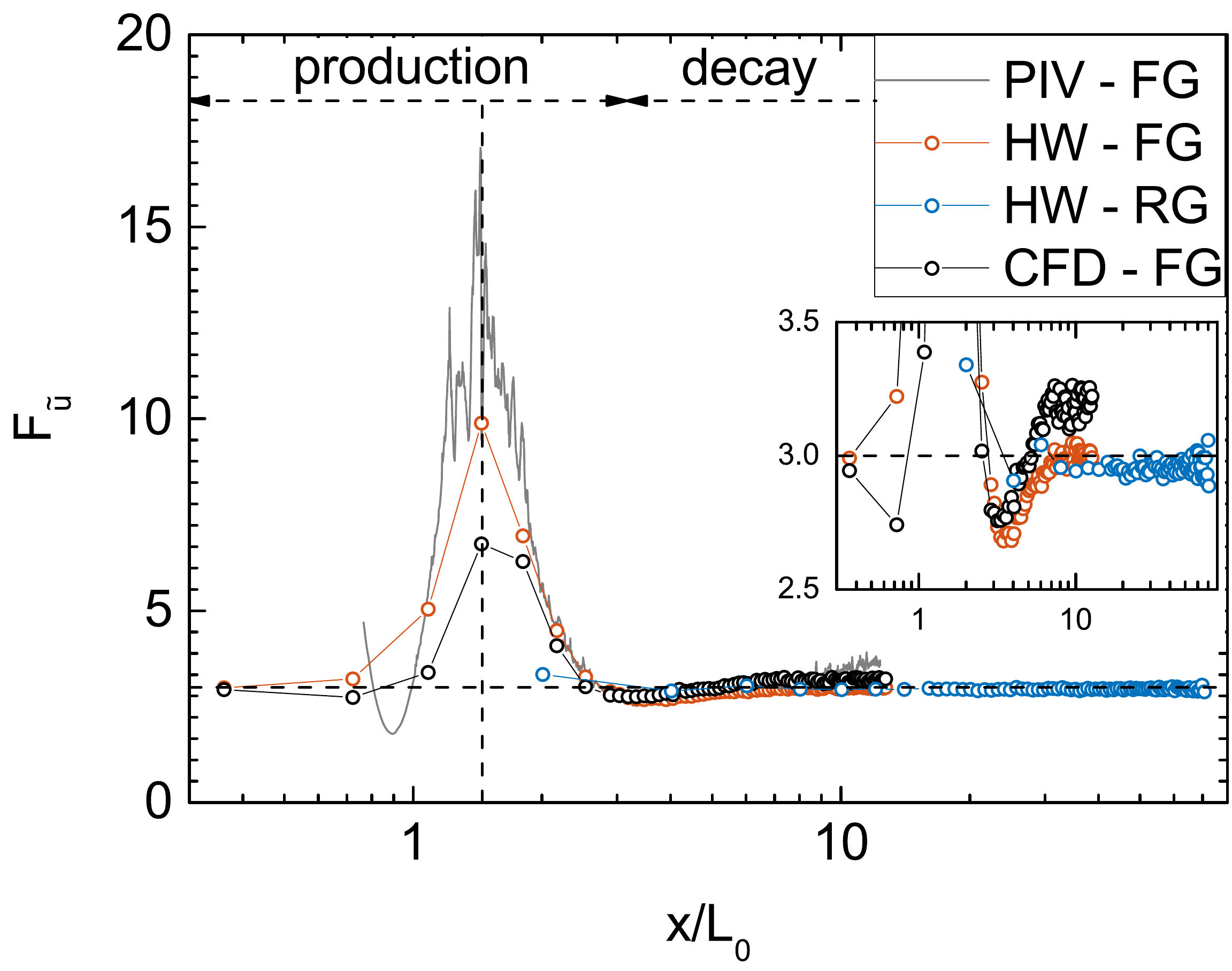}}  
	\subfloat[]{\label{fig:flatness_off}\includegraphics[width=0.425\textwidth]{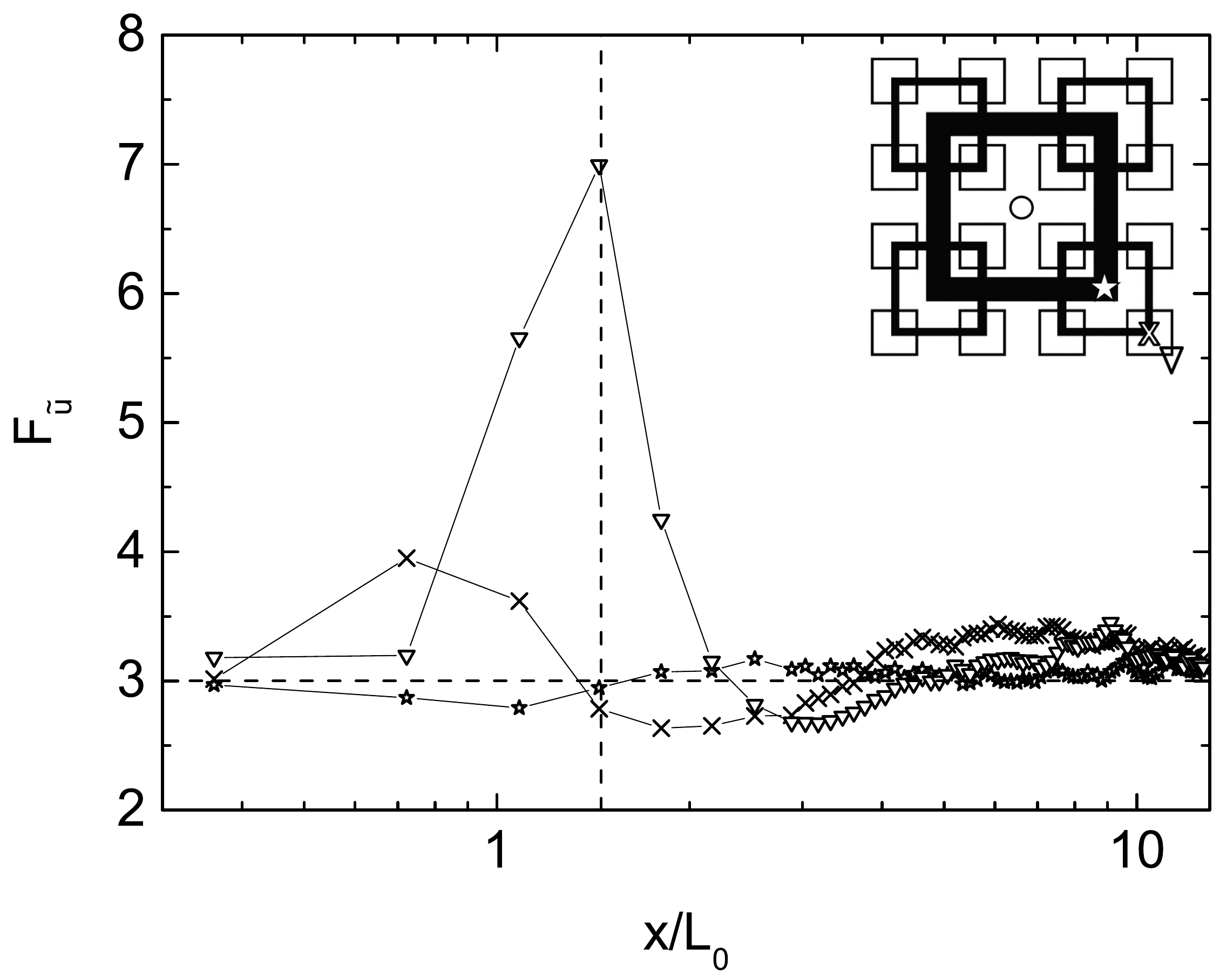}}  
	\caption{Skewness (top) and Kurtosis (bottom) of the longitudinal fluctuating velocity component $\widetilde{u}$ as functions of the distance $x$ downstream of the grid (on the left) along the centerline from hot-wire measurements and computational data and (on the right) along off-centerline positions from computational data. The horizontal dashed line in both upper plots is $S_{\widetilde{u}}=0$ and in the two lower plots is $F_{\widetilde{u}}=3$. The vertical dashed line represents the skewness and kurtosis peak position on the centerline.}
	\label{SundF}
\end{figure*}

\begin{figure*}
	\centering
	\subfloat[]{\label{fig:s_piv}\includegraphics[width=0.425\textwidth]{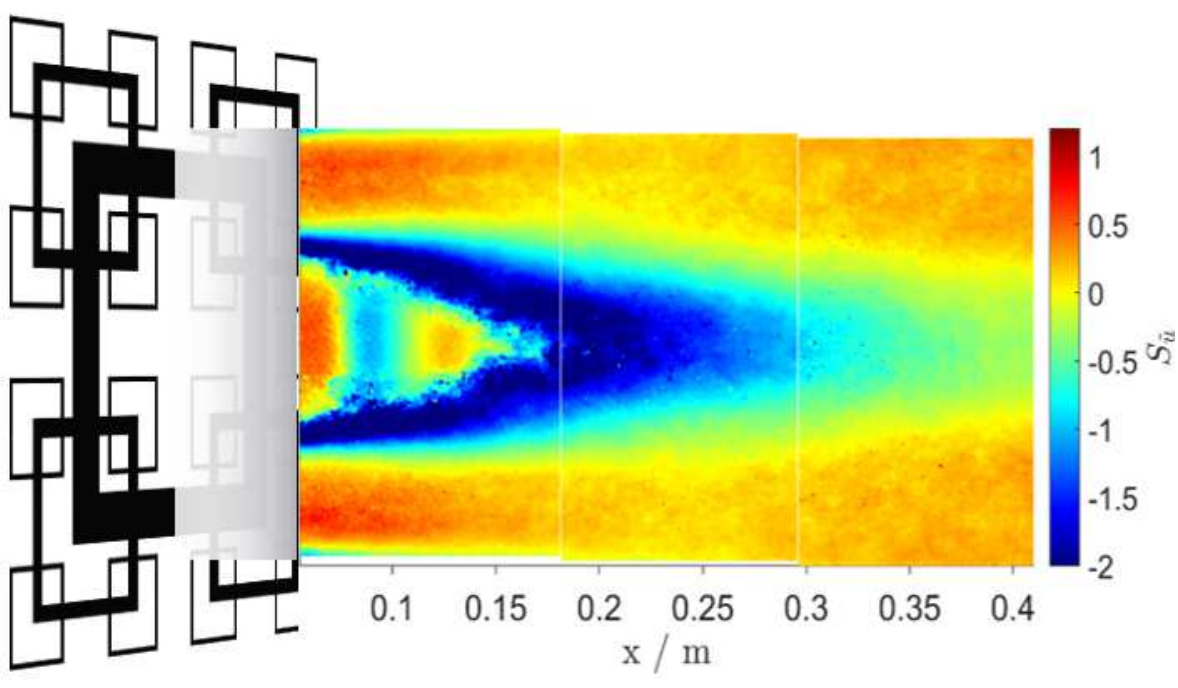}}
	\subfloat[]{\label{fig:f_piv}\includegraphics[width=0.425\textwidth]{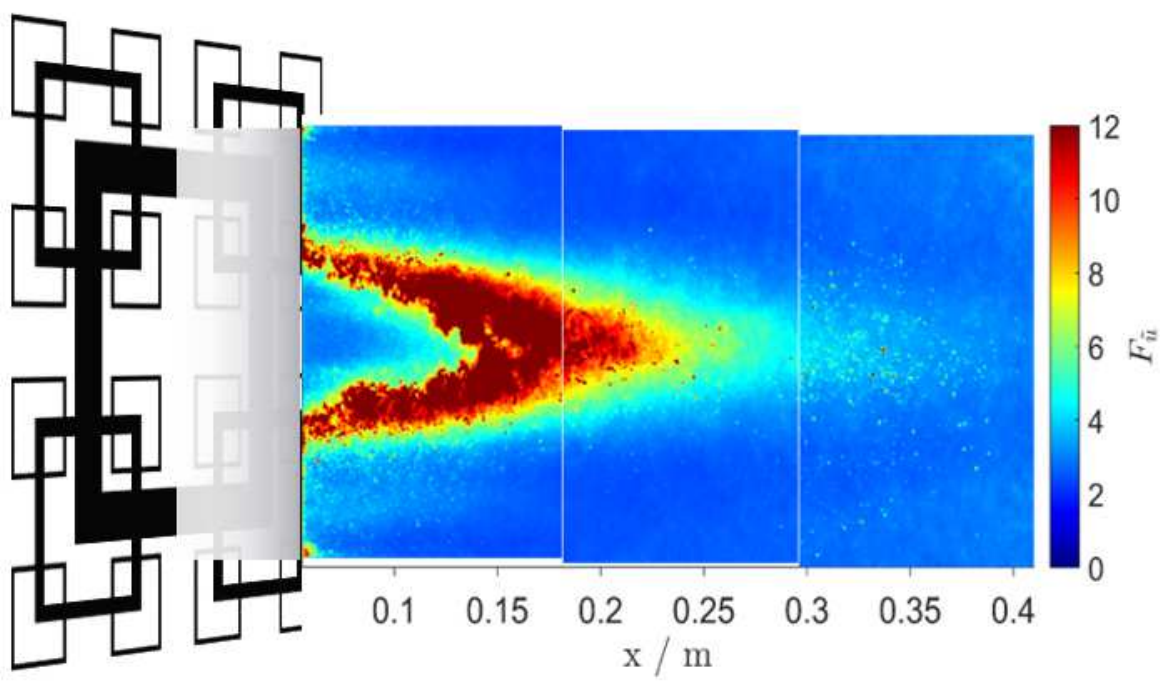}}  
	\caption{Distribution of (a) skewness and (b) kurtosis of the longitudinal fluctuating velocity component $\widetilde{u}$ in the x-y plane along the centerline. The data is obtained by PIV measurements.}
	\label{fig:S_F_PIV}
\end{figure*}

On the left side of Figure \ref{SundF} the evolution of skewness and kurtosis of the longitudinal fluctuating velocity component $\widetilde{u}$ as functions of the distance $x$ downstream of the grid along the centerline is plotted from experimental and computational data. On the right side of Figure \ref{SundF} the evolution of skewness and kurtosis along off-centerline positions is plotted from the computational data. Note that a value of $S_{\widetilde{u}}=0$ and $F_{\widetilde{u}}=3$ corresponds to a Gaussian distribution of the velocity fluctuations. In both figure pairs in addition $S_{\widetilde{u}}=0$ respectively  $F_{\widetilde{u}}=3$ (horizontal dashed line) is shown for comparison. The vertical dashed line represents the skewness and kurtosis peak position on the centerline.
Figure \ref{fig:skewness} and \ref{fig:flatness} show that the evolution of the skewness or kurtosis along the centerline for the regular and the fractal grid are significantly different. For the regular grid, the skewness in the near-grid region is positive and decays with increasing distance to small, yet nonzero, positive values, which is in good agreement with the literature \cite{Bennett_Corrsin_1978,Mohamed_1990}. 
The kurtosis remains almost equal to 3 for the whole length of the test section.
In the case of the fractal grid, the skewness evolution decreases in the production region ($x<0.45$ m) to values smaller than -1 and in the decay region ($x>0.45$ m), the skewness grows along the centerline to even positive values, which saturate for larger distances to small positive values. In contrast to the regular grid, the evolution of the kurtosis is non-constant when passing the fractal square grid. In the production region, the kurtosis reaches values significantly higher than 3 and in the decay region, the flow behind a fractal grid is characterized by having kurtosis values that are close to three. 
As shown in Figure \ref{fig:skewness} and Figure \ref{fig:flatness}, the experimental and computational results are in a fairly good agreement. Although, the computational results show some small deviations from the experimental results. This is due to the short computational time series, collected only for 16 seconds compared to 60 seconds of experimental data, which influences the quality of the calculated higher-order moments. As will be shown later the unusual behaviour of $S_{\widetilde{u}}$ and $F_{\widetilde{u}}$ indicate that extreme velocity fluctuations occur in the production region of the fractal grid (see time series in Figure \ref{fig:time20}). These extreme fluctuations are rare and longer time series for the CFD simulation are needed to obtain convergence. In this context, we also compared the results to an evaluation of only 10 seconds of simulated data and we have clearly seen that the longer the simulated time series is the more suitable the match.
For the evolution of the skewness and kurtosis along three off-centerline positions, (as indicated in Figure \ref{fig:positions_probes}) of the simulated data, shown in Figure \ref{fig:skewness_off} and \ref{fig:flatness_off}, roughly the same tendency as for the centerline is found. However, due to the sequential interactions of the single wakes generated by the different fractal grid bars, as shown in Fig. \ref{fig:inst_velocity_cfd}, the peaks of the skewness and kurtosis evolutions are shifted on the x-axis and less pronounced compared to the evolution along the centerline. The flow behind the biggest grid bar (\FiveStarOpen) shows the same behavior but smaller values compared to the other investigated off-centerline positions. 
Note that the skewness and kurtosis peaks occur approximately at about the same distance from the grid even though it is clearly different from the turbulence intensity peak position (see Figure \ref{fig:Tu}). This applies both for centerline and off-centerline positions. Most interestingly, we see that the evolution of the normalized third and fourth central moments along the centerline and behind the positions $\times$ and $\triangledown$ collapse when plotted against the distance to the grid normalized by the corresponding turbulence intensity peak position.
For a better understanding of the unusual behavior of $S_{\widetilde{u}}$ and $F_{\widetilde{u}}$ as well as the dependence on the position in the flow behind the fractal grid, we examine in more details in Figure \ref{fig:S_F_PIV} their spatial distribution in the production region ($x<0.45$ m) in the x-y plane downstream.  The data (obtained by PIV measurements) is taken on a plane along the centerline. 
Corresponding to the inhomogeneous local blockage of the grid and therefore the inhomogeneous distribution of the flow velocities, both Figures \ref{fig:s_piv} and \ref{fig:f_piv} show a clearly non-constant distribution of the normalized third and fourth central moments, explaining the shift of the peak positions shown in Figure \ref{SundF}. When analyzing the skewness and kurtosis distribution in the plane along the centerline, Figure \ref{fig:S_F_PIV} exhibits clearly that especially in the interaction region between the wake-like region where the flow velocity is low (corresponding to the wakes that are created in the vicinity of the biggest fractal grid bar) and the jet-like region where the flow velocity is relatively high, the skewness reaches values significantly lower than 0 and the kurtosis is for this region high. For the skewness, we find again like in the anisotropy pattern of Figure \ref{fig:iso_piv} fine structures in the region close to the grid, which can be seen in relation to SSU mechanism.
Overall, both the investigation of the normalized third and of the fourth central moments of the fluctuating velocities show a highly non-Gaussian character in the production region of the fractal grid and a return to Gaussian distribution ($pdf\left(\widetilde{u}/u'\right)$) in the decay region. This behavior is highly unusual compared to the turbulent flow generated by the regular grid. Especially in the interaction region between the wake-like and the jet-like-structure extreme velocity fluctuations occur. These extreme fluctuations will be studied in more detail in the next Section.

\subsubsection{Probability density function of velocity fluctuations}
\label{sec:pdf}
After having presented and discussed the analysis of higher order moments, first the computational and experimental acquired time series are further compared using the $pdf\left(\widetilde{u}/u'\right)$ and then a detailed examination of the above-mentioned extreme velocity fluctuations will be presented. These efforts aim to understand the production region in the wake of the fractal grid where the turbulence is still developing. 

\begin{figure*}
	\centering
	\subfloat[]{\label{fig:time20}\includegraphics[width=0.425\textwidth]{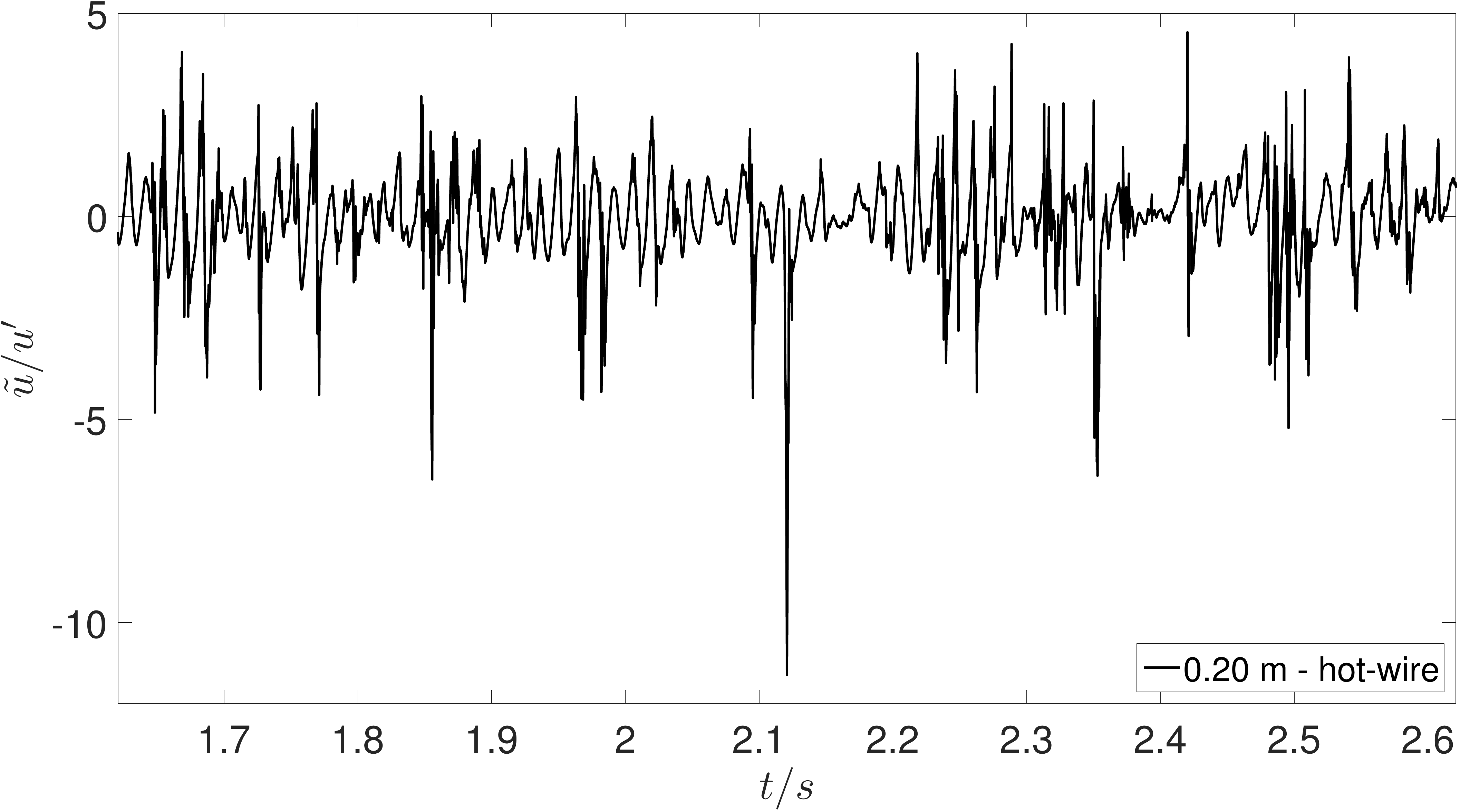}} 
	\subfloat[]{\label{fig:time176}\includegraphics[width=0.425\textwidth]{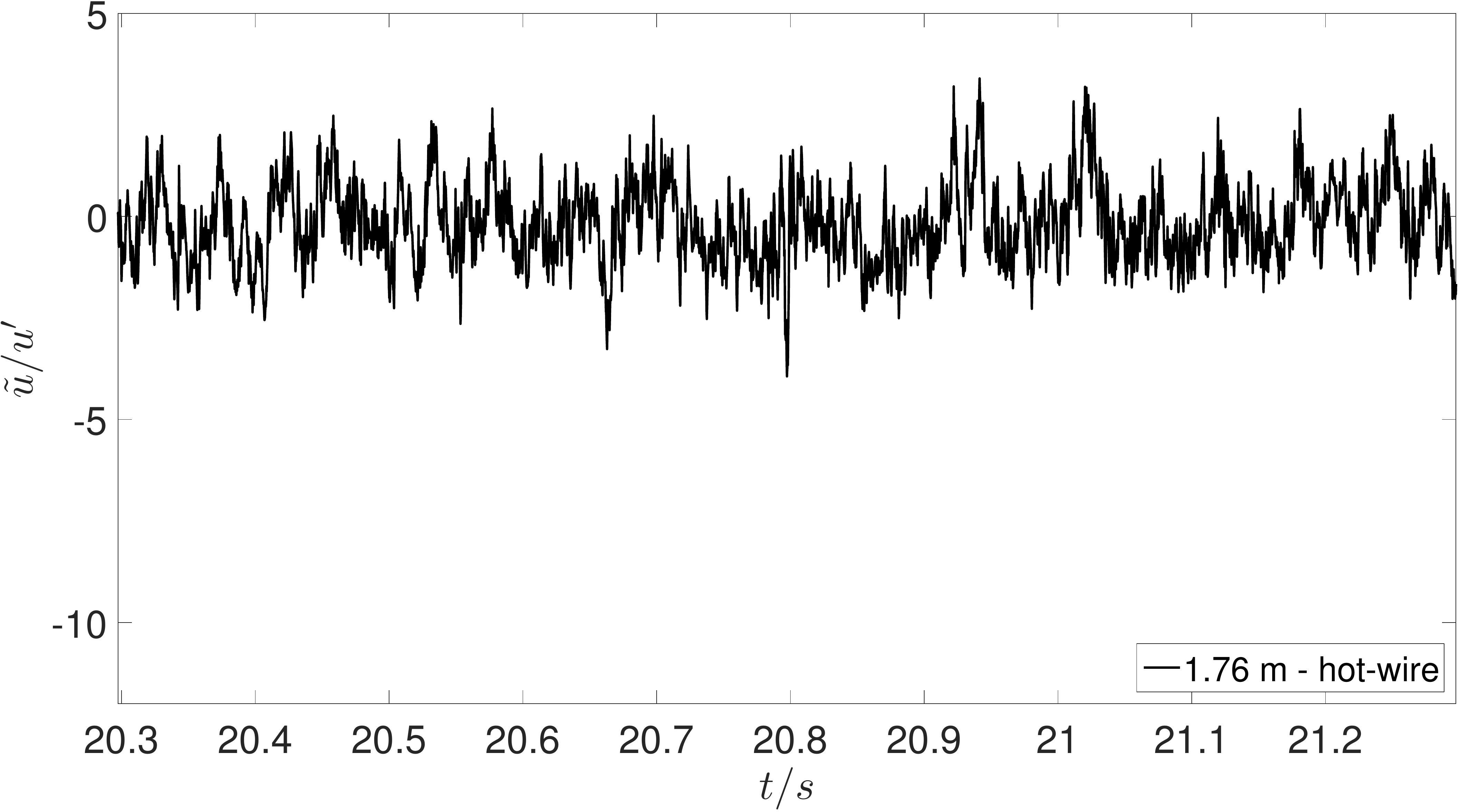}}\\
	\subfloat[]{\label{fig:PDF_exp}\includegraphics[width=0.425\textwidth]{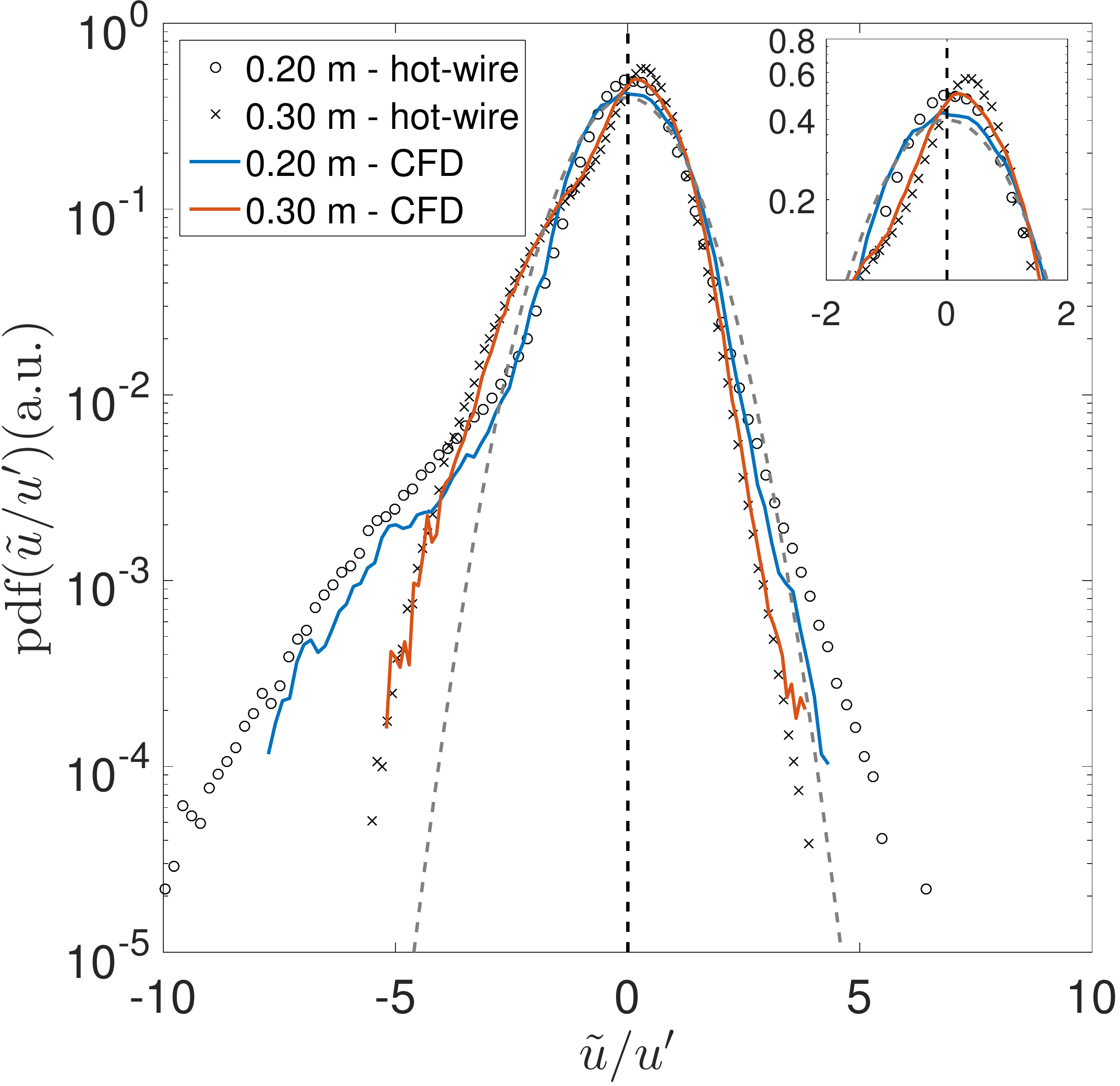}}  
	\subfloat[]{\label{fig:PDF_sim}\includegraphics[width=0.425\textwidth]{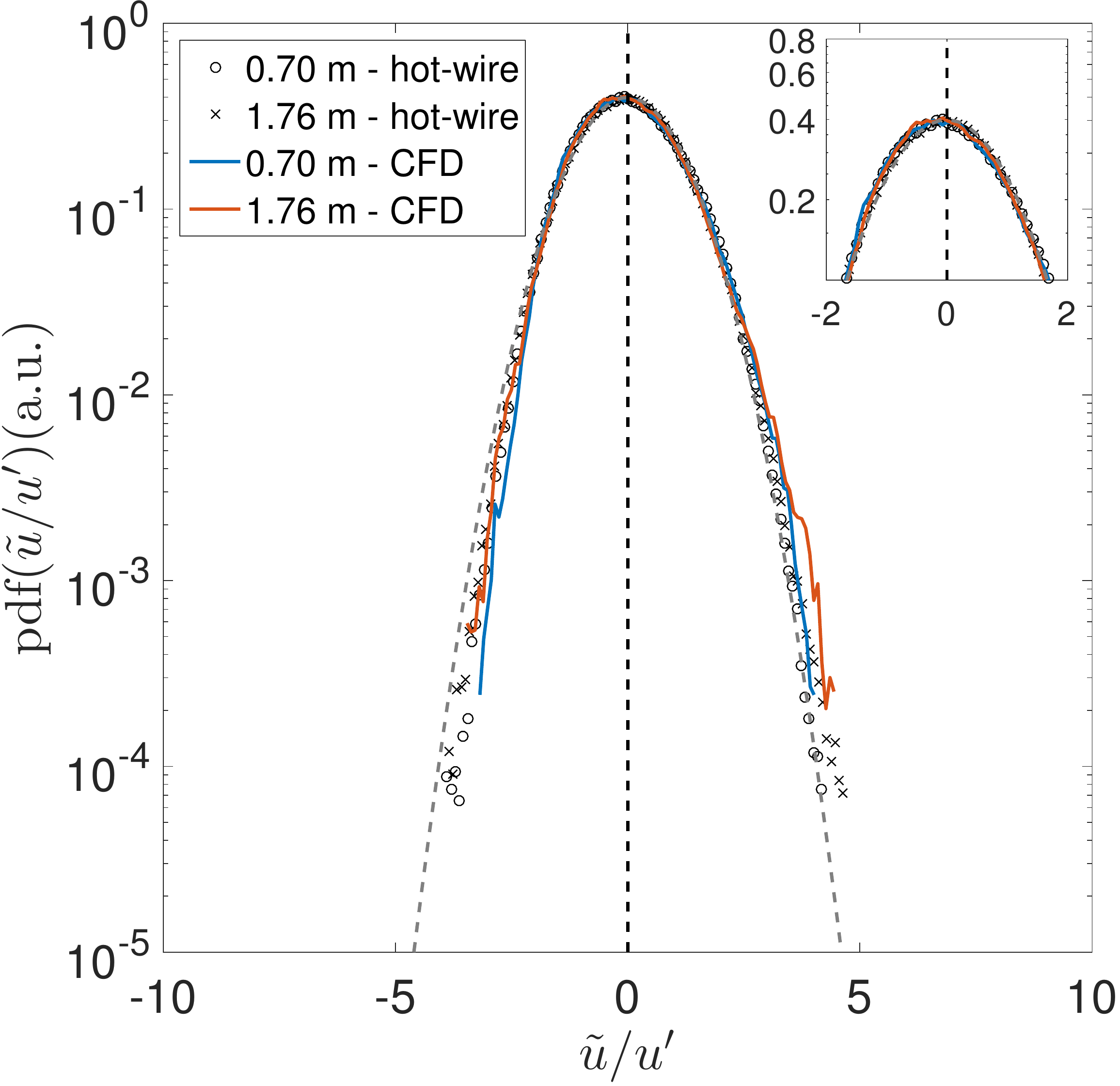}}  
	\caption{A section of 1 second of the experimental acquired time series is plotted for $x=0.20$ m (a) and $x= 1.76$ m (b). Probability density function along the centerline at two positions in the production region $x= [0.20, 0.30]$ m (c) and two $x= [0.70, 1.76]$ m positions in the decay region (d) from experimental data and the numerical simulations data compared to the Gaussian curve (dashed line). The insets represent a zoomed view.}
	\label{fig:PDF}
\end{figure*}

In Figure \ref{fig:time20} and \ref{fig:time176} a section of 1 second of the experimental acquired time series is plotted for one position in the production region $x=0.20$ m and one position in the decay region $x= 1.76$ m (along the centerline). By comparing the two time series it is clearly visible by eye that the turbulent flow at $x=0.20$ m is characterized by significantly larger velocity fluctuations and that these extreme fluctuations occur more frequently compared to the turbulent flow at $x= 1.76$ m. In the previous Section \ref{sec:Skewness and flatness} it has been demonstrated that the normalized third and the fourth central moment peaks at a distance of approximately $x=0.20$ m to the fractal grid. As also reported in \cite{Vassilicos_Mazellier_2010}, the reason for the skewness being non-zero and also for the kurtosis being much higher than the Gaussian value of 3 is the presence of extreme events or highly energetic bursts throughout the production region. For example, our measurements at the kurtosis peak position along the centerline show that when one of these extreme events occurs, the velocity drops by 11 times the standard deviation of the velocity within less than 5 ms (see Figure \ref{fig:time20}). Using Taylor's hypothesis, this corresponds to structures of about 0.08 m. These extreme fluctuations are also reflected in the probability density function of velocity fluctuations.
In Figure \ref{fig:PDF_exp} and \ref{fig:PDF_sim} the $pdf\left(\widetilde{u}/u'\right)$ normalized to the standard deviation for two positions in the production region $x= [0.20, 0.30]$ m and two positions in the decay region $x= [0.70, 1.76]$ m are plotted for the measurement position on the centerline from the experimental and the computational data. Gaussian distribution with a mean of 0 and standard deviation of 1 is plotted for comparison (dashed line). 

In accordance with the analysis of the normalized third and fourth central moments, the pdfs are highly non-Gaussian and are left-skewed (towards negative $\widetilde{u}$ values) in the production region, represented by the asymmetric pdfs at the positions $x= [0.20, 0.30]$m. A pronounced negatively skewed distribution indicates that the left tail is longer or fatter than the right tail and the peak of the pdf is slightly shifted toward positive velocity fluctuations so that the mass of the distribution is concentrated on the right of the mean as indicated in the inset of Figure \ref{fig:PDF_exp}. Furthermore, this behavior indicates that the extreme negative fluctuation values are far more likely than positive ones throughout the production region. The negatively skewed distribution and the extreme fluctuations on the time series correspond to a locally decelerating flow due to the interaction of jet-like (high velocity) and the wake-like flow (low velocity). Compared to the Gaussian distribution, the pdfs in the production region are characterized by heavy tails and a sharper peak near the mean, which further proves the extreme nature of the events occurring in this region. Thus, higher values of $\left|\widetilde{u}\right|$ are much more likely to occur than the Gaussian distribution. Comparing the pdfs at the locations of $x=0.20$ m and $x=0.30$ m it becomes obvious that the region with extreme bursts is concentrated only on a small region of the wake flow, see also Figure \ref{fig:S_F_PIV}.

The pdfs are approximately Gaussian when plotted for positions along the centerline in the decay region, at the positions $x= [0.70, 1.76]$m. In this region the distributions are almost symmetric and $S_{\widetilde{u}}$ and $F_{\widetilde{u}}$ takes values close to the Gaussian values. 
To summarize, the analysis of the pdfs confirms the results indicated by the skewness and kurtosis evolution (see Section \ref{sec:Skewness and flatness}).  
When comparing the pdfs relating to the experimental data with the pdfs relating to the simulated data, it becomes obvious that, at least for the qualitative behavior, the two techniques agree quite well. Both CFD and experiments show that the generated flow behind the fractal grid is characterized by non-Gaussian properties close to the grid and becomes nearly Gaussian in the decay region. The slight deviations are primarily attributable to the short computational time series.

\begin{figure*}
	\centering
	\subfloat[]{\label{fig:pro_20_fit}\includegraphics[width=0.33\textwidth]{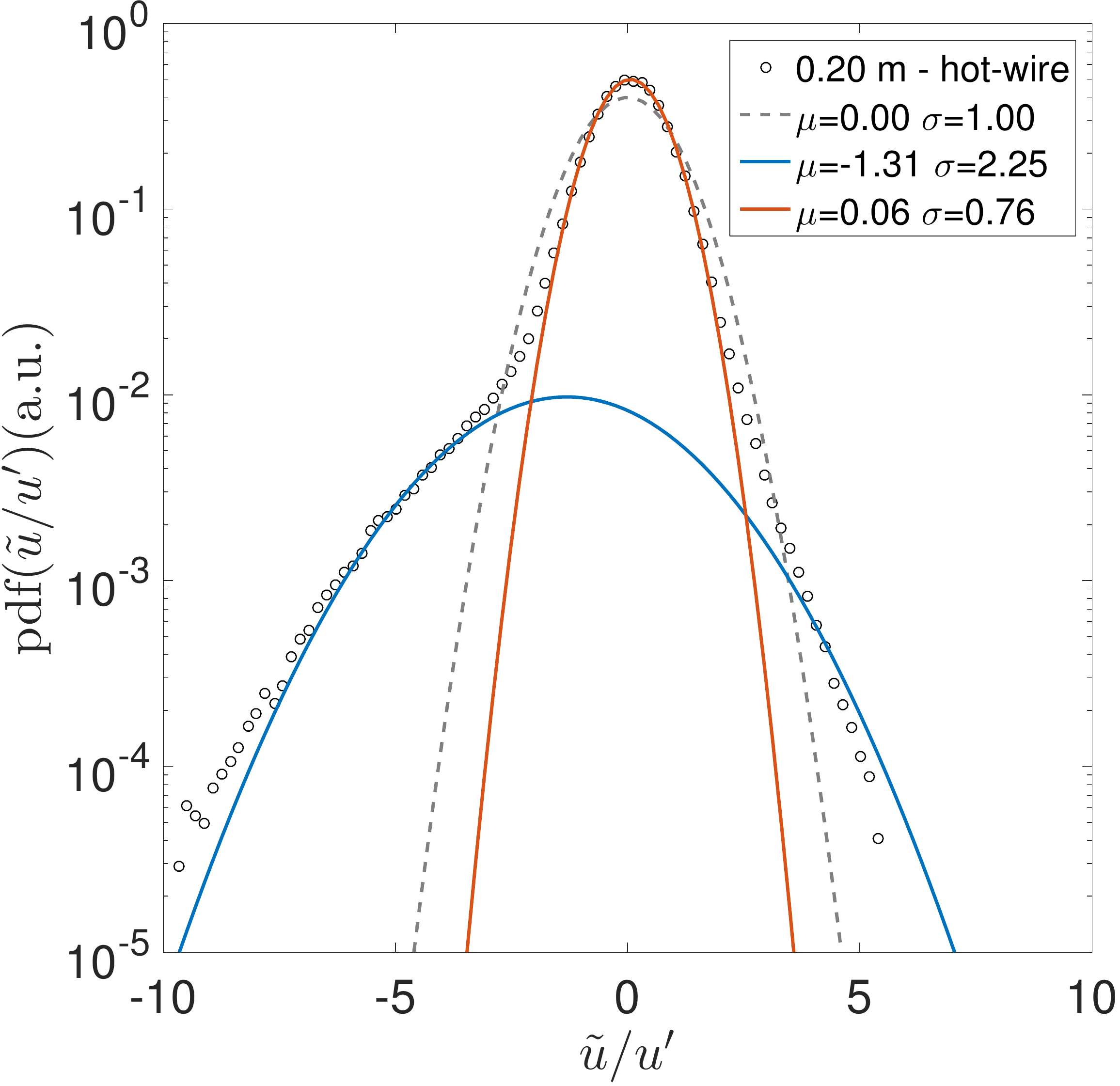}}
	\subfloat[]{\label{fig:pro_25_fit}\includegraphics[width=0.33\textwidth]{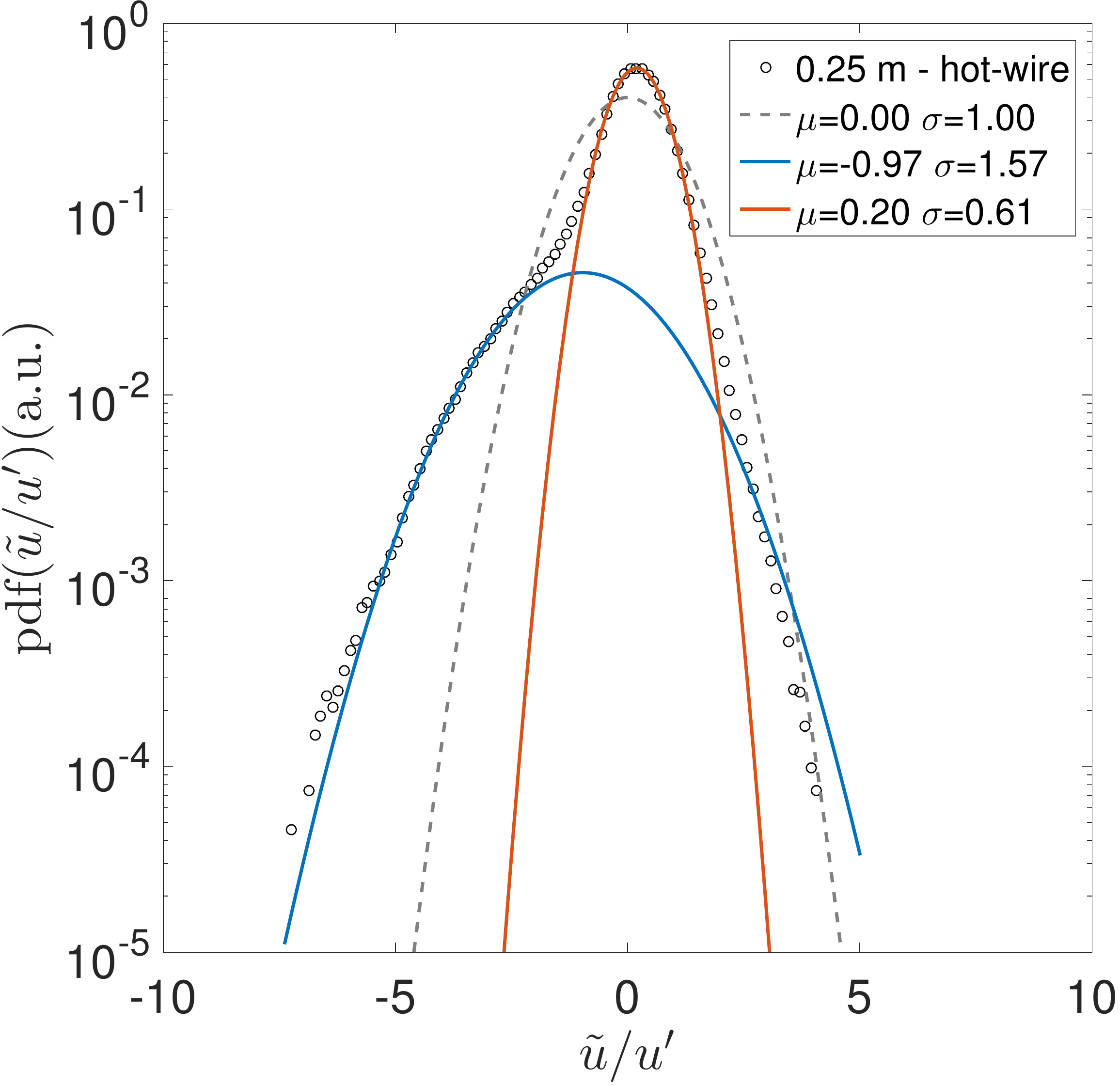}}
	\subfloat[]{\label{fig:pro_30_fit}\includegraphics[width=0.33\textwidth]{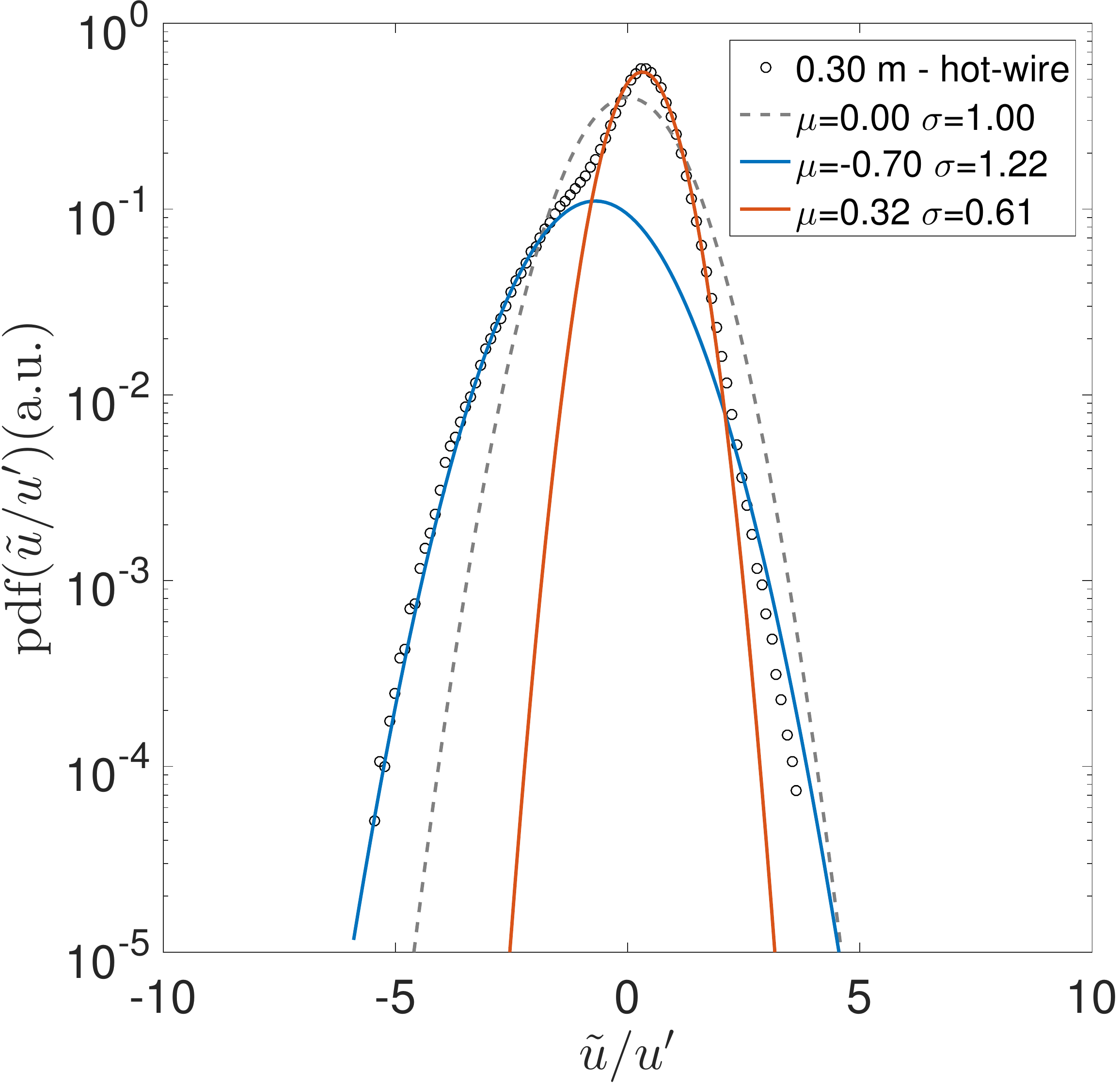}}
	\caption{Probability density function in the production region at $x= [0.20, 0.25, 0.30]$ m from experimental data compared to a Gaussian distribution with mean of 0 and standard deviation of 1 (dashed line). Solid lines indicate a nonlinear least-squares error fit using a Gaussian model.}
	\label{fig:pro_fit}
\end{figure*}

As shown in Section \ref{sec:Streamwise turbulence intensity distribution} the turbulent flow generated by a fractal grid changes its turbulent state from developing to fully developed and to decayed states of turbulence with an increase in the distance to the grid. To get an enhanced insight in the complexity of the fractal grid wake, it is necessary to get a basic understanding of the generated turbulence in the near-grid region where the turbulence is still developing. Here we look at $\widetilde{u}$ because the velocity fluctuations are dominated by the energy-containing large-scale motions, which are especially pronounced in this region, as described in Section \ref{chap:Instantaneous velocity magnitude distribution in the x-y plane}.
Based on the results shown in Figure \ref{fig:PDF}, we assume that the distribution of velocity fluctuations can be approximated by the superposition of different Gaussian distributions. Under this assumption we plotted in Figure \ref{fig:pro_fit} the $pdf\left(\widetilde{u}/u'\right)$ at $x= [0.20, 0.25, 0.30]$ m from experimental data. In addition, the solid lines indicate a nonlinear least-squares error fit to different regions of velocity fluctuations to find the parameters (mean $\mu$ and width $\sigma$ in terms of $u'$ of the distribution modeling) giving the best-fitting Gaussian distribution. 
A Gaussian distribution with mean of 0 and a standard deviation of 1 is plotted for comparison (dashed line).

In Figure \ref{fig:pro_fit} it can be seen clearly that the distribution indicated by the solid blue lines corresponds to a Gaussian distribution with a negative mean $\mu$ and a significantly larger width $\sigma$ compared to the distribution indicated by the dashed line. Whereas the distribution indicated by the solid red lines corresponds to a Gaussian distribution with a slightly positive mean $\mu$ and a smaller width $\sigma$ compared to the standard deviation of 1 (see legend added to plots). Figure \ref{fig:pro_fit} demonstrates that the $pdf\left(\widetilde{u}/u'\right)$ can be well approximated by the superposition of two different Gaussian distributions which converge to one Gaussian distribution with increasing distance to the grid (see also Figure \ref{fig:PDF_sim}). In connection with the space-scale unfolding (SSU) mechanism introduced by \cite{laizet2012fractal} one possible explanation for this behavior is that at small distances from the grid ($x<0.45$ m), where the turbulence is still developing, the turbulent flow consist of two different turbulent structures; the jet-like flow induced by the large opening at the center and a flow with a complex wake interaction and mixing. In this sense, the pdfs can be approximated by the distributions indicated by the red and blue solid lines, respectively. Taking also Figure \ref{fig:CL-instvel} and Figure \ref{fig:LIC_middle} into account, it can be argued that in the near-grid region, the presence of large-scale spatial structures (originating from the wakes of the grid bars), which intermittently pass along the centerline implies large deviation from the average value $\left\langle u \right\rangle$ represented by a distribution that is more outlier-prone (indicated by solid blue lines) than the Gaussian distribution with a mean of 0 and standard deviation of 1 (dashed line). Further downstream, the large-scale wakes of the grid bars evolved to energy-containing eddies that are approximately random and independent of their initial conditions. 
Therefore the turbulent flow in the production region is in a transition state from bimodal motions to developed turbulence with an increase in the distance to the grid.

\subsection{Results in terms of two-point statistics}
\subsubsection{Energy spectrum density}
\label{sec:Energy_spectrum_density}
\begin{figure*}
	\centering
	\subfloat[]{\label{fig:frac_pro}\includegraphics[width=0.425\textwidth]{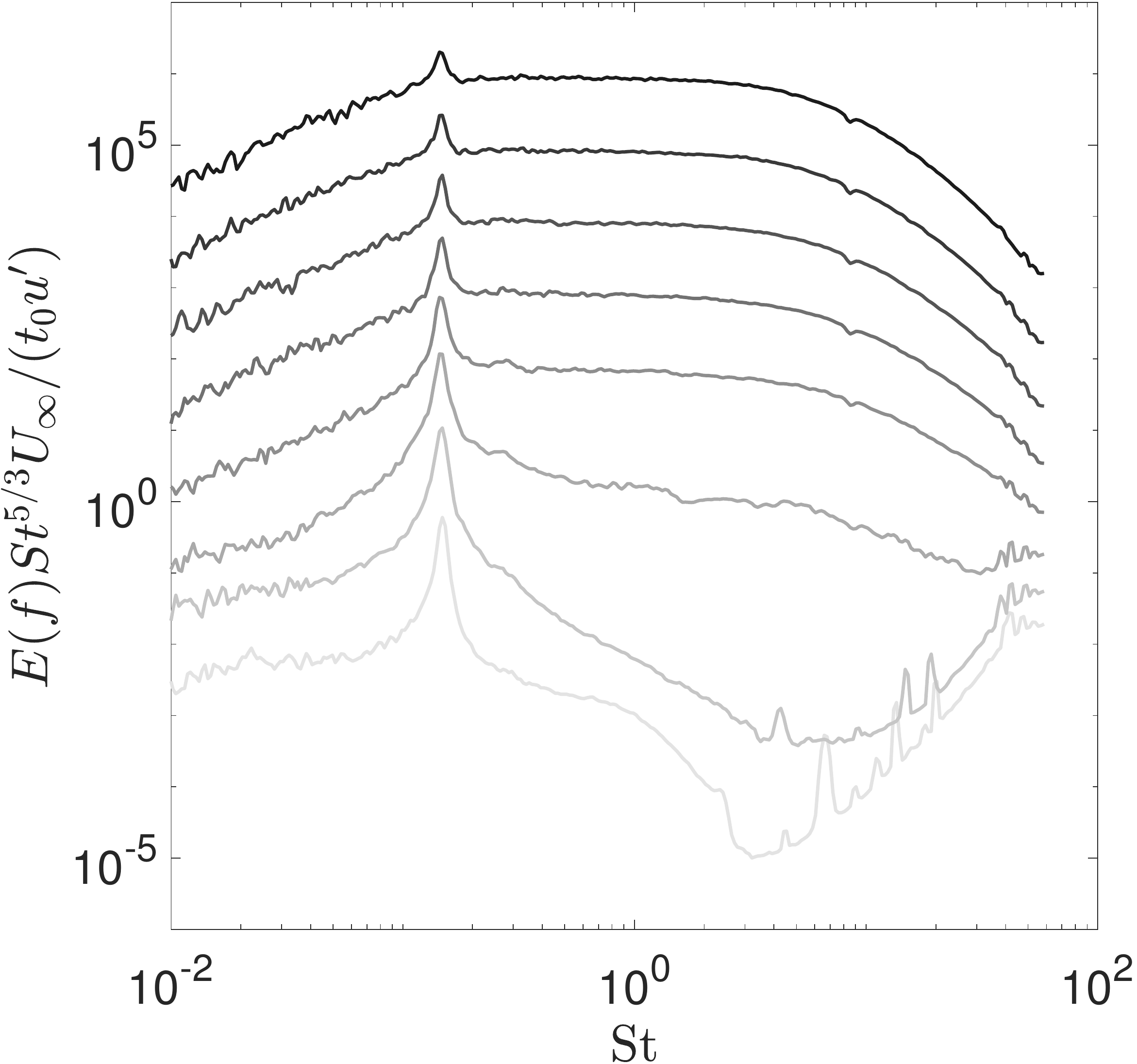}} 
	\subfloat[]{\label{fig:frac_pro_decay_norm}\includegraphics[width=0.425\textwidth]{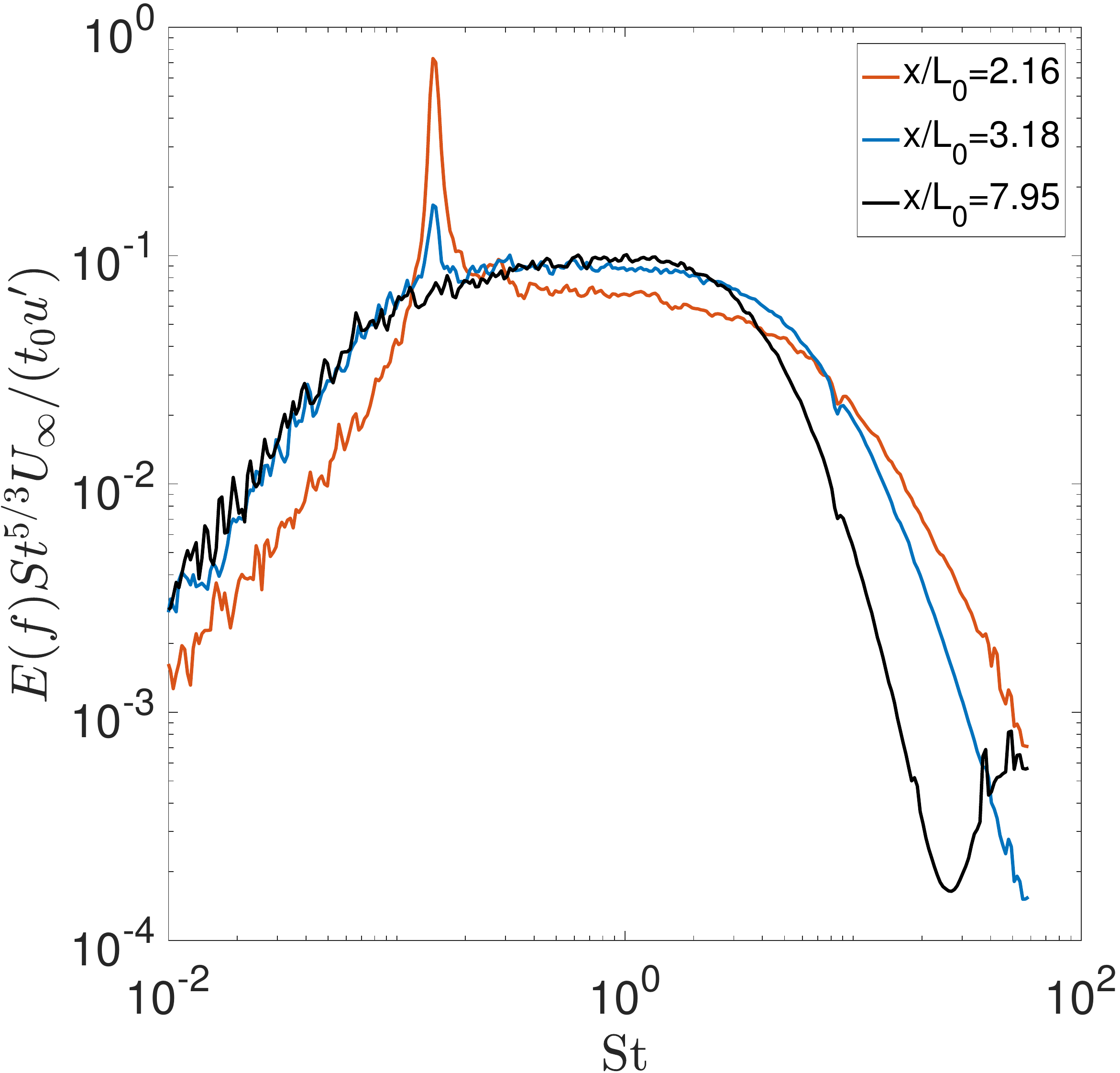}}
	\caption{(a) Spatial evolution of compensated one-dimensional energy spectrum density of $\widetilde{u}$ at five different centerline positions in the production region of the fractal grid. From bottom to top $x/L_{0}=$0.36, 0.72, 1.08, 1.45, 1.80, 2.16, 2.53, 2.89. 
	(b) Compensated energy spectrum plotted at one position in the production region $x/L_{0}=2.17$, at the position where the turbulence intensity peak occurs $x/L_{0}=3.18$ and at one position in the decay region $x/L_{0}=7.95$ from experimental data in a log-log plot.}
	\label{fig:frac_spec}
\end{figure*}

Next, an examination of streamwise fluctuating velocity component energy spectra along the centerline is presented. Conversion from frequency $f$ to wavenumber $k$ domain (wavenumber in the direction of mean velocity) was done utilizing Taylor hypothesis:
\begin{equation}
	k=2\pi f/\langle u\rangle \hspace{0.1cm} ; \hspace{1cm}  E(k)=E(f)\langle u\rangle/2\pi.
\end{equation} 
\noindent The resulting energy spectrum $E(k)$ has the interpretation that $\int^{\infty}_{0} E(k)dk=\left\langle \widetilde{u}^2\right\rangle=u'^2$, accordingly the area under the energy spectrum is equal the value of the variance of the time series (Parseval’s theorem). Figure\ref{fig:frac_pro} shows the spatial evolution of the compensated one-dimensional energy spectrum density 
at eight different streamwise positions in the production region ($0.3<x/L_{0}<3$) of the fractal grid. Frequencies have been converted to Strouhal numbers $St$ using $t_0$ (thickness of the biggest gird bar) and the inflow velocity $U_\infty$. The vertical axis has been normalized with $St^{5/3} U_{\infty}/(t_0 u')$. Each spectrum has been shifted by one decade to show the increase of the flat region in the compensated spectra. In Figure \ref{fig:frac_pro_decay_norm} compensated spectra is plotted at one position in the production region $x/L_{0}=2.16$, at the position where the turbulence intensity peak occur $x/L_{0}=3.18$ and at one position in the decay region $x/L_{0}=7.95$ from experimental data in a log-log plot. 

In Figure \ref{fig:frac_pro} it can be seen clearly that the spectra have a single peak very close to the fractal grid. This distinct peak can be associated with the Kármán vortex shedding signature of the biggest grid bar with a normalized frequency $f_{peak} t_0/U_\infty=0.14$ (Strouhal number). 
Furthermore, this Figure reveals how turbulence is generated in the production region. As the distance to the grid increases, there appears to be a transition in the nature of the spectra. Energy progressively increases for $f>f_{peak}$ further downstream from the grid. As the turbulence approaches the energy peak at $x/L_o=3.2$ 
 the compensated spectra become flat. Thus there exist a region with a well-defined power-law signature close to $-5/3$ over a significant range in the wavenumber domain. Interestingly, the $-5/3$ shape of the spectra originate relatively near the fractal grid, where the turbulence is still developing and the statistics of fluctuating velocity are intermittent and non-Gaussian. Clearly, in the production region of the fractal-generated turbulent flow, non-equilibrium characteristics are present (none of the assumptions made in the Kolmogorov theory is valid), but the spectra still show a wide scaling behavior related to $-5/3$ Kolmogorov law.    
These results are in agreement with observations made in previous studies \cite{laizet2015spatial,gomes2015energy}, which also find a $-5/3$ scaling region in the near-field of the fractal grid. 
Figure \ref{fig:frac_pro_decay_norm} shows that the intensity of the single peak in the spectra gets attenuated as one proceeds downstream in the decay region. Comparing this with results presented in section \ref{sec:Skewness and flatness} it becomes clear that there is a relationship between the vortex shedding and the normalized third and fourth central moment of the fluctuating velocities $\widetilde{u}$. In the region where vortex shedding is detectable, the skewness and flatness indicate a highly non-Gaussian character, whereas when vortex shedding is not significant, a return to the Gaussian distribution of velocity fluctuations is observable. Moreover, in Figure \ref{fig:frac_pro_decay_norm} all presented spectra, both in the production region and in the decay region, show in the inertial range the common scaling behavior of homogeneous isotropic turbulence. As the distance to the grid increases, the extent of the inertial range diminishes due to the decay of $Re_{\lambda}$ with downstream distance (see Section \ref{sec:turbulent length scales}), this is in agreement with the observations made in \cite{laizet2015spatial}.

\begin{figure*}[htbp]
	\centering
	\subfloat[]{\label{fig:norm_spec}\includegraphics[width=0.425\textwidth]{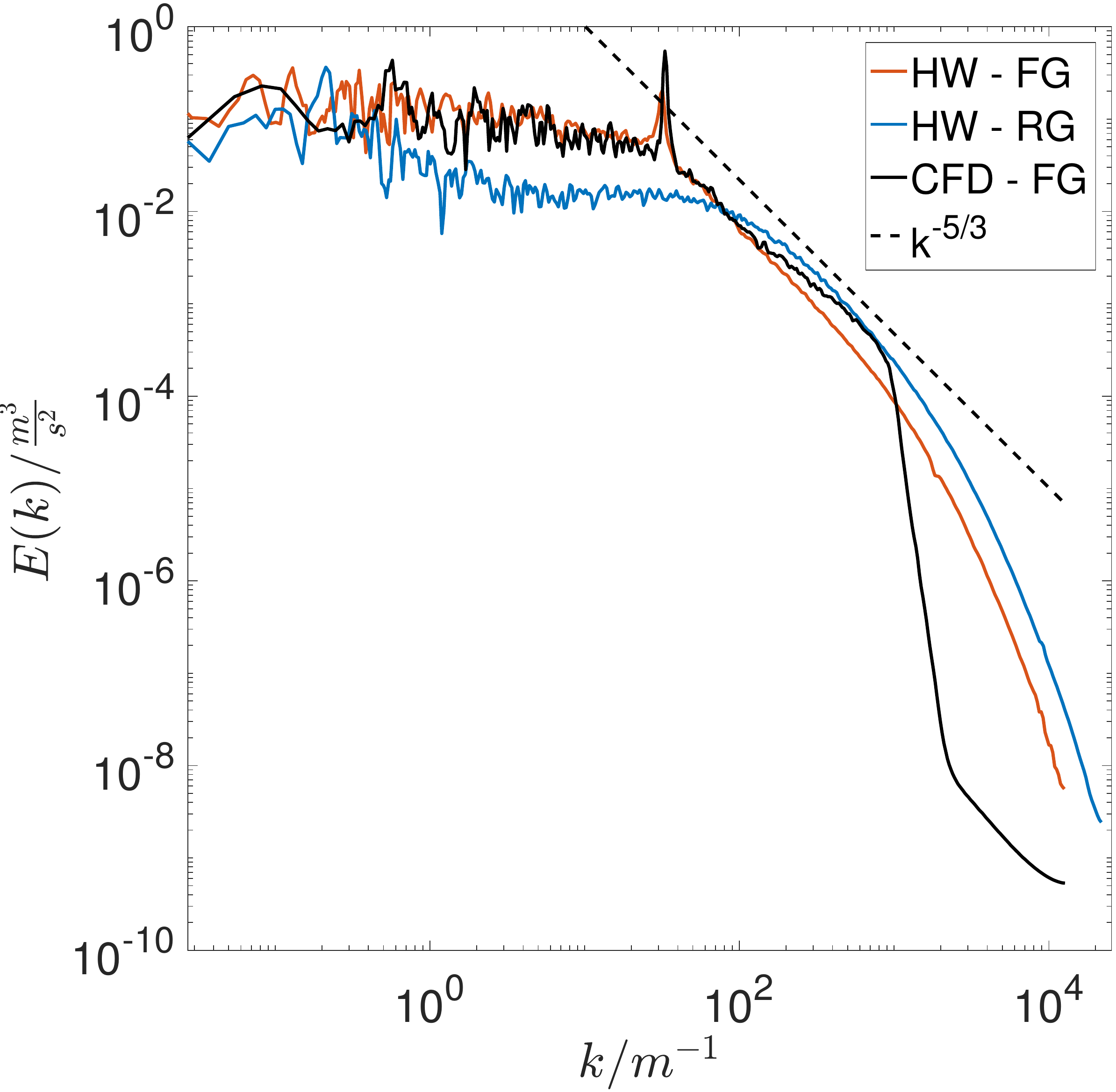}} 
	\subfloat[]{\label{fig:autokorr_vortex}\includegraphics[width=0.42\textwidth]{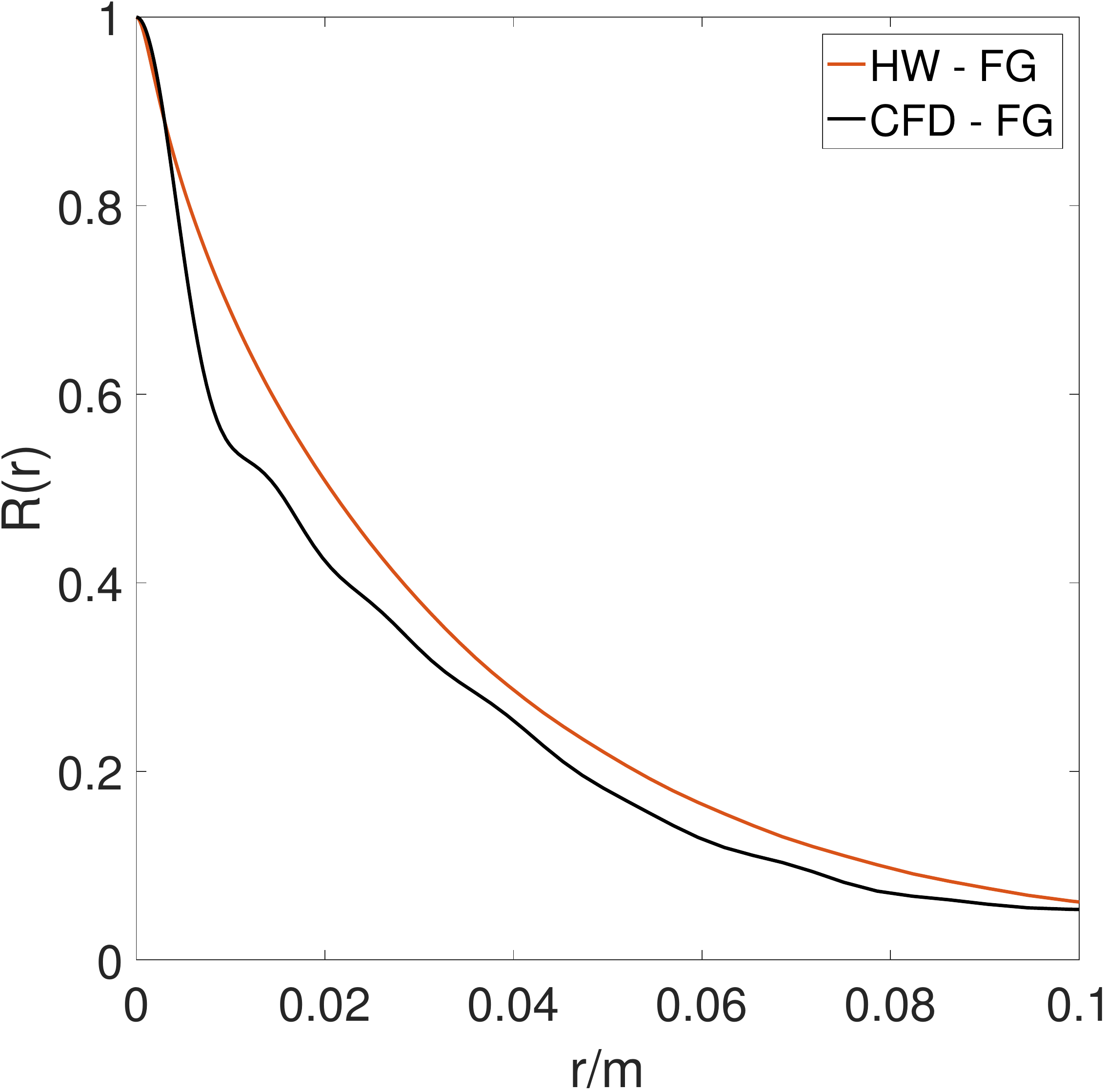}} 
	\caption{One-dimensional energy spectrum density (a) and autocorrelation function (b) of $\widetilde{u}$ obtained computationally and experimentally at the same position (fractal grid $x/L_0=2.17; Re_{\lambda}==$, regular grid $x/L_0=2.01; Re_{\lambda}=?$). The black dotted-line indicates a slope of $k^{-5/3}$.}
\end{figure*}

Finally in Figure \ref{fig:norm_spec} we compare energy spectrum $E(k)$ obtained computationally and experimentally at the same position (fractal grid $x/L_0=2.17$, regular grid $x/L_0=2.01$) generated by the three different grids.

In contrast to the regular grid spectra, both the numerically and experimentally acquired spectra show in the near-grid region a distinct peak. The effect of vortex shedding is not visible in the spectra corresponding to the regular grid. Since the single peak is detectable for both the experiment and the simulation, it can be assumed that it is not an effect of the wind tunnel or experimental setup. In addition, it can be seen that the extent of the inertial range for the regular grid is much smaller than the fractal grid. This observation is confirmed when calculating the local Reynolds number based on Taylor length scale $Re_{\lambda}$ (see Section \ref{sec:turbulent length scales}). The effect of vortex shedding is evident in the autocorrelation function plotted in Figure \ref{fig:autokorr_vortex} (see eq. \eqref{eq:integralLength}). As mentioned by \cite{tennekes1972first} a distinct peak in the spectrum generates a decaying oscillation in the correlation. Close to the grid $R(r)$ shows a periodic behavior, whereas this characteristic is damped for a larger distance to the grid. When considering instead the measurements performed for the regular grid $R(r)$ exhibit a different form depending on the different grid geometry.
Since, Delayed Detached Eddy Simulation is a hybrid method, which involves the use of Reynolds Averaged Navier-Stokes Simulation (RANS) at the wall and Large Eddy Simulation (LES) away from it, three regions can be identified in the spectra corresponding to the simulated data. 
Up to a wavenumber of approximately 150 $m^{-1}$ (frequency of 350 Hz), the spectra corresponding to the simulation are in a good agreement with the experimental results. In this region, the large, energy-containing scales of the turbulent flow are fully resolved. The range between 150 $m^{-1}$ or 350 Hz and 2600 $m^{-1}$ or 6000 Hz can be identified as a "transition region" between the LES region at low frequencies (fully resolved large scales) and the RANS region at high frequencies (modeled small scales). A large deficit in the energy spectra is measurable in this region compared to the experimental spectra. This is a signature of unresolved subgrid scales or structures in the simulated turbulent flow field. At wavenumbers above 2600 $m^{-1}$ or frequencies above 6000 Hz, the spectrum of turbulent motions of small scales is solved using a turbulence model, which leads to the gap between the experimental and computational spectra. Within the framework of this investigation, a significant filtering effect resulting from the RANS modeling on the spectra is clearly visible in the simulation. This filtering effect (for approximately $r<1$ cm) is also visible in the autocorrelation function plotted in Figure \ref{fig:autokorr_vortex}.

\subsubsection{Turbulent length scales}
\label{sec:turbulent length scales}
We now report on measurements of longitudinal integral length scale $L_u$ and Taylor microscale $\lambda$. $L_u$ and $\lambda$ are important quantities that can deliver valuable information about the turbulent flow. It should be noted here that only in the far-field of the lee of the fractal grid in the region around the centerline turbulence is approximately homogeneous and isotropic (see Section \ref{chap:Mean velocity distribution in the x-y plane} and \ref{sec:Large-scale anisotropy distribution in the x-y plane}). The determination of longitudinal integral length scale $L_u$ is based on the definition in \cite{frisch1996turbulence, Pope_book2000}, where the autocorrelation function $R(r)$ of the fluctuating velocity component $\widetilde{u}$ is integrated from zero to infinity 
\begin{equation}
L_u=\int^{\infty}_{0} R(r) dr = \int^{\infty}_{0}\frac{\left\langle \widetilde{u}(x) \widetilde{u}(x+r) \right\rangle}{u'^2(x)} dr,
\label{eq:integralLength}
\end{equation}
\noindent where $x$ indicates the spatial variable and is obtained from time $t$ by means of the local Taylor hypothesis. Here, $r$ is the separation in the spatial domain between two velocities. Ideally, $R(r)$ should decay monotonically to zero for large $r$, however in experiments, the correct determination of $L_u$ is not straightforward and is often associated with evaluation errors for example due to a weak oscillatory tail and several zero-crossings of $R(r)$. This, for example, is the case if vortex shedding is perceptible in a turbulent flow. To overcome this problem the integration domain, thus a threshold value $r_{up}$ as the upper limit, can be specified in a number of ways (see \cite{o2004autocorrelation}). An alternative possibility to eq \eqref{eq:integralLength}, is to use one-dimensional energy spectrum (see \cite{hinze1975turbulence,roach1987generation}). This study uses the following expression, valid for truly homogeneous and isotropic turbulence
\begin{equation}
L_u= \lim\limits_{f \rightarrow 0}\left[\frac{E(f) \langle u\rangle}{4u'^2} \right].
\label{eq:integralLength_spec}
\end{equation} 

\noindent Parameter $\lim\limits_{f \rightarrow 0}E(f)$ is obtained by an extrapolation of  the energy spectrum. Although such a procedure leads to a smaller integral length scale compared to the standard procedure of \cite{Batchelor1953},  it was applied to the data from experiment and simulation so that the comparison across all is possible.

The longitudinal Taylor microscale $\lambda$ is evaluated using its isotropic definition 
\begin{equation}
\label{eq:TaylorLength}
\lambda^2 = 15 \nu \frac{u'^2}{\epsilon} = \frac{u'^2}{\langle(\partial \widetilde{u} / \partial x)^2 \rangle},
\end{equation}  

\noindent where $\nu$ is kinematic viscosity and $\epsilon$ the turbulent kinetic energy dissipation rate per unit mass. Assuming isotropy at small scales and Taylor's frozen turbulence hypothesis, $\epsilon$  is estimated from 
\begin{equation}
\label{eq:kinetic energy dissipation rate}
\epsilon = 15 \nu \left\langle\left(\frac{\partial{\widetilde{u}}}{\partial{x}}\right)^{2}\right\rangle.
\end{equation} 

\noindent For a discrete time series, the calculation of derivative of $\widetilde{u}$ is often associated with noise and requires a high resolution. In order to account for this, the derivative is determined using eq. \eqref{eq:derivative approximation}
\begin{equation}
\label{eq:derivative approximation}
\left\langle\left(\frac{\partial{\widetilde{u}}}{\partial{x}}\right)^{2}\right\rangle = \int^{k_{max}}_{k_{min}} k^2 E(k) dk,
\end{equation} 

\noindent where $k_{min}$ is the lowest wavenumber determined from the specific time series and the sampling frequency and $k_{max}$ is the cutoff wavenumber. 
This method is valid for truly isotropic turbulence. However, this approximation is very sensitive to the spectral shape at high wave numbers. 
The frequency response of the hot-wire measurement is high enough to resolve energy spectrum up to $k\eta = 1$ and above. $\eta$ is the Kolmogorov length scale which can be determined from the isotropic estimate of dissipation rate in eq. \eqref{eq:kinetic energy dissipation rate}:
\begin{equation}
\label{eq:Kolmogorov length scale}
\eta = \left(\frac{\nu^3}{\epsilon}\right)^{\frac{1}{4}}.
\end{equation} 

\noindent On the other hand, the resolution of the numerical simulation is significantly reduced up to $k\eta = 0.1$ due to the spatial filtering effect resulting from the RANS modeling. Therefore, in the case of simulation $\epsilon$ and consequently also the estimated length scale $\eta$ is significantly underestimated by up to 35\%. 

\begin{figure*}
\centering
\subfloat[]{\label{fig:IntegralLength_CL}\includegraphics[width=0.425\textwidth]{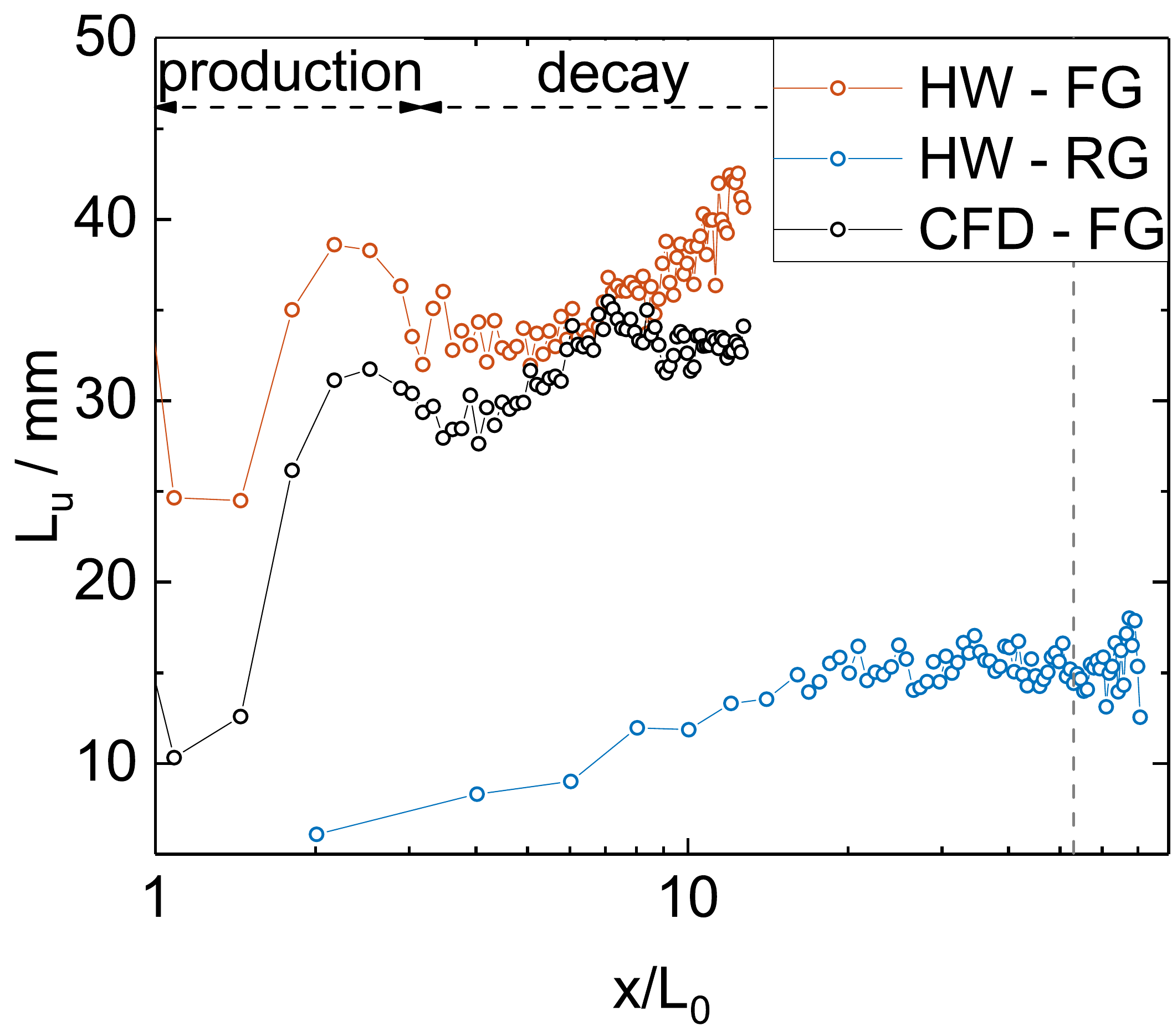}}  
\subfloat[]{\label{fig:TaylorLength_CL}\includegraphics[width=0.422\textwidth]{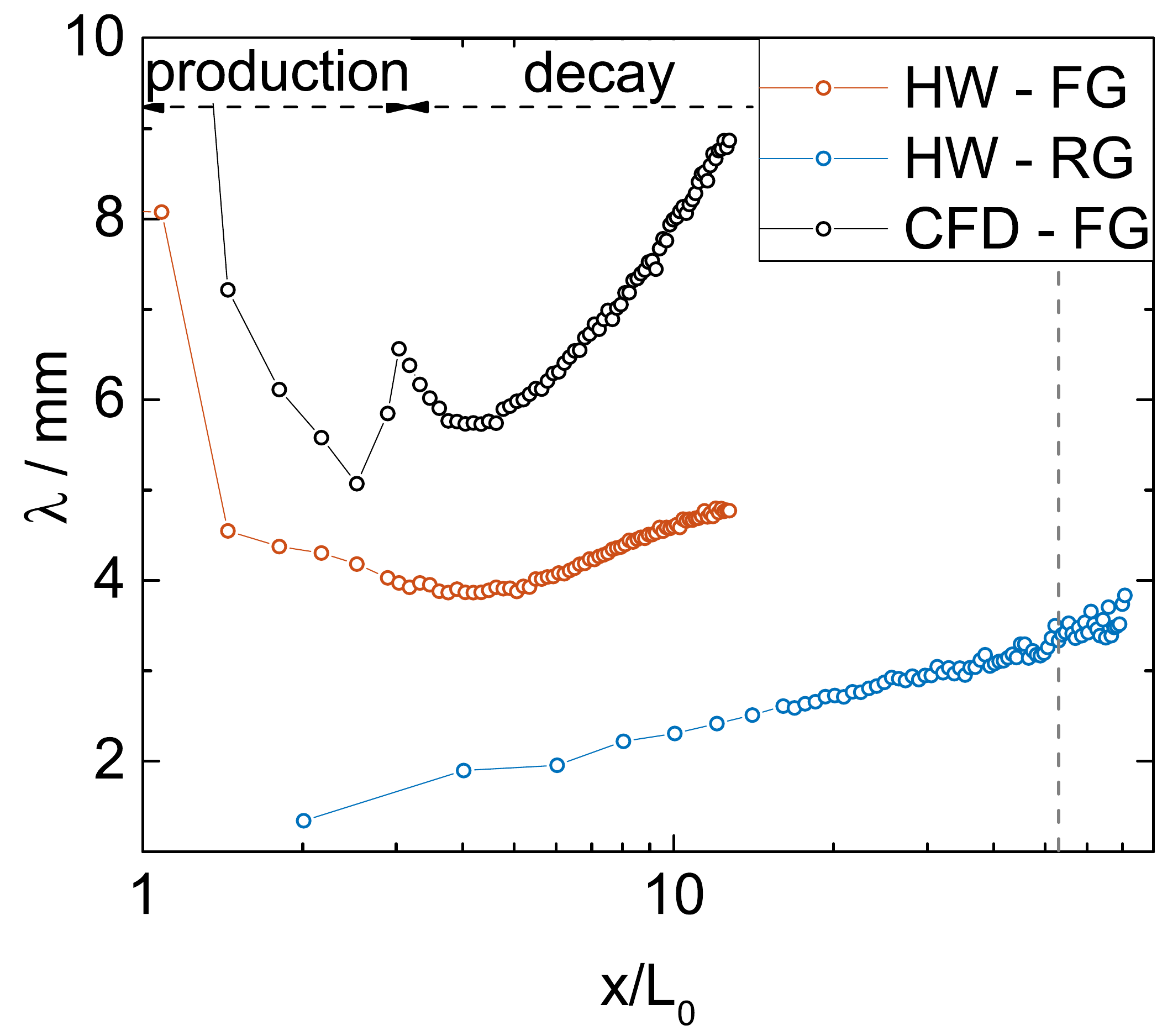}}
\caption{Spatial evolution of longitudinal integral scale $L_u$ (a) and Taylor microscale $\lambda$ (b) along the centerline. For $x/L_0$ larger than the position marked with the vertical dashed line the turbulence generated by the regular grid is dominated by the background turbulence of the wind tunnel.}
\label{fig:turb_length_scales}
\end{figure*}

In Figure \ref{fig:turb_length_scales} the spatial evolution of the longitudinal integral length scale $L_u$ (eq. \eqref{eq:integralLength_spec}) and Taylor microscale $\lambda$ (eq. \eqref{eq:TaylorLength}) as functions of the distance $x/L_0$ downstream of the different grids along the centerline is plotted from experimental and computational data.

Figure \ref{fig:turb_length_scales} show that the spatial evolution of $L_u$ and $\lambda$ for the regular and fractal grid are significantly different. Both $L_u$ and $\lambda$ generated by the fractal grid are consistently larger than the regular grid because the biggest grid bar length $L_0$ is smaller for regular than for fractal grid. 
For the regular grid profile of $\lambda$ show a monotonic increase in the downstream direction. $L_u$ exhibit the same behavior as $\lambda$ for $x/L_{0}<20$, whereas for $x/L_{0}>25$ the integral length scale appears independent of the distance to the grid. For $x/L_0$ larger than the position marked with the vertical dashed line $L_u$ shows considerably more scatter due to the wind tunnel's dominant influence of background turbulence.
For the fractal grid Figure \ref{fig:turb_length_scales} displays that, there are two regions characterizing the profiles of $L_u$ and $\lambda$. In the production region, the integral length scale increases and the Taylor length scale decreases until both reach a local maximum respectively minimum at a position close to ($x/L_{0}=3.2$), where turbulent kinetic energy peaks. Since the flow is highly intermittent and has not organized itself from distinct wakes into homogeneous turbulence in the production region, the estimated length scales have limited significance. Further downstream in the turbulence decay region, Figure \ref{fig:turb_length_scales} indicates that $L_u$ and $\lambda$ may be constant for some distance behind the turbulent kinetic energy peak and then slightly increase in the downstream direction.   
The same observation about the course of $L_u$ and $\lambda$ made for the experimental measurements can be made about numerical simulation. Note that the estimated integral length scale from the computation is consistently lower than the values corresponding to the experiment. Although the spectra predicted by CFD show a good agreement with the experiment in the low-frequency region, but due to the slightly lower mean velocity (\ref{chap:Mean streamwise velocity distribution}) and higher turbulent kinetic energy (\ref{sec:Decay power law of the turbulence intensity}) $L_u$ will be underestimated by eq. \eqref{eq:integralLength_spec}. Moreover, the CFD results predict significantly larger and faster increasing $\lambda$ in the decay region. The reason for this is, that due to the RANS modeling $\langle(\partial \widetilde{u} / \partial x)^2 \rangle$ is underestimated by eq. \eqref{eq:derivative approximation} and therefore $\lambda$ is overestimated by eq. \eqref{eq:TaylorLength}. Nevertheless, the basic development of $L_u$ and $\lambda$ is similar to the experiment.

\subsubsection{Normalized energy dissipation rate $C_\varepsilon$}
\begin{figure*}
	\centering
	\subfloat[]{\label{fig:ratio}\includegraphics[width=0.32\textwidth]{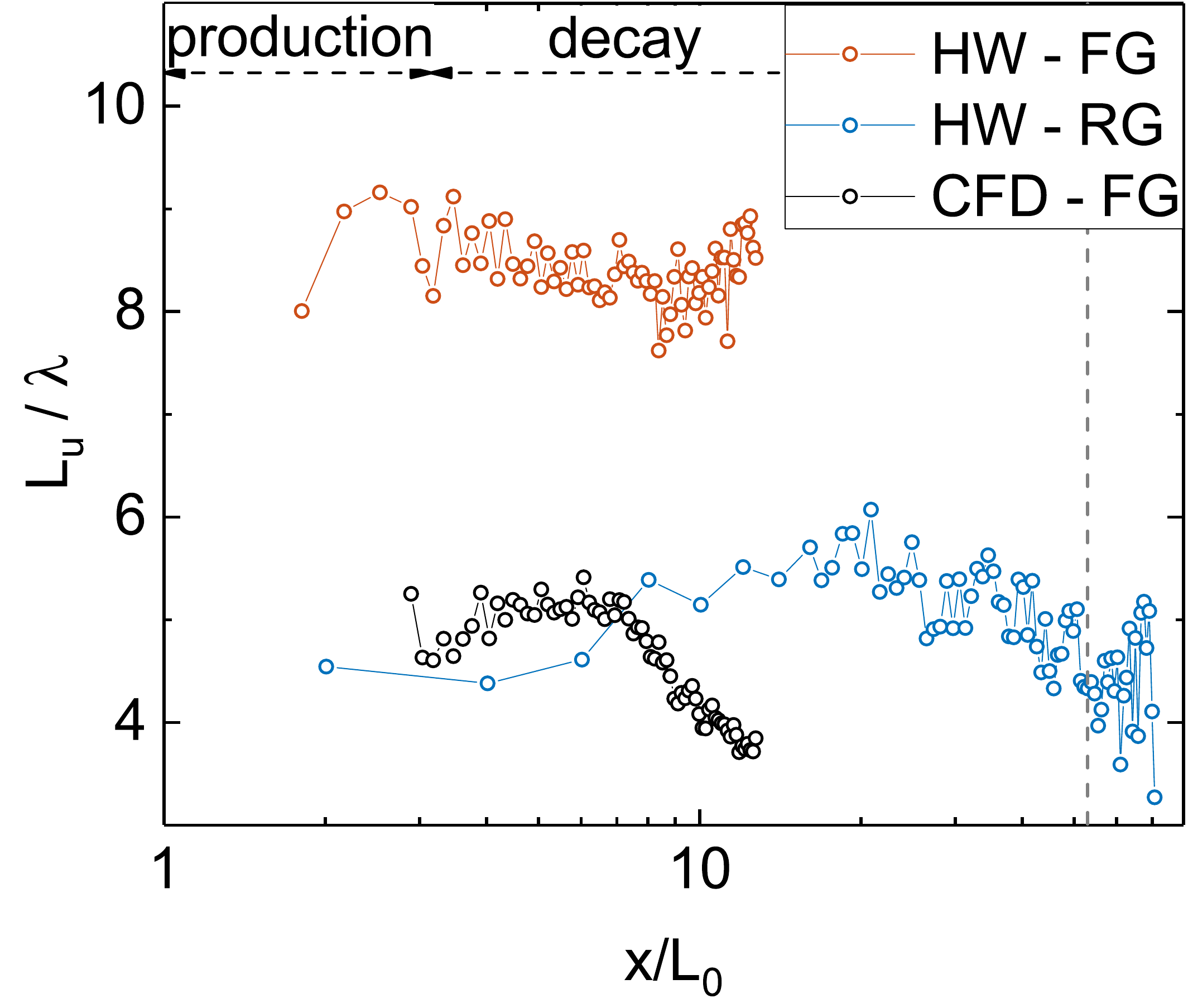}}  
	\subfloat[]{\label{fig:Re_lambda}\includegraphics[width=0.335\textwidth]{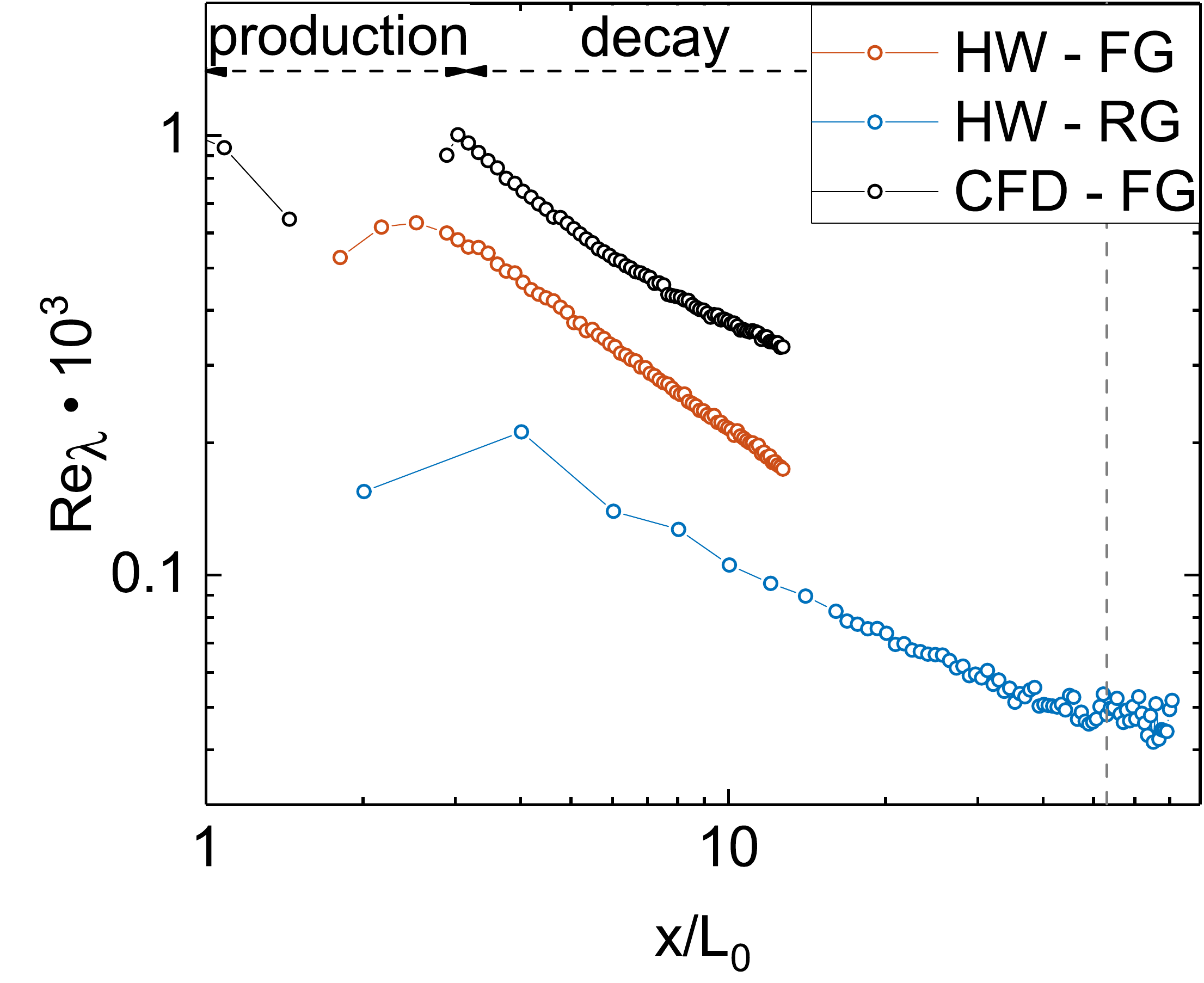}}
	\subfloat[]{\label{fig:Cepsi}\includegraphics[width=0.33\textwidth]{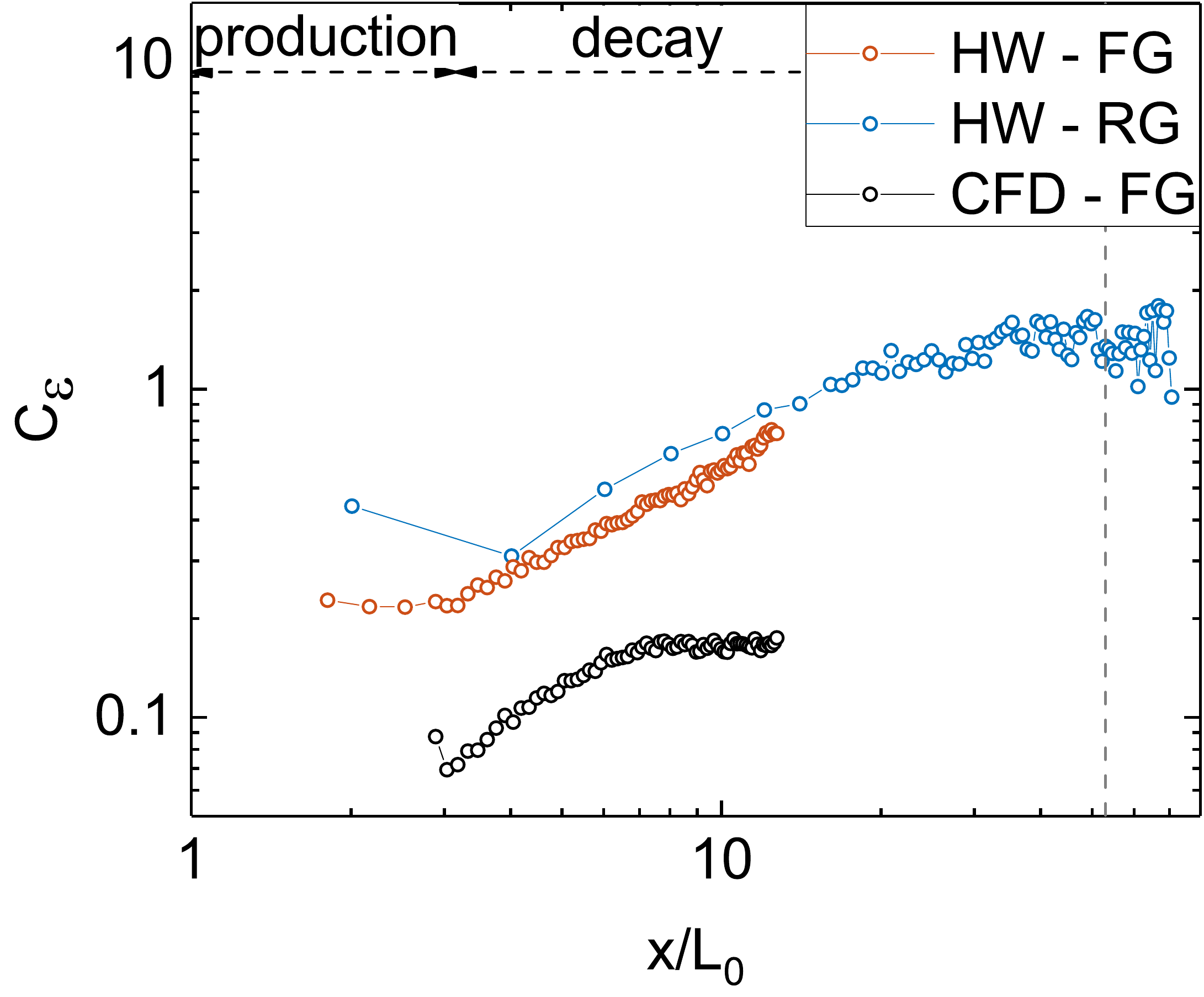}}
	\caption{(a) ratio between integral and Taylor length scales, (b) downstream profiles of Reynolds number based on Taylor length scale and (c) normalized energy dissipation rate $C_\varepsilon$ over $x/L_0$ along the centerline are plotted. For $x/L_0$ larger than the position marked with the vertical dashed line the turbulence generated by the regular grid is dominated by the background turbulence of the wind tunnel.}
	\label{fig:ratio_cepsi}
\end{figure*}

In this section, non-equilibrium characteristics and normalized energy dissipation rate $C_\varepsilon$ (see eq. \eqref{eq:dissipation coefficient}) are investigated. Figure \ref{fig:ratio} shows $L_u$ in units of $\lambda$ over $x/L_0$. This ratio indicates the separation between large and small scales of turbulent fluctuations. For the hot-wire measurements performed for the fractal grid, the ratio of $L_u$ and $\lambda$ is almost constant in the decay region, and reaches a value of approximately 8.5.
When considering instead the measurements performed for the regular grid when $x/L_0 < 20$ values of $L_u/\lambda \approx 5$ can be detected. For larger distances to the grid the ratio of integral length scale to Taylor microscale slowly decreases with $Re_{\lambda}$.
For the simulation values of $L_u/\lambda$ are consistently lower due to the fact that the Taylor length scale is overestimated. Also the region ($4<x/L_0<7$) where the ratio is almost constant is significantly smaller compared to the experimental results and decreases further downstream. Thus, the experimental results seem to agree quantitatively better with the observations made in \cite{Hurst_2007,Seoud_2007, Vassilicos_Mazellier_2010} which also report a constant ratio $L_u/\lambda$ in the decay region. 

In isotropic turbulence $C_\epsilon$ is a dimensionless constant that is independent of the Reynolds number and is evaluated from (using eq. \eqref{eq:TaylorLength} and eq. \eqref{eq:Re_lambda})
\begin{equation}
\label{eq:dissipation coefficient}
C_\epsilon=\frac{\epsilon L_u}{u'^3}=15 \frac{L_u/\lambda}{Re_\lambda}, 
\end{equation}  

\noindent where $Re_{\lambda}$ is the local Reynolds number defined as 
\begin{equation}
\label{eq:Re_lambda}
Re_{\lambda} = \frac{u'\lambda }{\nu}.
\end{equation}  

\noindent $Re_{\lambda}$ depends on the intrinsic properties of the flow and is often used to characterize the states of turbulent flow. Figure \ref{fig:Re_lambda} demonstrates the local Reynolds number as function of distance $x/L_0$ to the grid. It can be seen that the fractal grid produces a much larger local Reynolds number compared to the regular grid. For the fractal grid, an $Re_{\lambda}$ range of 170 - 630 is attained by varying downstream locations from the fractal grid (45 - 210 for the regular grid). The profile of $Re_{\lambda}$ for the simulation is similar to the experimental one, $Re_{\lambda}$ deceases as the turbulence decays, whereas the magnitude of $Re_{\lambda}$ differ considerably. This is due to the fact that, the Taylor length scale is significantly overestimated for the computational data (see Section \ref{sec:turbulent length scales}). In this context, the abovementioned evaluation errors in estimating $L_u$ and $\epsilon$ should be kept in mind when evaluating $C_\epsilon$. Moreover, the results presented in \cite{bos2007spectral,puga2017normalized} show that even a small deviation from isotropy or homogeneity could cause inconsistencies in the computation of $C_\epsilon$. Furthermore, this analysis shows that the fractal grid generated turbulence is characterized by $Re_{\lambda}>170$ for all measurements. Due to the large $Re_{\lambda}$ it can be reasonably assumed that the small dissipative scales are isotropic. For the regular gird, there may be low Reynolds number effects in the far-grid region, where a sufficiently large enough separation of large and small scales is not present.

Figure \ref{fig:Cepsi} presents evolution of the energy dissipation coefficient $C_\varepsilon$ against $x/L_0$ for the different flows. In the region where $C_\varepsilon$ is constant the relation $L_{0}/\lambda \propto Re_{\lambda}$, which is a direct expression of the Richardson-Kolmogorov cascade, must be fulfilled (see eq. \eqref{eq:dissipation coefficient}). This relation illustrates the enlargement of separation between large and small scales with increasing Reynolds number.
Comparing the plots in Figure \ref{fig:ratio_cepsi}, two distinct behaviors in the evolution of $C_\varepsilon$ are visible. In the far-grid region of the regular grid ($x/L_0 > 25$) there is good agreement with $L_{0}/\lambda \propto Re_{\lambda}$. In this region, $C_\varepsilon$ approach approximately a constant value, where $Re_{\lambda}$ is sufficiently high and is decreasing. 
This assumption of turbulence theory does not hold in the near-grid region ($x/L_0 < 20$), where the turbulence decays. Instead, both fractal and regular grid profiles of $C_\varepsilon$ increases with downstream distance when $x/L_0 < 20$. In this region, $Re_{\lambda}$ decreases, however the ratio $L_{0}/\lambda$ remains almost constant and therefore the measurements obtained in the non-equilibrium region of turbulence and do not conform with the scaling eq. \eqref{eq:dissipation coefficient}. Note that the energy spectra are characterized by a wide scaling behavior close to the $-5/3$ Kolmogorov law in this region. These findings are in good quantitative agreement with those of previous studies \cite{valente2012universal,hearst2014decay}.
Note that, for the simulation the same feature can also be observed, but the non-equilibrium range ($3<x/L_0<7.5$) is smaller compared with the experimental results. The assumption $C_\varepsilon = const$ is valid in the region $x/L_0>8$.

\section{Conclusions and outlook}
The turbulent flow generated by a space-filling square N3 fractal grid was investigated numerically and experimentally.
DDES simulations of the fractal grid with a low number of fractal iterations were conducted for the first time. In addition, we did an extensive statistical study and a direct comparison between the experimentally and numerically acquired time series in order to investigate and compare one-point statistics. 
The CFD simulations agree well with the experimental results obtained by PIV and hot-wire measurements. The agreement is qualitative and quantitative and extends to higher moments of fluctuating velocities. Both CFD and experiments show that the generated turbulent flow behind the fractal grid is characterized by non-Gaussian properties in the production region and becomes nearly Gaussian in the decay region. This strong non-Gaussian behavior distinguishes the fractal grid from the regular grid flow. 
The validation of the computational results is a first step in determining the important scales and mechanisms governing the turbulent flow in the wake of fractal grids. Furthermore, our results indicate that the turbulent flow in the production region is in a transition state from bimodal motions to developed turbulence with an increased distance to the grid.
An interesting point for investigation in the future will be to set these findings of the near-grid region in the context of the non-traditional turbulent features of the decaying region. It was found that in the region of decaying turbulence just after $x_{peak}$ turbulence has special properties, like independencies of the turbulent structure from Reynolds number \cite{Vassilicos_Mazellier_2010,  vassilicos2015dissipation,Vassilicos_Stresing_2010}. Thus the open question has to be investigated how different turbulent structures and their complex interaction will pass over in fully developed turbulence and how far a developing turbulent state differs due to its generating mechanism. Definitely, such questions are of high relevance for applications. If turbulence is used in applications, it has to pass usually through a transition mechanism. For such a future investigation, it will be necessary to extend the analysis of the numerically and experimentally acquired time series by higher-order point statistics. These analyses will provide an enhanced insight into the statistical properties of multi-scale generation of turbulence using a fractal grid and the suitability of CFD models (in our case DDES: Delayed Detached Eddy Simulation with a Spalart-Allmaras background turbulence model) to characterize this turbulence.

\begin{acknowledgments}
We gratefully acknowledge the computer time provided by the Facility for Large-Scale Computations in Wind Energy Research (FLOW) of the University of Oldenburg. The authors also thank the German Bundesministerium f\"r Umwelt, Naturschutz, Bau und Reaktorsicherheit (BMUB), Bundesministerium f\"ur Wirtschaft und Energie (BMWi) and the Ministerium f\"ur Wissenschaft und Kultur (MWK) which financed this project.
\end{acknowledgments}

\nocite{*}
\bibliography{biblio}

\end{document}